\documentclass[useAMS,usenatbib]{mn2e}

\newcommand{\be}{\begin{equation}}
\newcommand{\ee}{\end{equation}}
\newcommand{\gs}{\;\raisebox{-.8ex}{$\buildrel{\textstyle>}\over\sim$}\;}
\newcommand{\ls}{\; \raisebox{-.8ex}{$\buildrel{\textstyle<}\over\sim$}\;}

\newcommand{\apj}{{\it ApJ, }}

\newcommand{\icar}{{\it Icarus, }}
\newcommand{\mnr}{{\it MNRAS, }}

\newcommand{\ana}{{\it A\&A, }}

\title[Migration-induced resonances]
{On the migration-induced resonances in a system of two planets
with masses in the Earth mass range}
\author[J. C. B. Papaloizou and E. Szuszkiewicz]
{J. C. B. Papaloizou$^{1,2}$
\&
E. Szuszkiewicz$^{3}$ \\
$^{1}$Astronomy Unit, Queen Mary,  University of London, Mile End Rd,
London E1 4NS, England,\\
$^{2}$Department of Applied Mathematics and Theoretical Physics,
Centre for Mathematical Sciences,\\
Wilberforce Road,
Cambridge CB3 0WA,
United Kingdom\\
$^{3}$Institute of Physics and CASA*, University of Szczecin, Wielkopolska 15,
70-451 Szczecin, Poland}

\begin{document}

\date{Accepted;  Received; in original form}

\pagerange{\pageref{firstpage}--\pageref{lastpage}} \pubyear{2005}

\maketitle

\label{firstpage}

\begin{abstract}
We investigate orbital resonances expected to arise when a system of
two planets, with masses in the range  1-4 $M_{\oplus}$, undergoes convergent
migration while embedded 
in a section of gaseous disc where the flow is laminar. 
We consider surface densities
corresponding to $0.5 -4$ times that expected for a minimum mass solar 
nebula at $5.2$~AU.
For the above mass range the planets undergo type I migration. 
Using hydrodynamic simulations we find
that when the configuration is such that convergent migration occurs
the planets can become locked in a first order 
commensurability for which the period ratio is $(p+1)/p$ with $p$ being an
integer and migrate together maintaining it for many orbits.
Slow convergent migration results in commensurabilities with small $p$ such
as 1 or 2. Instead,
when the convergent migration is relatively rapid as tends to occur
for disparate masses, higher $p$ commensurabilities are realized such as
4:3, 5:4, 7:6 and 8:7. However, in these cases the dynamics is found to 
have a stochastic
character with some commensurabilities showing long term instability
with the consequence that several can be visited during the course 
of a simulation.
Furthermore the successful attainment of commensurabilities is also 
a sensitive function
of initial conditions. When the convergent migration is slower,
such as occurs in the equal mass case, lower $p$ commensurabilities 
such as 3:2
are obtained which show much greater stability. 
 
 \noindent  Resonant capture leads to a rise in eccentricities
that can be predicted using a simple analytic model, 
that assumes the resonance is isolated,  constructed  in this paper.
We find that, once the  commensurability has been established,
 the system with an  8:7 commensurability 
is fully consistent with this 
prediction. 
 
 \noindent We find that very similar behaviour is found when the systems 
are modeled
 using an N body code with simple prescriptions for the disc planet 
interaction.
 Comparisons with the hydrodynamic simulations indicate reasonably 
good agreement
 with predictions for these prescriptions obtained 
 using the existing semi-analytic theories of type I migration.
 
 \noindent We have run our hydrodynamic simulations
for up to $10^4-10^5$ orbits of the inner planet. Longer times
could only be followed in the simpler  N-body approach. Using that, 
we found that on the one hand  an  8:7
resonance established in a hydrodynamic simulation could be maintained 
for more than $10^6$ orbits. On the other hand  other similar cases  
show instability leading to 
another resonance and ultimately a close scattering. 
 
\noindent There is already one known example of a system with nearly
equal masses in the several Earth mass range, namely  the two pulsar
planets in PSR B1257+12 which are intriguingly, in  view of the results 
obtained here,
 close to a 3:2 commensurability.
This will be considered in a future publication.

\noindent Future detection of other systems with masses in the Earth 
mass range that display 
orbital  commensurabilities will  give useful information on the role  
and nature of orbital  migration
in planet formation.

\end{abstract}


\section{Introduction}

The increasing number of extrasolar multi-planet systems, their diversity,
and dynamical complexities provide a strong motivation to study the 
evolution and stability of such systems. One of the important features
connected with  planetary system evolution is the occurrence of mean
motion resonances, which may relate to conditions
at the time of or just after the process of  formation.
There are  some well known examples of systems of celestial bodies exhibiting 
mean motion resonances both within our Solar System (Neptune and Pluto; Io,
Ganymede and Europa (see eg. Goldreich, 1965)~) 
and outside such as   Gliese 876 (Marcy et al., 2001),
 HD 82943 (Mayor et al., 2004) and 55 Cancri (McArthur et al., 2004).
As the latter examples involve planets with
masses in the Jovian mass range, most of the investigations
appropriate to extrasolar planets  have
focused on giant planets.  
However, it is likely that planetary  systems around other stars  
may harbor planets with masses in the Earth mass range  as well. These 
should be  revealed by  future space-based missions, such as Darwin,
COROT, Kepler, SIM and TPF.

Meanwhile, it is important to establish the main  features of the 
evolution of low mass planets embedded in  a gaseous disc and  in 
particular  to determine the types of resonant  configurations that 
might arise when a pair of such planets evolves together.
The disc planet interaction naturally produces orbital migration 
through the action of tidal torques \citep{gt80,lpap86}
which in turn  may lead to an orbital resonance in a many planet system 
eg. \citep{spn01,lp02}.
For low mass planets the disc undergoes small linear 
perturbations that induce density waves that propagate away from the
planet. The angular momentum these waves transport away 
results in rapid orbital  migration  called type I migration \citep{wa97}. 
In this type of migration, when the disc is laminar
and inviscid,  the planet is embedded
and  the surface density profile of  the disc remains
approximately unchanged. The rate of migration is proportional to mass
of the planet  and the timescale of inward  migration on a circular
orbit  can be estimated 
for a disc with  constant surface  density   
to be given by (see \citet{ttw02})
 
\begin{equation}
\tau_r = \left|{r_p\over {\dot r_p}}\right |
=W_m{M_* \over m_{planet}} {M_* \over \Sigma_p r_p^2}
\left({c \over r_p \Omega_p}\right)^2 \Omega_p^{-1} 
\label{MIG}
\end{equation}
Here $M_*$ is the mass of the central star, $m_{planet}$ is the mass 
of the planet orbiting  at distance $ r = r_p$, $\Sigma _p$ is the 
disc  surface density at $r = r_p,$ $c$ and $\Omega_p$  are the
the local sound speed and 
angular velocity, $\Omega,$  at $r = r_p$ respectively.
The numerical coefficient $W_m =0.3704.$

\noindent It is important to note that type I migration appropriate
to a laminar disc may  lead to short migration times in standard
model discs, that may  threaten
the survival of protoplanetary cores  (\citet{wa97}).
However, in a  disc with turbulence driven by
the magnetorotational instability, the migration may be stochastic
and accordingly less effective (\citet{np04}).
Nonetheless there is considerable uncertainty as to the extent
of turbulent regions in the disc resulting from uncertainties
in the degree of ionization (eg. Fromang, Terquem \& Balbus, 2002)
so that type I migration appropriate to a laminar disc may operate
in some regions. We also note that 
because it is inversely proportional to
the disc surface density, the migration
time becomes long in low surface density regions.
 Accordingly, for this first study of resonant interactions
of planets in the mass range $1 - 30 M_{\oplus}$ embedded
in a gaseous disc, we shall consider
only  migration induced by a laminar disc.

\noindent 
The density waves excited by a low mass planet with small eccentricity
also lead to orbital circularization (eg. \citet{ar93,pl00})
at a rate that can be estimated to be given
by \citet{tw04}

\begin{equation}
t_c = {\tau_{r}\over W_c} \left({c\over r_p \Omega_p}\right)^2.
\label{CIRC}
\end{equation} 
Here, the numerical coefficient $W_c =0.289.$

It is expected  from equation (\ref{MIG}) that
two planets with different masses will migrate at different rates.
This has the consequence that their period ratio will evolve
with time and may accordingly attain and, 
in the situation where the migration is such that the orbits  
converge, subsequently become  locked in a mean motion resonance 
(eg. \citet{np02,kpb04}).

\noindent In the simplest case of nearly 
circular and coplanar orbits the strongest resonances are 
the first-order resonances which occur at locations
where the ratio of the two orbital periods can be expressed 
as the ratio of two consecutive integers, $(p+1) / p$,  with $p$   
being an integer.
As $p$ increases,  the two  orbits  approach each other and
the strength of the resonance increases. In addition, the 
distance between successive resonances  decreases as $p$ increases. 
The combination of these effects ultimately causes successive  
resonances to overlap and so, in the absence of gas,  leads to the  
onset of chaotic motion.

\noindent   
Resonance overlap  occurs when the difference of the semi-major 
axes of the two planets is below a limit  with  half-width given, 
in the case of two equal mass planets, by \citet{gl93} as
\begin{equation}
{\Delta a\over a} \sim {2\over 3p} \approx 2 
\left({m_{planet} \over M_*}\right)^{2/7},
\label{chao}
\end{equation}
with $a$ and $m_{planet}$ being the mass and  semi-major axis 
of either one of them respectively.
Thus for a system consisting a two equal four Earth mass planets 
orbiting a central solar mass we expect resonance overlap for 
$p \gs 8.$ Conversely we might expect isolated resonances in 
which systems of planets can be locked and migrate together 
if  $p  \ls 8.$

\noindent 
However, note that the above discussion does not
incorporate the torques producing  convergent migration 
or eccentricity damping and thus may not give a complete account 
of the forms of chaos that might be expected.
\citet{klg93} have discussed the case of small particles
migrating towards a much more massive planet and indeed 
conclude that chaotic behaviour is more extensive in the non 
conservative case.
This occurs because the higher $p$ resonances can be unstable
thus preventing long term trapping. The instability arises 
because, if locked in resonance, the orbit of a planet has some 
eccentricity. Making a slight perturbation to the orientation
of the orbit  produces an impulsive change to the semi-major 
axis at the next conjunction.
This in turn affects the phase of the next conjunction and 
the next impulsive change to the semi-major axis. For resonances 
of high enough $p$ this sequence results in instability and 
chaotic behaviour. \citet{klg93} give an estimate
of when this form of instability occurs as

\be 
p > 0.667\left({8\pi e m_{planet}\over 3 M_*}\right)^{-1/5} .
\label{chao1}
\ee
Here $m_{planet}/M_*$ refers to the mass ratio for the most 
massive planet, and $e$ is the orbital eccentricity for the planet 
with the smallest mass. Taking 
values appropriate to our simulations of $e\sim 0.04,$ and 
$m_{planet}/M_* \sim 10^{-5},$ gives $p \gs 8.$
This is similar to the non dissipative estimate.

\noindent 
When this form of instability operates, long term  stable trapping 
in commensurabilities is not possible.  However, a system can remain 
in one for a long time before moving into another higher $p$ 
commensurability. Furthermore detailed outcomes are very sensitive
to input parameters, which is characteristic of chaotic motion. Slight 
changes can alter the sequence of commensurabilities a system resides 
in making the issue of their attainment acquire a probabilistic character.
Both our hydrodynamic simulations and N body calculations show evidence 
of this behaviour.
Although there may be islands of apparent stability, a system in this 
regime  may ultimately undergo a scattering and exchange of 
the orbits of the two planets so that we should focus on stable 
lower $p$ commensurabilities as being physically possible planetary 
configurations.
The arguments given above suggest that these must have $p \ls 8.$
Our calculations indicate that the limit is even smaller.

\noindent 
Hydrodynamic simulations of disc planet interactions, in which 
the discs are  modelled as flat  two dimensional objects with laminar 
flow governed by the Navier Stokes equations  and which incorporate 
a migrating  two giant planet system  which evolves into
a 2:1 commensurability have been performed by \citet{kl00}, 
\citet{spn01} and successfully applied to the GJ876 system.
In addition to performing additional simulations,
\citet{pa03}  has developed an analytic model describing
two planets migrating in resonance with arbitrary eccentricity.
In this model the eccentricities are determined as a result of the
balance between migration and orbital circularization. However, disc 
tides were considered to act on the outer planet only.
Capture of  giant planets into resonance  has also
been recently  investigated numerically  by \citet{kpb04}.  

\noindent 
In this paper we  extend the types of two planet systems
considered to include those with orbital resonances formed  
when the masses of the planets are $1 - 30M_{\oplus}.$
Then  for discs with aspect ratio $H/r = c/(r\Omega) \sim 0.05$ 
as in standard  protostellar disc models, the  orbital  migration  
will be induced by type I migration. 

\noindent 
In Section 2 we describe initial set up for our
calculations giving the properties of the system in which  two 
planets interact with each other and a gaseous disc.

\noindent 
In Section 3 we present the results of two dimensional numerical 
calculations showing how  convergent 
differential migration of two planets can lead to  resonance 
trapping followed by migration in which the resonance is maintained
for the duration of the simulation which is characteristically
around $2\times 10^4$ orbits.
By considering simulations with a range of planet masses and 
initial conditions, we are able to explore a variety
of commensurabilities and show the sensitivity
of the attainment of higher $p$ commensurabilities
to the input parameters, this  being indicative of stochastic 
behaviour.

\noindent 
In Section 4 we compare our results with  simplified N-body 
integrations which model the effects of the disc through the 
addition of terms to cause migration and eccentricity damping
(see eg.  \citet{spn01,lp02,al03}). The comparison enables us 
to calibrate these terms and then apply the much faster N-body 
calculations   to consider a  wider range of planet masses  
and  disc surface densities for a  longer evolution time than 
is possible for the hydrodynamic simulations. 

\noindent 
We summarize and discuss our results in the context
of potentially observable commensurabilities in Section 6.

\noindent 
In addition, we generalize the analytic model of \citet{pa03} 
to incorporate migration and circularization effects for both 
planets and also to consider general first order commensurabilities 
in an Appendix. Thus this model can be applied
to systems such as those considered here for which the planets 
have comparable mass and disc interactions may be important for 
either of them.
  
\section{Numerical simulations of  migrating planets in resonance}

We have performed simulations of two interacting planets
together with an  accretion disc with which they also interact.
The simulations performed here are of the same general  type as 
that performed by \citet{spn01} of the resonant coupling in the GJ876 
system induced by orbital migration caused by interaction with the disc.
For details of the numerical scheme and code adopted see \citet{npmk00}.
For the simulations performed here we adopt $n_r = 384,$ and 
$ n_{\varphi} = 512$ equally spaced grid points in the radial and 
azimuthal directions respectively.

We  use a system of units in which the unit of mass is the central 
mass $M_*,$ the unit of distance is the initial   semi-major axis 
of the inner planet, $r_{2},$ and the unit of time is 
$2\pi(GM_*/r_{2}^3)^{-1/2},$ this
being  the orbital period on the initial orbit of the inner planet.
When, as adopted below,  $r_2$ corresponds to 1AU this unit of time 
is one  year.
In dimensionless units the inner boundary of the computational 
domain was at $ r = r_{min} =0.33,$ and the outer boundary at 
$r= r_{max} = 4.$ In all simulations, the disc model is locally
isothermal with aspect ratio $H/r = 0.05$ and the kinematic viscosity 
is set to zero.

\subsection{Initial configuration and computational set up}

\begin{figure}
\vskip 5cm
\includegraphics{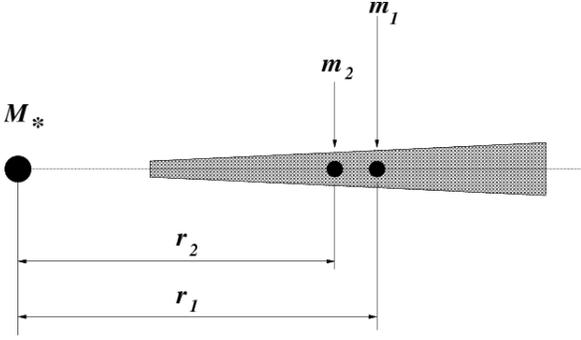}
\caption{\label{fig1}{The initial configuration: two planets with
masses $m_{1}$ and $m_{2}$, respectively, in circular orbit around
a central star with mass $M_*$ at distances $r_{1}$ and $r_{2}$
are embedded in a gaseous disc.
        }}
\end{figure}

Our initial set up shown in Figure \ref{fig1} includes the central 
star with a mass $M_*$ and two orbiting planets with masses $m_{1}$ 
and $m_{2}$ respectively. The two planets
are embedded in the disc which is a source of planet orbit  migration.
They are initialized on circular orbits with
the central mass  taken  to have a fixed value of 1 solar mass.
The gravitational potential was softened with softening parameter 
$b =0.8H.$ This results in the formation of an equilibrium atmosphere 
around the embedded planet which then does not accrete.
This softening also allows an adequate representation of type 
I migration in two dimensional discs
(see eg. \citet{np04}).
The disc in which planets are initially  embedded has the
initial surface density, $\Sigma(r),$ profile  specified by
\[ \Sigma(r) = \left\{
\begin{array}{ll}
0.1\Sigma_0(15(r-r_{min})/r_{min} +1) \  \mbox {if $r \geq r_{min}$ and }  \\
                             \  \mbox { \ \  $ r \leq 8r_{min}/5$} \\
\Sigma_0 \  \mbox{if $r > 8r_{min}/5  $ and } \\
         \  \mbox{ \ \   $r < 4.5r_{min}$} \\
\Sigma_0 \left(4.5r_{min} / r \right) ^{1.5} \ \mbox{if $r \geq 4.5r_{min}$}
\end{array}
\right . \]
where $r_{min}$ is the inner edge of the computational domain.
The planets are located in the flat part of this distribution. 
We use four different values for
the maximum value of the surface density, $\Sigma_0$, 
namely $2\times 10^3 (5.2 {\rm AU}/r_2)^2$ kg/m$^2$, the standard 
value attributed to the
minimum mass solar nebula at 5.2AU  which we denote $\Sigma_1$, 
then $\Sigma_{0.5}=0.5 \Sigma_1$, $\Sigma_{2}= 2 \Sigma_1$ and 
finally  $\Sigma_{4}= 4\Sigma_1$.
The radial boundaries were taken to be open.

\section{Two dimensional hydrodynamic simulations: A survey}

We have performed simulations of two interacting planets   embedded 
in a disc with which they also interact. This interaction leads to 
spiral wave excitation, energy and angular momentum  exchange 
between the planetary orbits and  the gaseous disc which in turn   
results in orbital migration and  eccentricity damping. We have  
carried out simulations for a variety of planet masses, initial 
orbital separations  and initial surface density scaling
in order to explore the possible outcomes. Because these simulations
are very computationally demanding, the extent of the survey is limited.
Nonetheless some characteristic features emerge.
 We describe some of 
these results below. 

\subsection{One  pair of planets in  three different resonances}

In order to investigate how the orbital evolution
depends on the surface density of the disc in which the planets 
are embedded we have considered   two planets with  masses 
 $m_1 = 4M_{\oplus}$  and  $ m_2 = 1M_{\oplus}.$  
We assume that  previous evolution brought the  two planets to 
a  configuration with circular orbits with  $r_{1}/r_{2} =1.2.$ This
separation is slightly smaller than that required for a strict 4:3
commensurability. The faster migration   expected for 
the outer planet (see equation (\ref{MIG})) will make
the distance between the  planets smaller and as a consequence 
of such convergent migration   it is possible 
that  they may become  locked in a  first order mean motion 
resonance (eg. \citet{go65}) such as 5:4, 6:5, 7:6, 
8:7, ..., $(p+1):p$, ...  and subsequently migrate together 
maintaining the commensurability for a considerable period of 
time or perhaps indefinitely, depending on how close the system 
is to a stochastic regime.

\subsection{Attainment of commensurability}

\citet{pa03} has given a simple approximate analytic solution for two
migrating planets locked in a 2:1 commensurability.
Only the outer planet was presumed to interact with the disc.
As a result of the resonant interaction,
the eccentricities of both planets grew with time until the effects 
of circularization due to disc tides balanced this.
For the situation studied here, both planets are embedded
in the disc and may have significant interaction with it.
Furthermore, higher $p$ commensurabilities  than 2:1 occur
for planet masses in the Earth mass range.
Thus in the appendix we generalize the solution so that it applies
to general first order commensurabilities and allows for disc 
tides to act on both planets. In particular we show that
when the eccentricities stop growing, provided they are not too
large,  they must satisfy

\begin{eqnarray}
{e_1^2\over t_{c1}} + {e_2^2\over t_{c2}}{m_2n_1a_1
\over m_1n_2a_2}
 -\left({e_1^2\over t_{c1}}-{e_2^2\over t_{c2}}\right)f
 =  \nonumber \\
\left({1\over t_{mig1}}-{1\over t_{mig2}}\right){f\over 3},    
\label{ejcons0}
\end{eqnarray}
where  $ f =m_2a_1/((p+1)(m_2a_1+m_1a_2)).$

Here the semi-major axes and  eccentricities of the two planets 
$(i=1,2)$ are $a_i$ and $e_i.$
The migration rates (assumed directed inwards) and circularization times 
induced by the disc tides are $t_{migi}  =  |n_i/{\dot n_i}|
=  |2a_i/(3{\dot a_i})| = 2\tau_r/3,$ and
 $t_{ci} =  |e_i/{\dot e_i}|$ respectively.
The mean motions are $n_i.$

In addition near a   $p+1:p$ resonance,  the resonant angles 
$\phi = (p+1)\lambda_1-p\lambda_2-\varpi_1, $ 
$\psi = (p+1)\lambda_1-p\lambda_2-\varpi_2$
and $\varpi_1 -\varpi_2$ librate about equilibrium values 
which could be near to
$0$ or $\pi$ mod $2\pi.$
Dissipative effects may be responsible for   some shift.  
Here the mean longitudes and the longitudes
of pericentre  of the two planets $(i=1,2)$ are
$\lambda_i$ and $\varpi_i$ respectively.
For giant planets like GJ876 these all librate about 
values close to zero.
However, for lower mass planets the libration of
 $\varpi_1 -\varpi_2$ in particular may be about  $\pi$ mod $2\pi$
(eg. \citet{spn01,lp02}).
All the simulations carried out here indicate libration about a value
closer to $\pi$ than $0.$

\noindent 
Which resonance  is established depends on the rate of relative 
migration. Roughly speaking one expects that locking occurs only 
if the relative migration is slow enough that the time to migrate 
through the resonance  is longer that a characteristic libration
period. The resonances become stronger as $p$ increases, 
so that increasing the relative migration rate tends to cause 
the attainment of higher $p$ commensurabilities.
However, if the $p$ becomes high enough
($p\sim 8$ for the planet masses considered)
the planets enter a chaotic  region (see 
equations(\ref{chao}-\ref{chao1})) of phase space
making commensurabilities unstable in the long term  (although 
a system may remain in the vicinity of one for a considerable time) 
and introducing sensitivity of detailed
outcomes to initial conditions.   In  such cases  a scattering may
ultimately occur that interchanges the positions of the planets.
Thus  very high surface densities with very rapid migration rates
do not favour attainment of very stable commensurabilities.

\begin{figure}
\vskip 8cm
\includegraphics{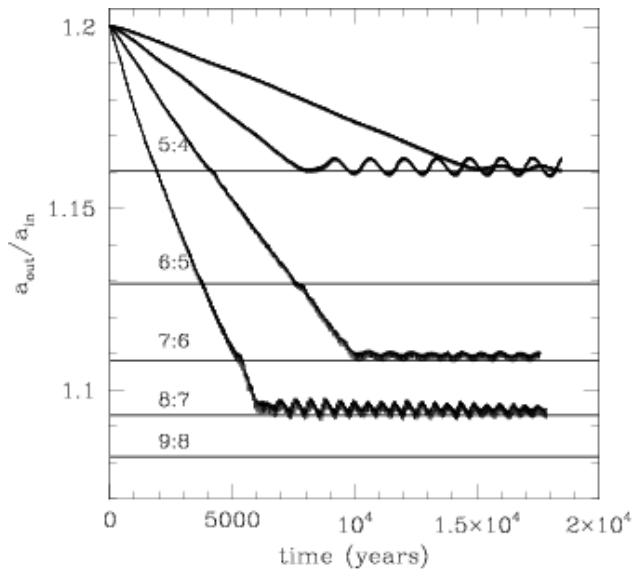}
\caption{\label{fig2}{The evolution of the ratio of semi-major 
axes for the two planets with masses $m_1= 4M_{\oplus}$ and 
$m_2 = 1M_{\oplus}.$ Starting from the lower curve and going 
upwards the curves correspond to the  initial surface
density scalings  $\Sigma_0 = \Sigma_{4}$, 
$\Sigma_0 =\Sigma_{2}$, $\Sigma_0 =\Sigma_{1}$, and  
$\Sigma_0 =\Sigma_{0.5}$ respectively.
 }}
\end{figure}

\noindent 
In order to study these effects, we have considered  the pair 
of planets evolving in discs with  $ \Sigma_0 = \Sigma_{0.5}$, 
$ \Sigma_0 = \Sigma_{1}$, $\Sigma_0 = \Sigma_{2}$ and 
$\Sigma_0 = \Sigma_{4}.$ The evolution was  followed
until a resonance  was established. The results are summarized 
in Figure \ref{fig2} where the evolution of the ratio of the   
semi-major axes, which starts from the value 1.2 for all four 
cases, is shown. 
The fastest migration (steepest
slope) corresponds to the  case   where
the two planets with $m_1 = 4M_{\oplus}$ and  $m_2 = M_{\oplus}$
are embedded in a disc with $\Sigma_0 = \Sigma_{4}$ and the slowest 
to a case with a disc with $\Sigma_0 = \Sigma_{0.5}$. The planets 
in the disc with $\Sigma = \Sigma_{4}$  become 
trapped in 8:7 resonance. If the disc surface density is two times
smaller then a 7:6 resonance is attained, if it is four or eight 
times smaller then the resonance attained is 5:4.
These results are fully consistent with the idea that higher
$p$ resonances are associated with faster relative migration
rates.

\noindent 
However, this situation corresponds to the evolution during first 
18000 years and it is not obvious that the planets will remain  in 
the  same resonances in their subsequent evolution.
This is because in addition to possibly being in a chaotic regime,
the ratio of migration time to local orbital period
varies with disc parameters and accordingly disc location.
Consequently,  the relative  migration might become relatively faster 
as the evolution proceeds resulting in a shift to a higher $p$ 
resonance. We comment that  a slowing of relative  migration, as long 
as it does not lead to a reversal,  will not result in 
a transition to a lower $p$ resonance. Therefore changes in
effective local migration rates will tend to  lead to
 higher $p$ commensurabilities and ultimately chaos.

\begin{figure*}
\begin{minipage}{175mm}
\vspace{220mm} 
\includegraphics{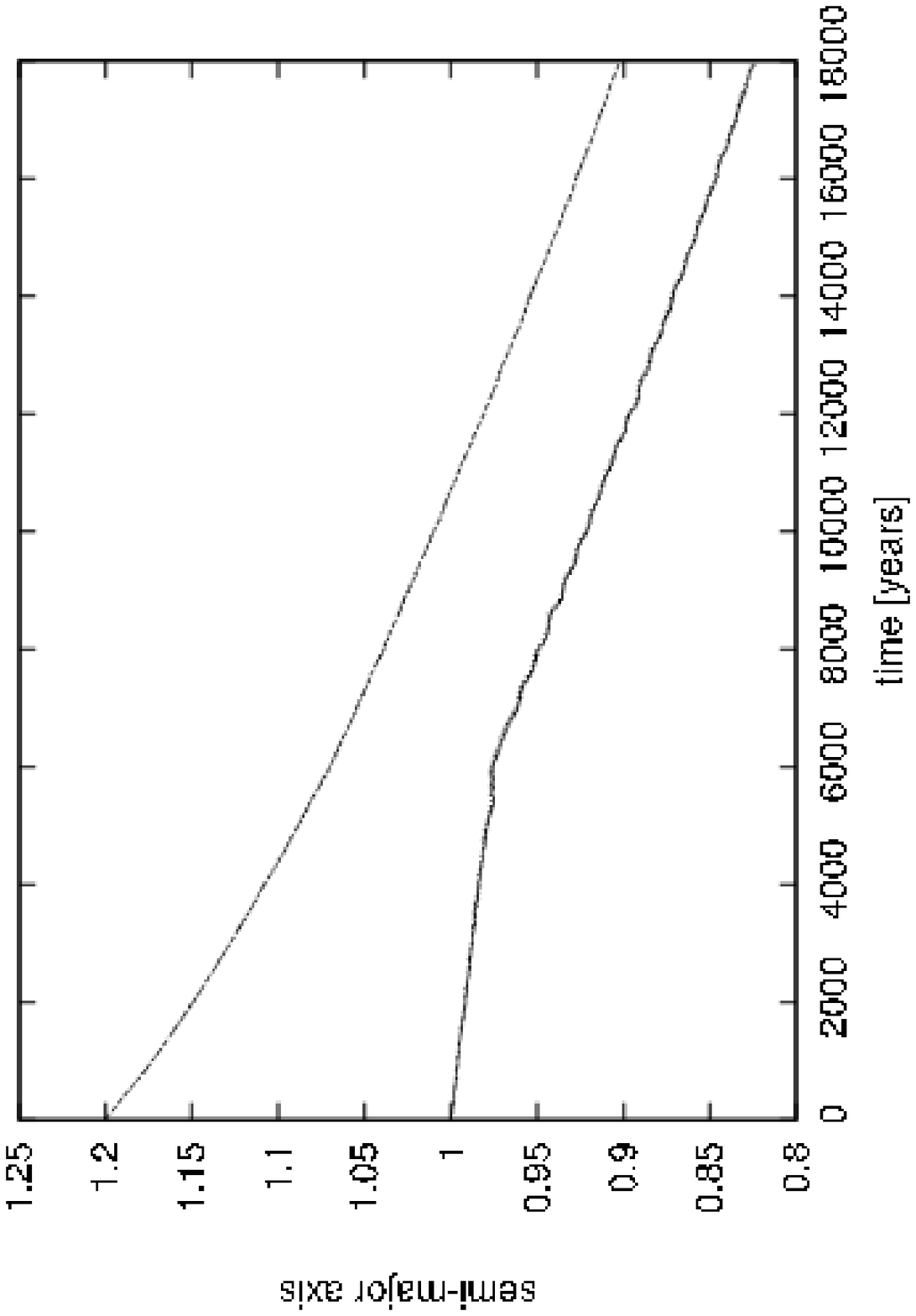}
\includegraphics{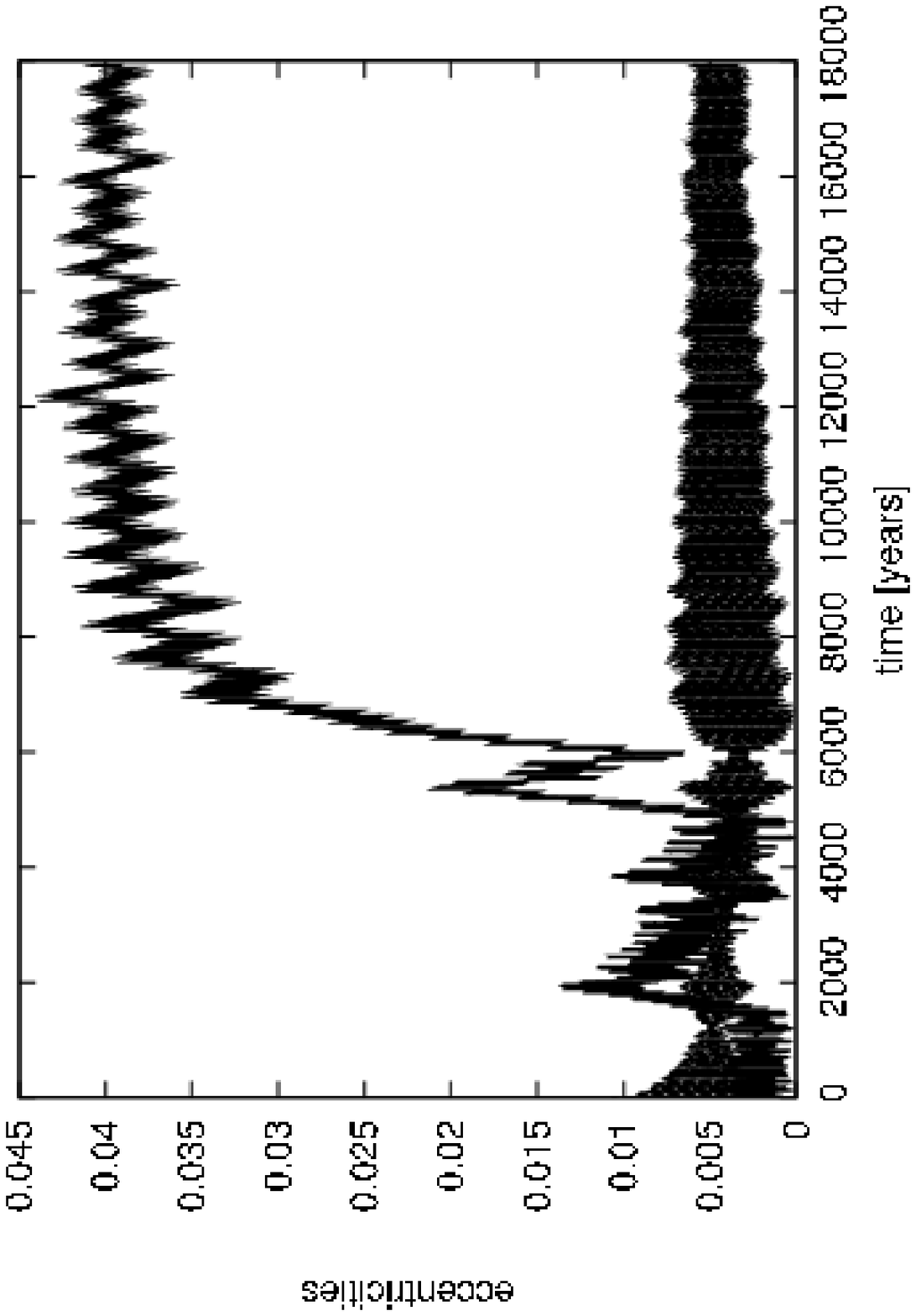}
\includegraphics{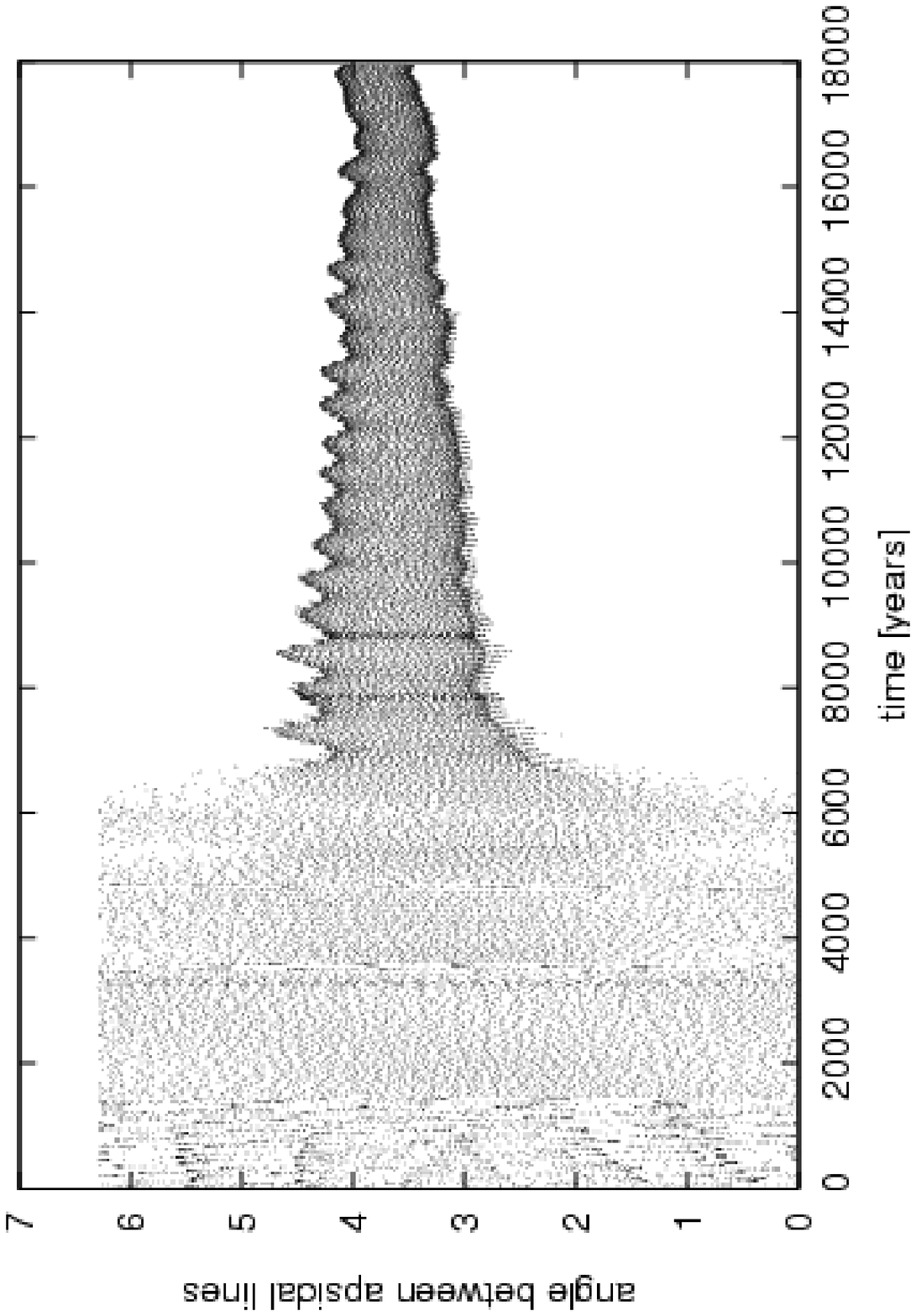}
\includegraphics{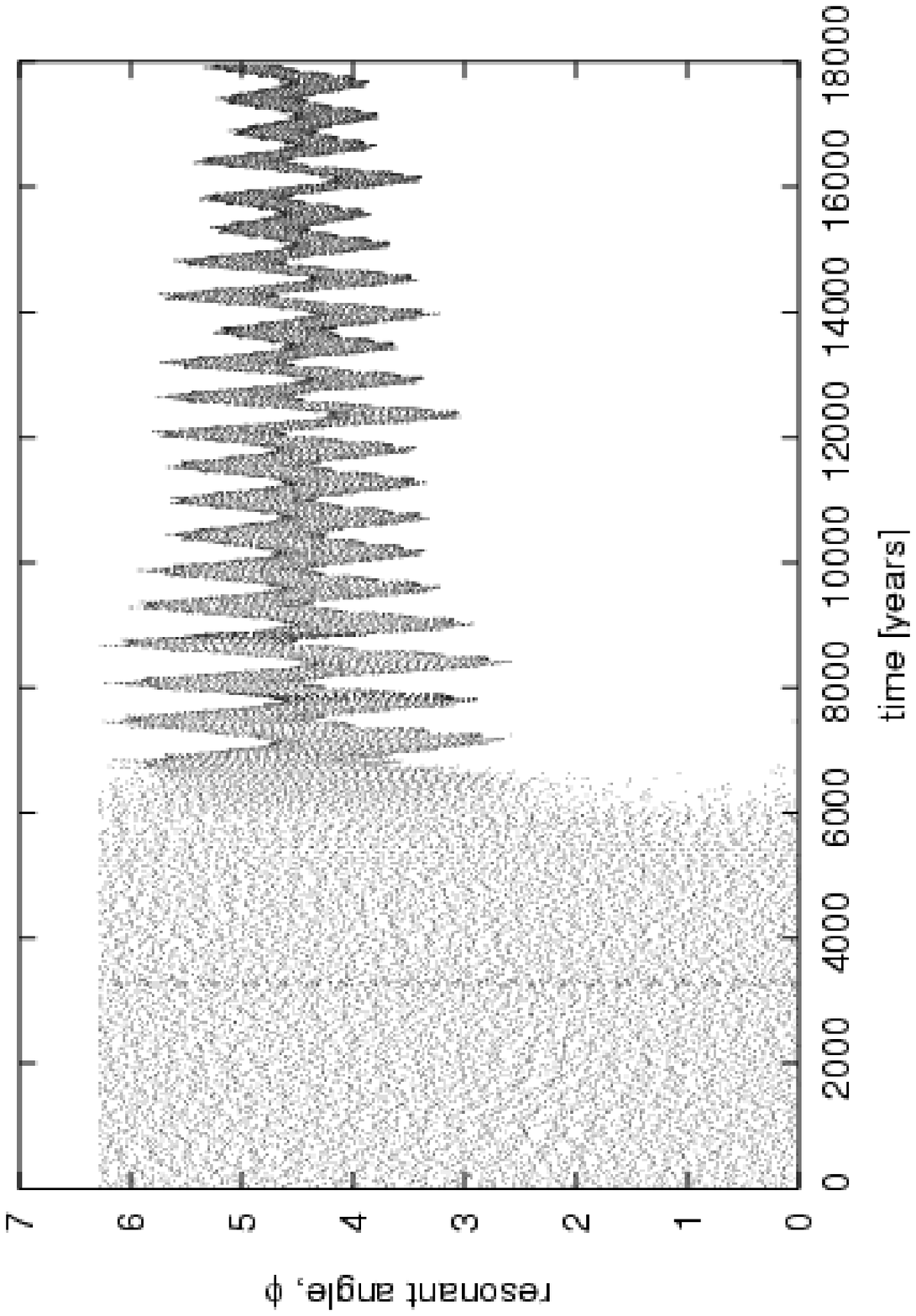}
\includegraphics{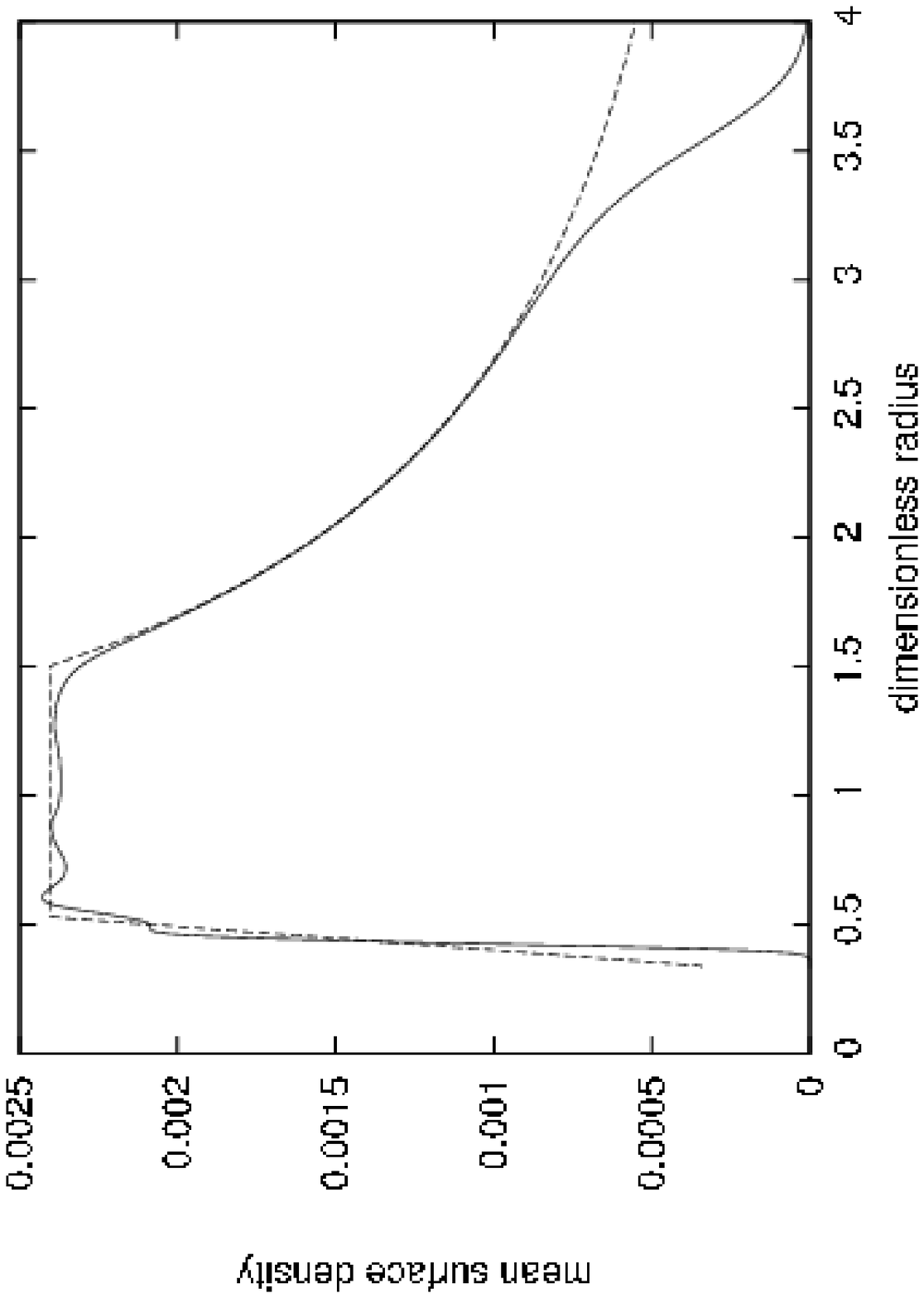}
\includegraphics{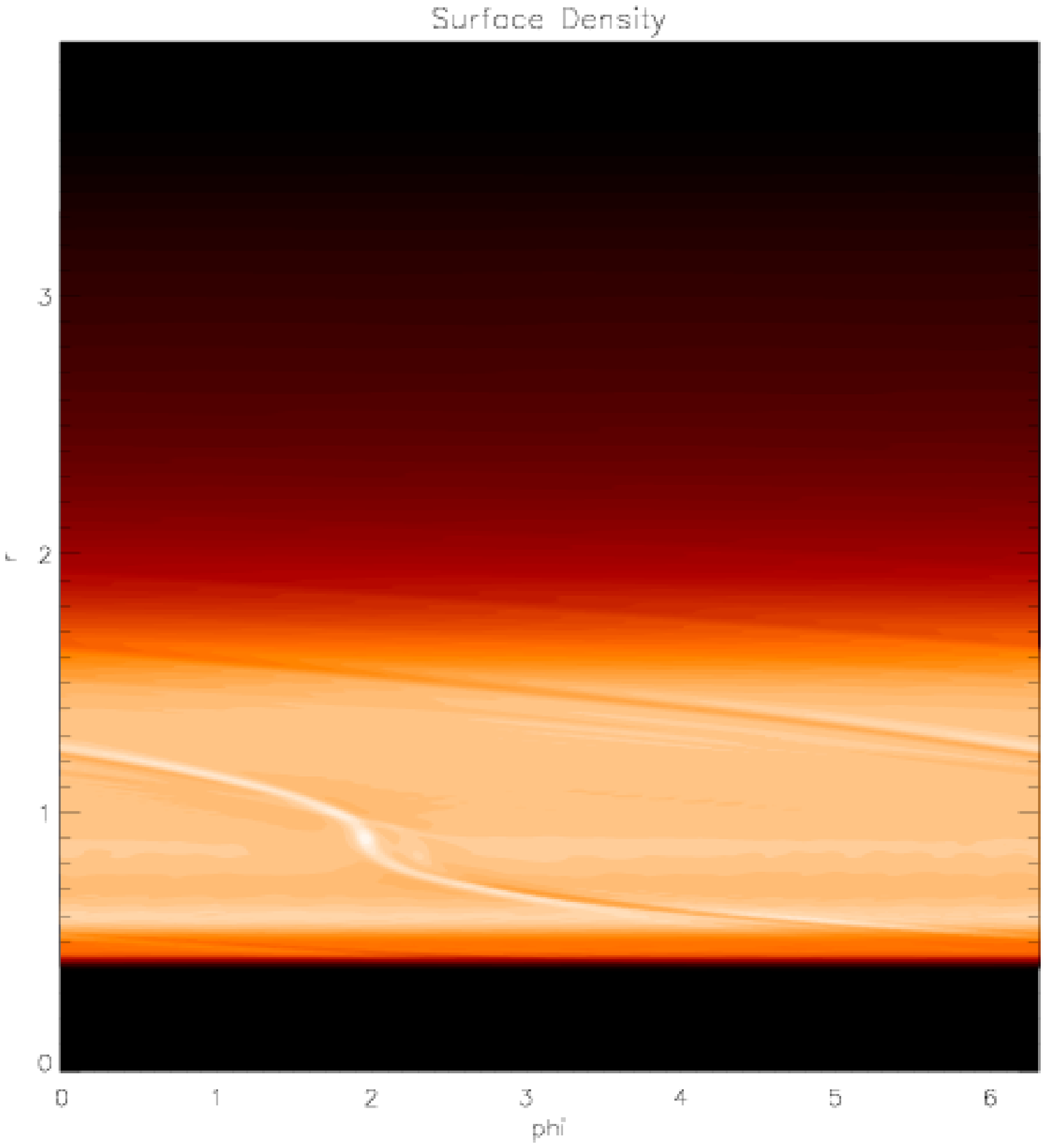}
\caption{ \label{fig3} { 
The evolution of the  semi-major axes, eccentricities, angle between 
apsidal lines and the resonant angle  
for the two planets with masses, $m_{1} = 4 M_{\oplus}$ and
$m_{2} = M_{\oplus}$ migrating towards
a central star, embedded in a disc with initial surface
density scaling  $\Sigma_0 =\Sigma _4,$ 
obtained by hydrodynamical simulations (four upper panels).
The mean surface density profile of the disc near the end of the
simulation (solid line)  and the initial surface density profile 
(dashed line) and a surface density  contour plot are given in 
two lower panels.
}}
\end{minipage}
\end{figure*}

\noindent 
The whole evolution for the fastest migration rate is illustrated
in  Figure \ref{fig3}. We show there the evolution
of the individual semi-major axes, eccentricities,
angle between apsidal lines and  one of the resonant angles.  
A surface density  contour plot of  the disc near the end of the
simulation is given in Figure  \ref{fig3}  together with    
a comparison between the initial and final surface density profiles.

\begin{figure*}
\begin{minipage}{175mm}
\vspace{220mm}
\includegraphics{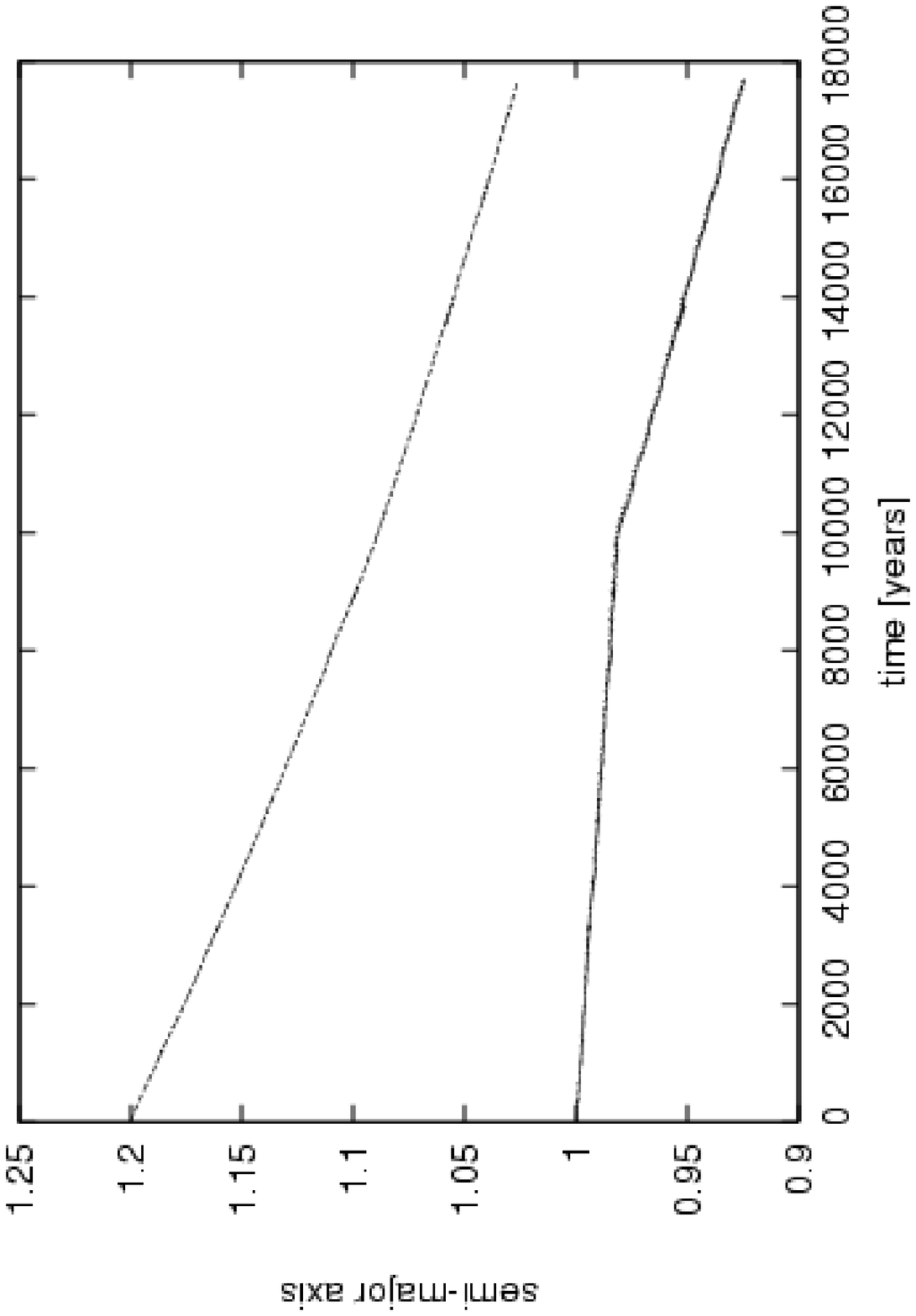}
\includegraphics{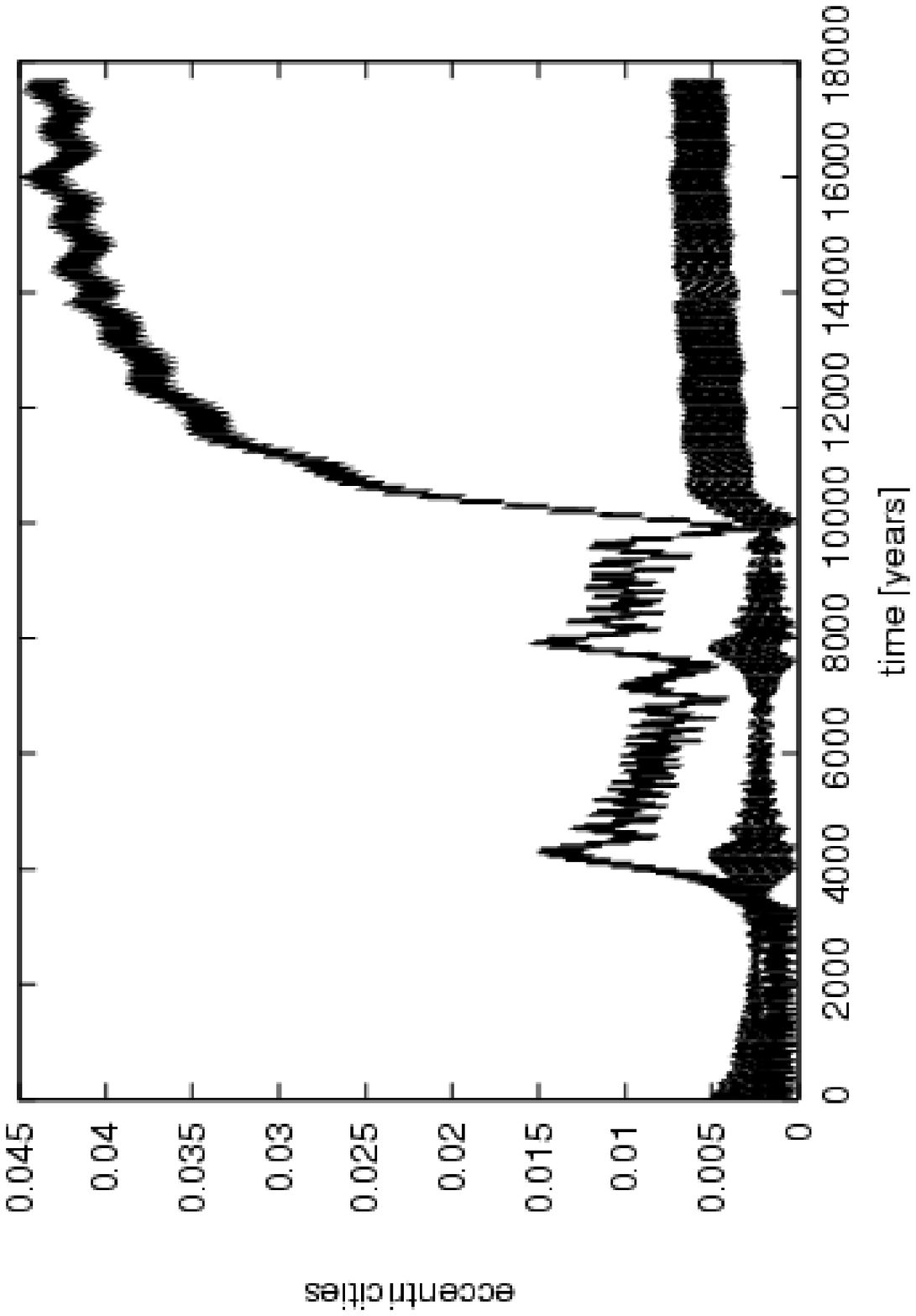}
\includegraphics{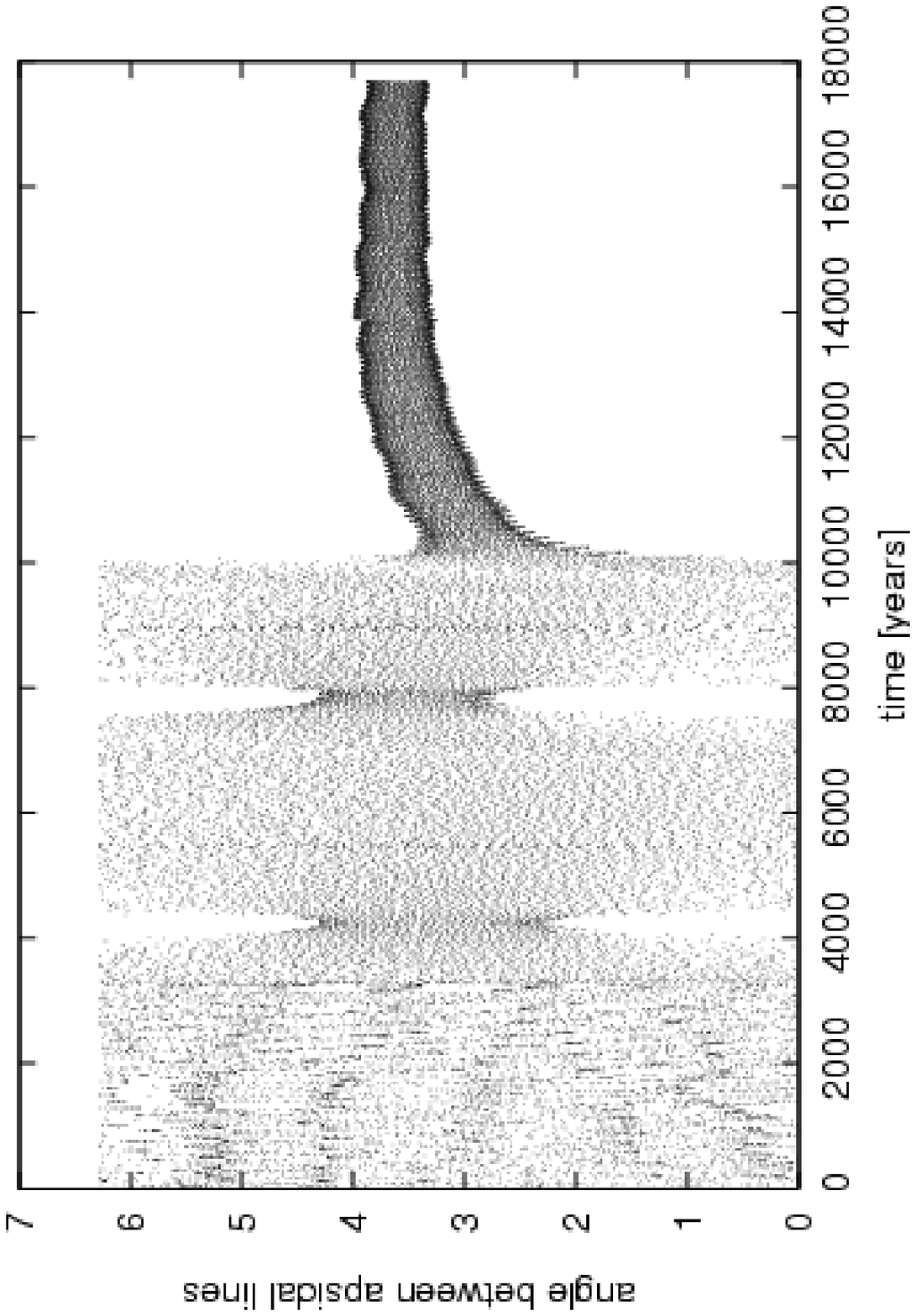}
\includegraphics{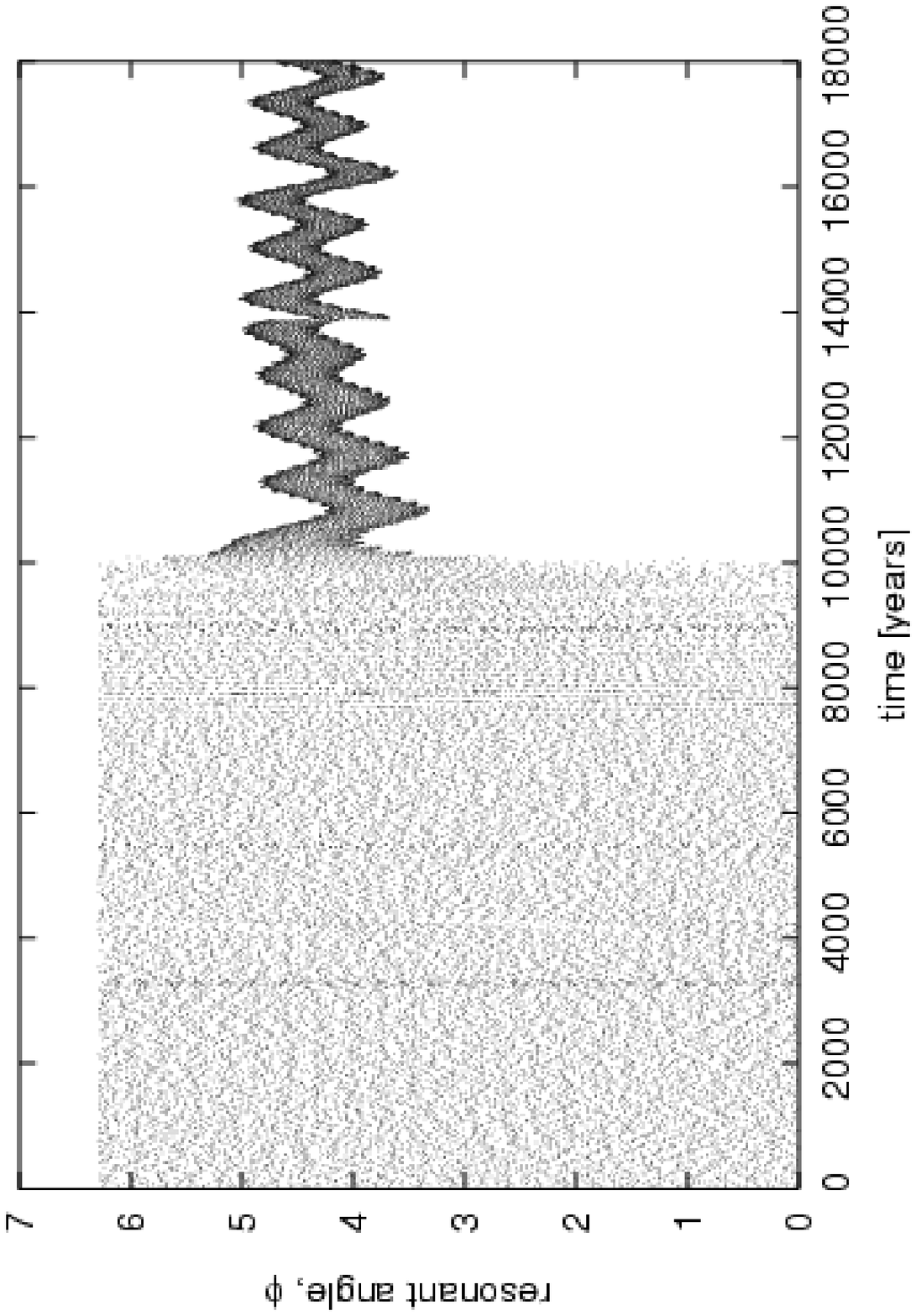}
\includegraphics{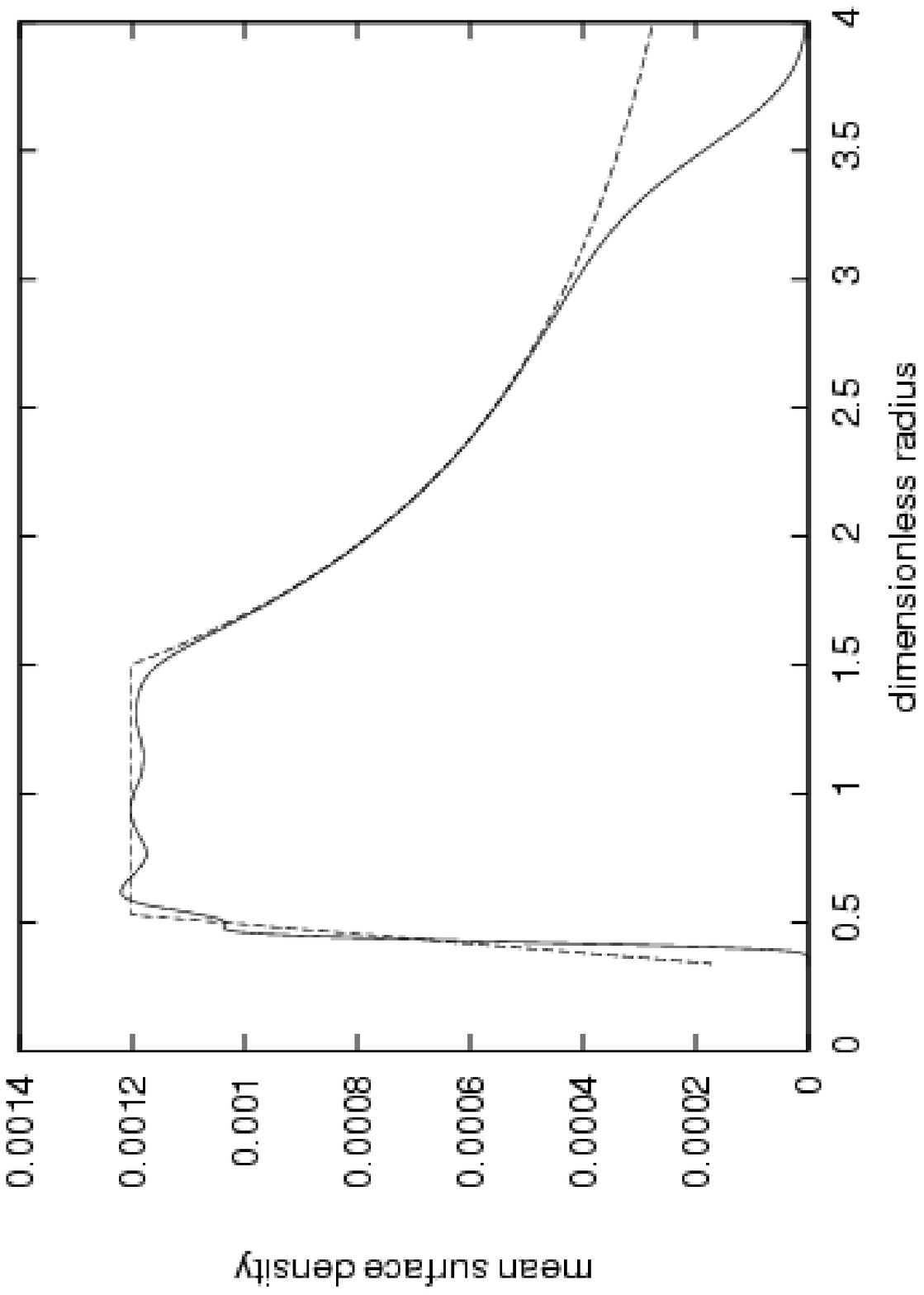}
\includegraphics{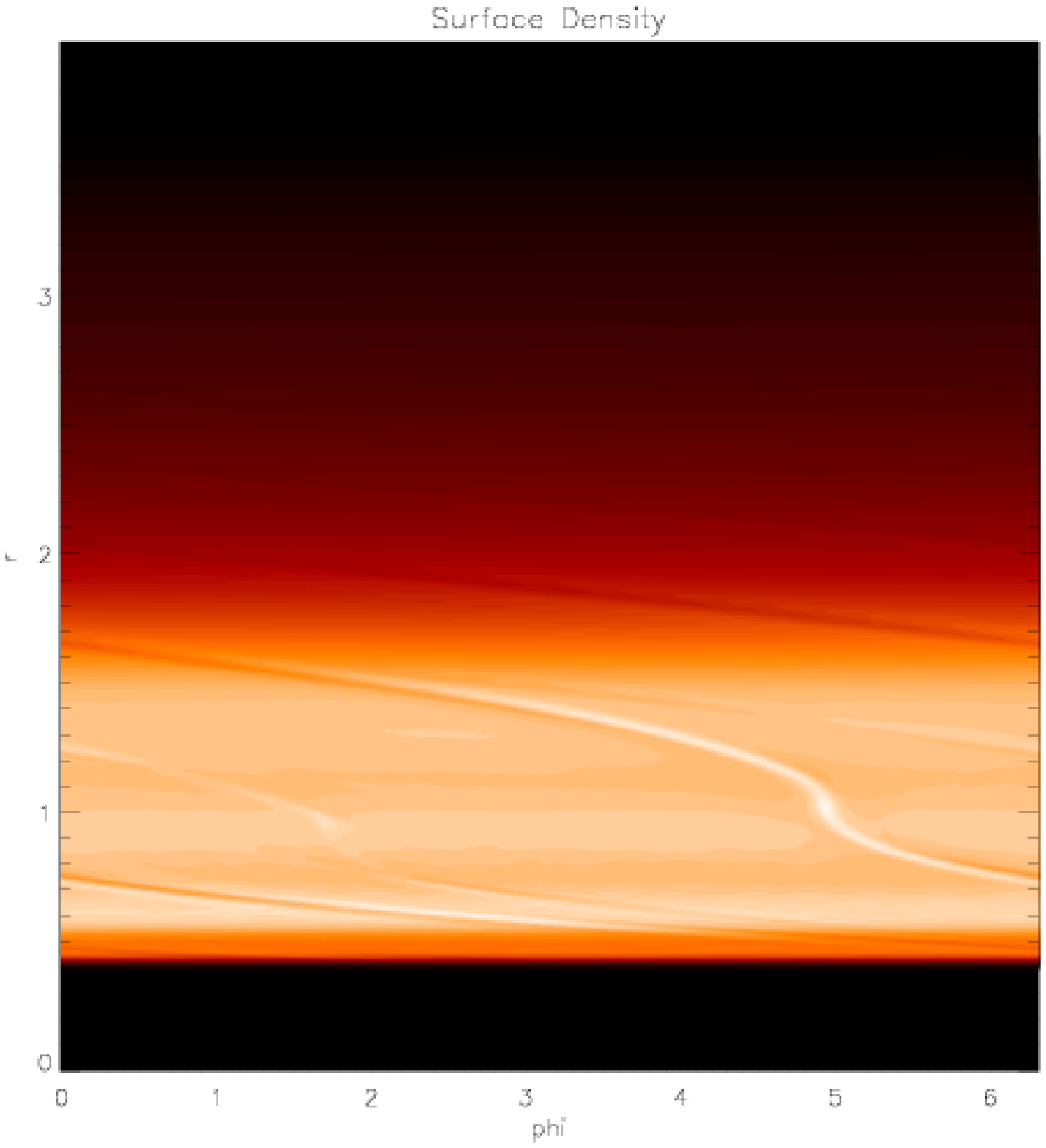}
\caption{ \label{fig4} {As for Figure \ref{fig3} but   when the planets
are embedded in a disc with $\Sigma_0 =\Sigma_2$.
Note the passage through the 5:4 and 6:5 resonances at
$t\sim 4000$ and $t\sim 8000$ years respectively. }}
\end{minipage}
\end{figure*}

\noindent 
The inner planet did not migrate significantly until 8:7 resonance was 
attained at about 6000 years. Subsequently the two planets migrated 
inwards together maintaining the commensurability. The  system appears 
to be close to the chaotic regime as estimated by  use of equations
(\ref{chao} - \ref{chao1}). During 18000 years
of evolution the outer planet changed its location from a dimensionless
radius of  1.2 to 0.9. The eccentricity  of the outer planet increased 
slightly and that of the inner planet substantially, reaching at 
around 10000 years the balanced value  of 0.04 and at  later times 
oscillating around this value. At the end of simulation
shown here the ratio of eccentricities, $e_{1}/e_{2}$, is equal
to 0.125. The local peaks in the values of inner planet eccentricity
occurring at 2000, 4000 and around 5400 years correspond to the 
planet passing through the 5:4, 6:5 and 7:6 resonances respectively. 
After about 8000 years the angle between the  apsidal lines is 
oscillating around 212$^{\circ}$ and the resonant angle $\phi$ around 
264$^{\circ}$. The amplitude of the 
oscillations for both angles is decreasing   with time. 
Similar behaviour can be seen in the evolution of these planets 
when embedded in a disc with lower surface density (see Figure \ref{fig4}).

\begin{figure*}
\begin{minipage}{175mm}
\vspace{220mm}
\includegraphics{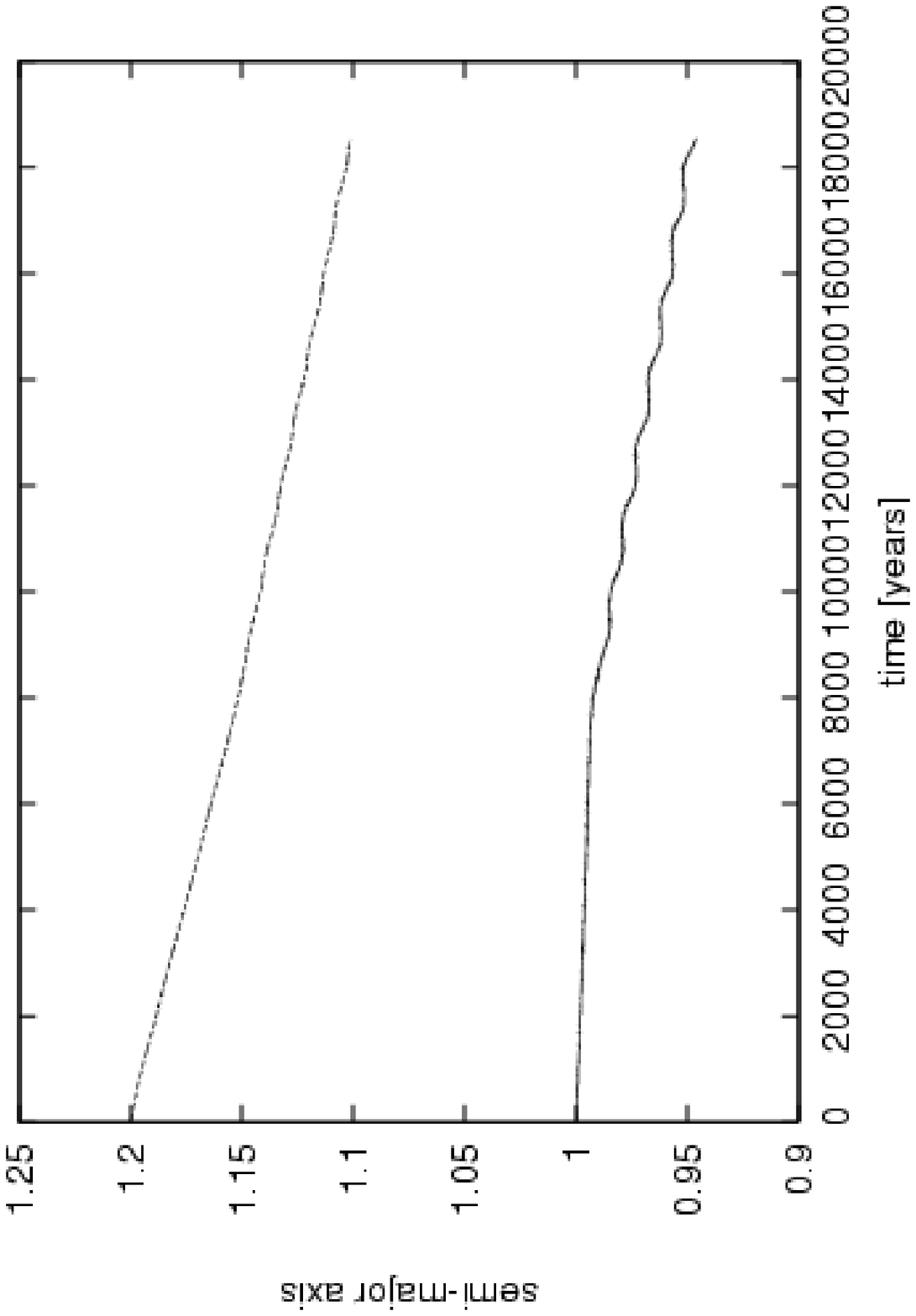}
\includegraphics{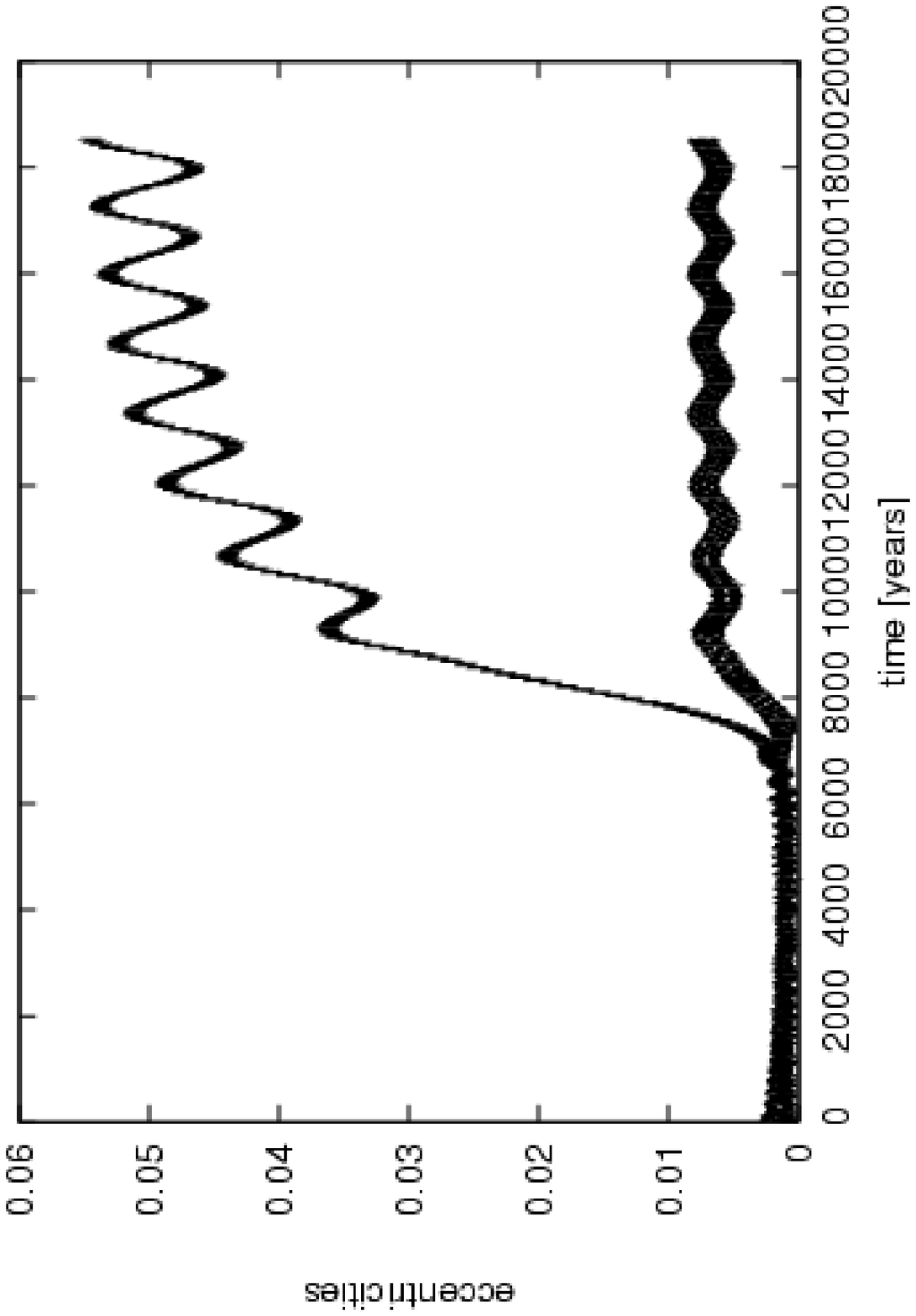}
\includegraphics{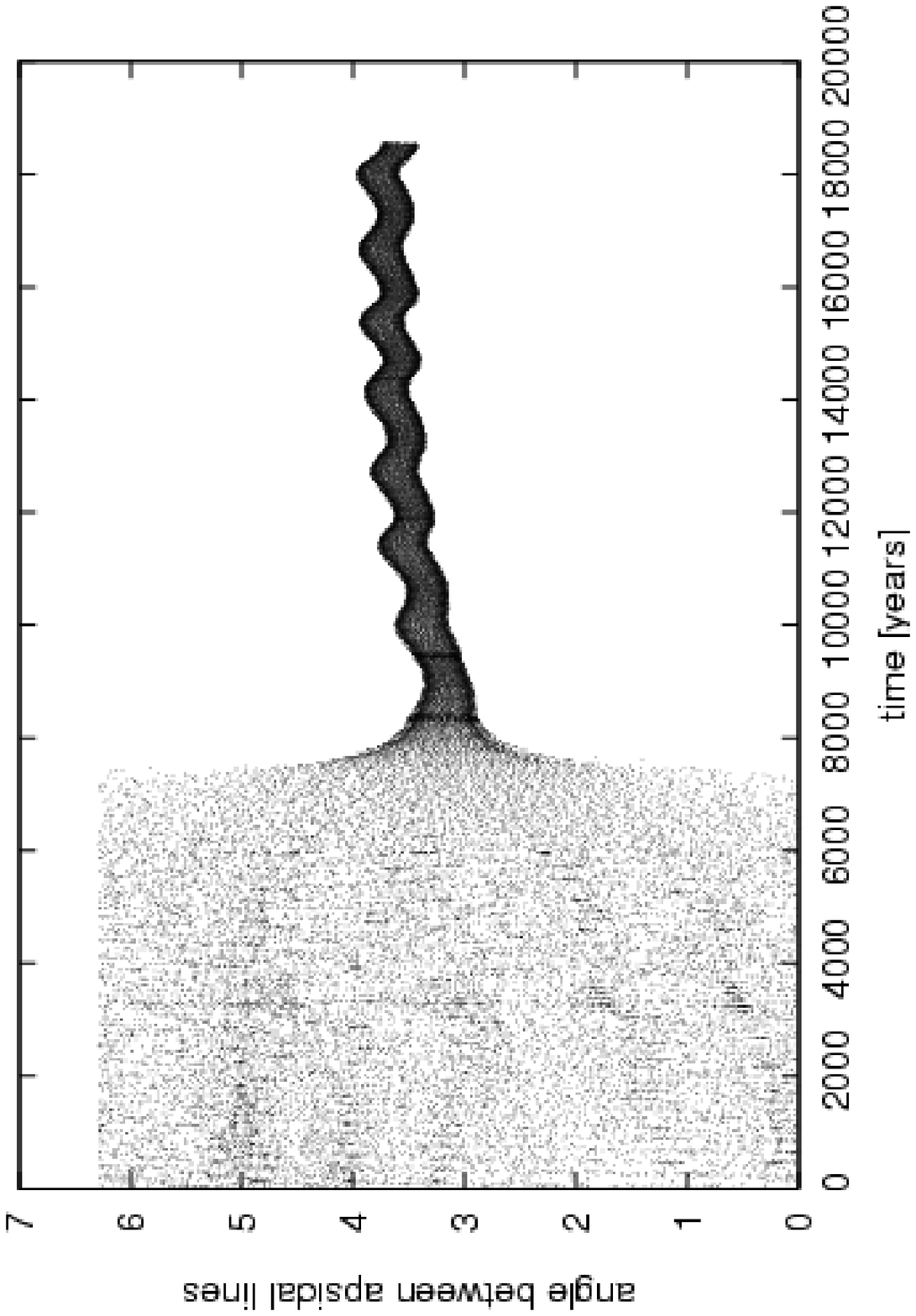}
\includegraphics{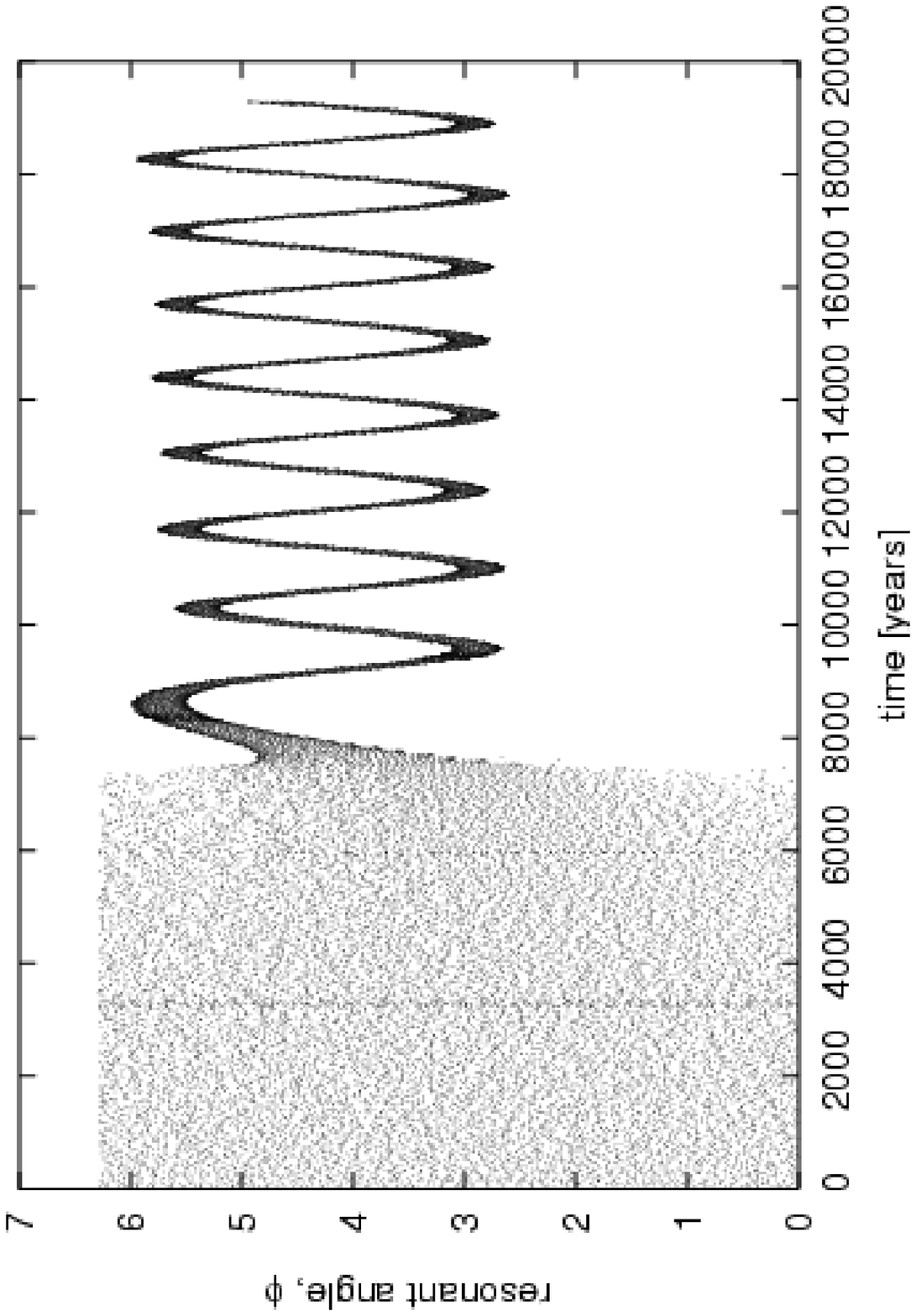}
\includegraphics{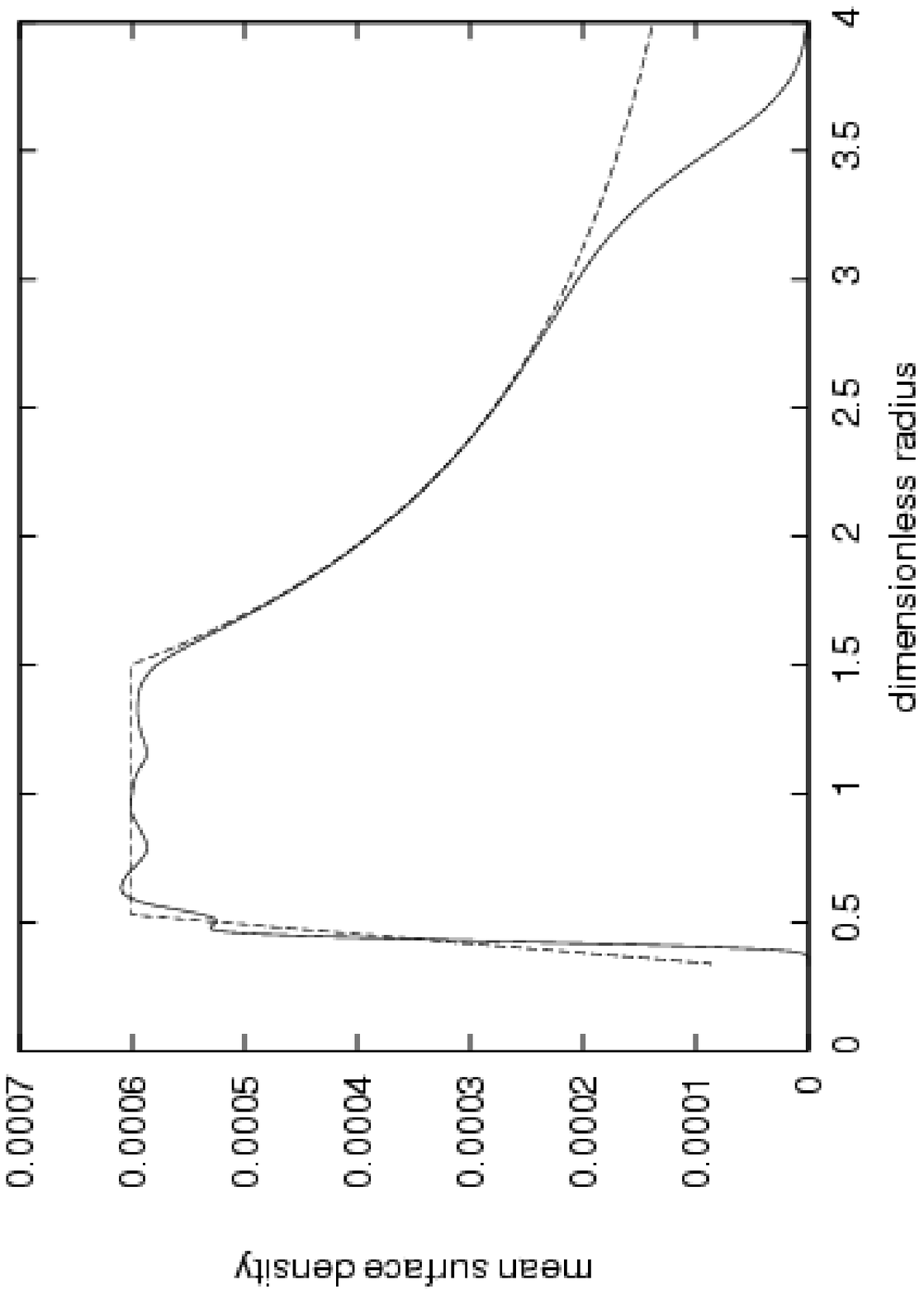}
\includegraphics{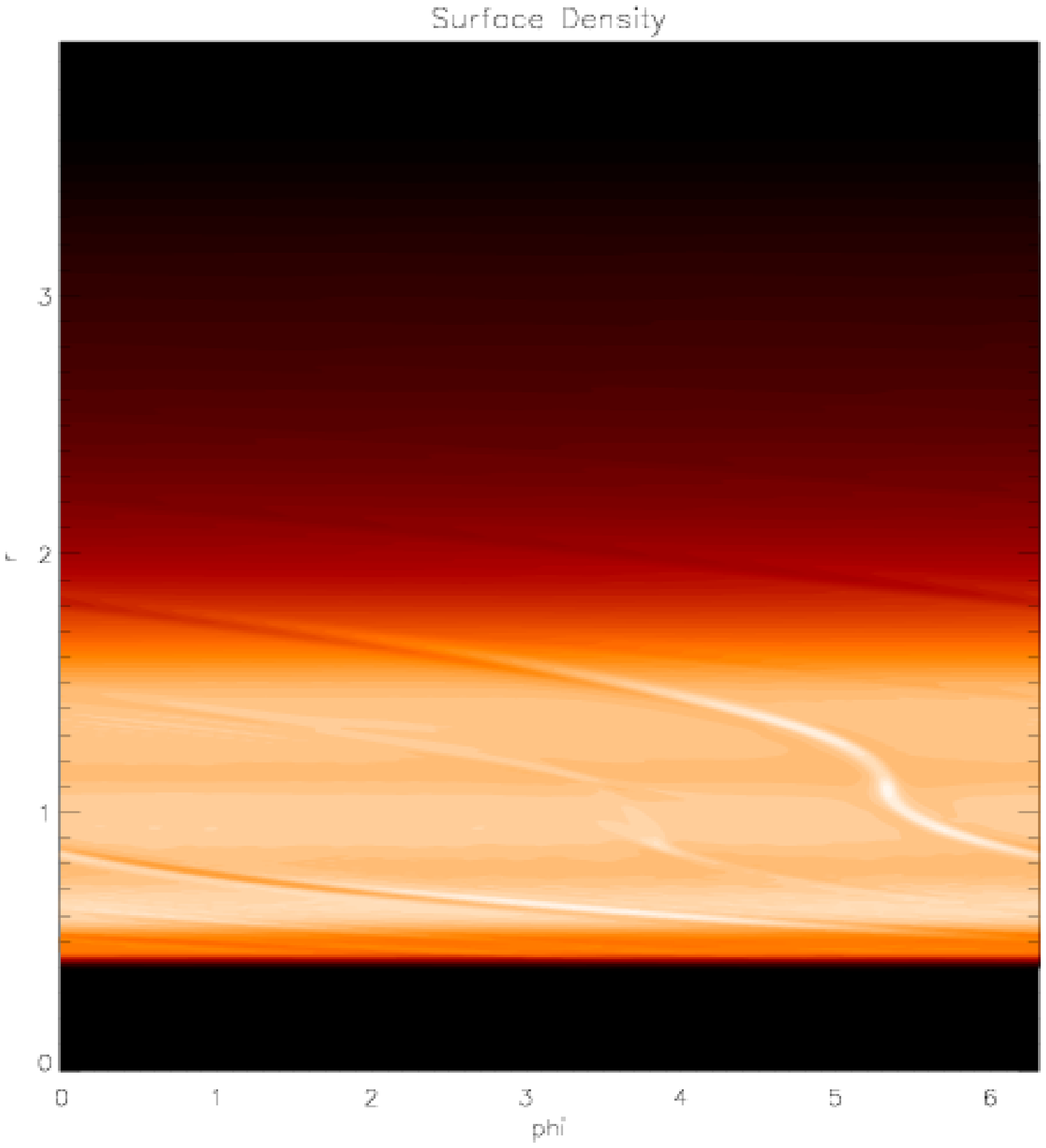}
\caption{\label{fig5}{ As for Figure \ref{fig3} but   for the case 
when the planets are embedded in the disc with initial surface 
density scaling  $\Sigma_0 =\Sigma_1$. 
}}
\end{minipage}
\end{figure*}

\begin{figure*}
\begin{minipage}{175mm}
\vspace{220mm}
\includegraphics{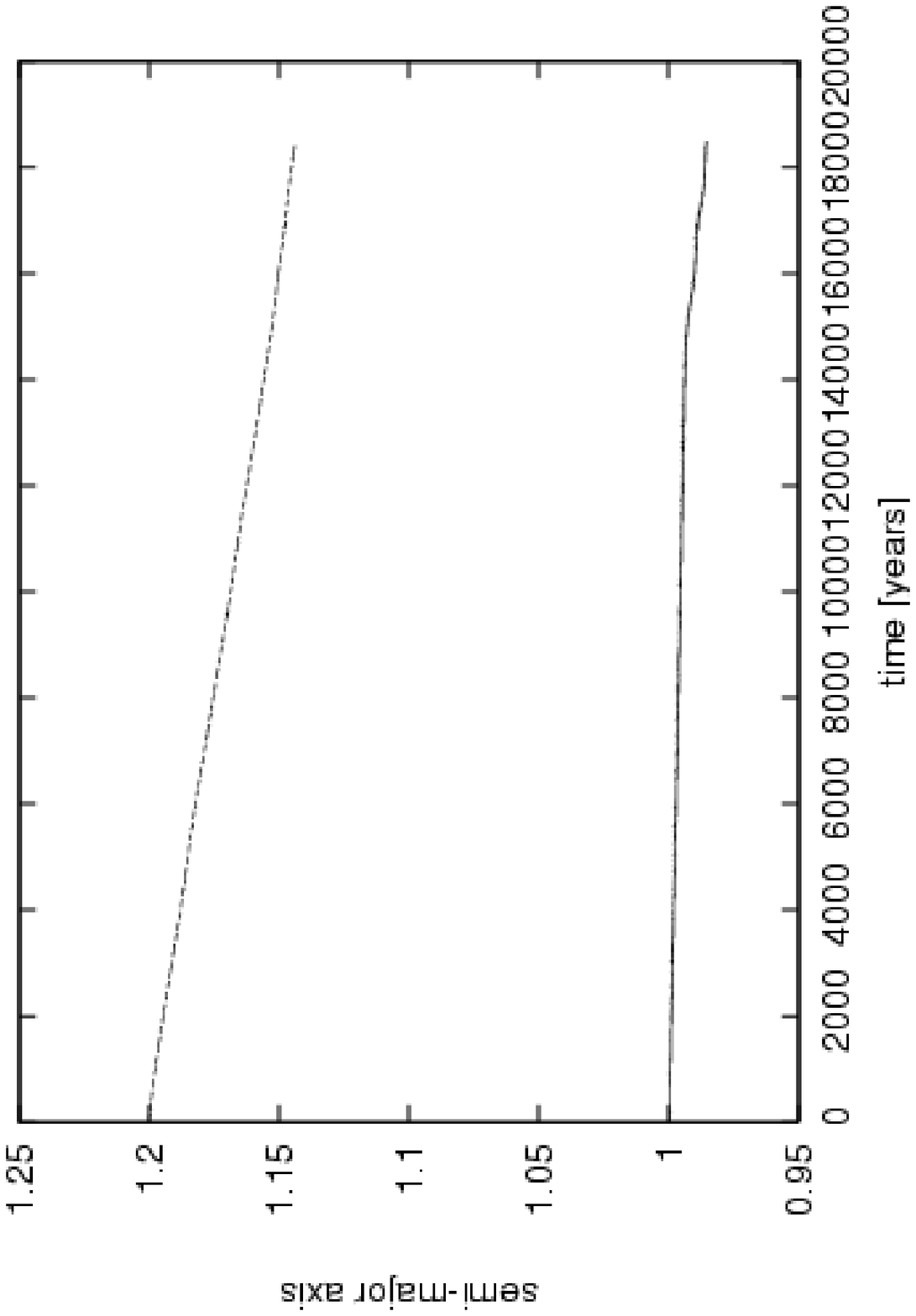}
\includegraphics{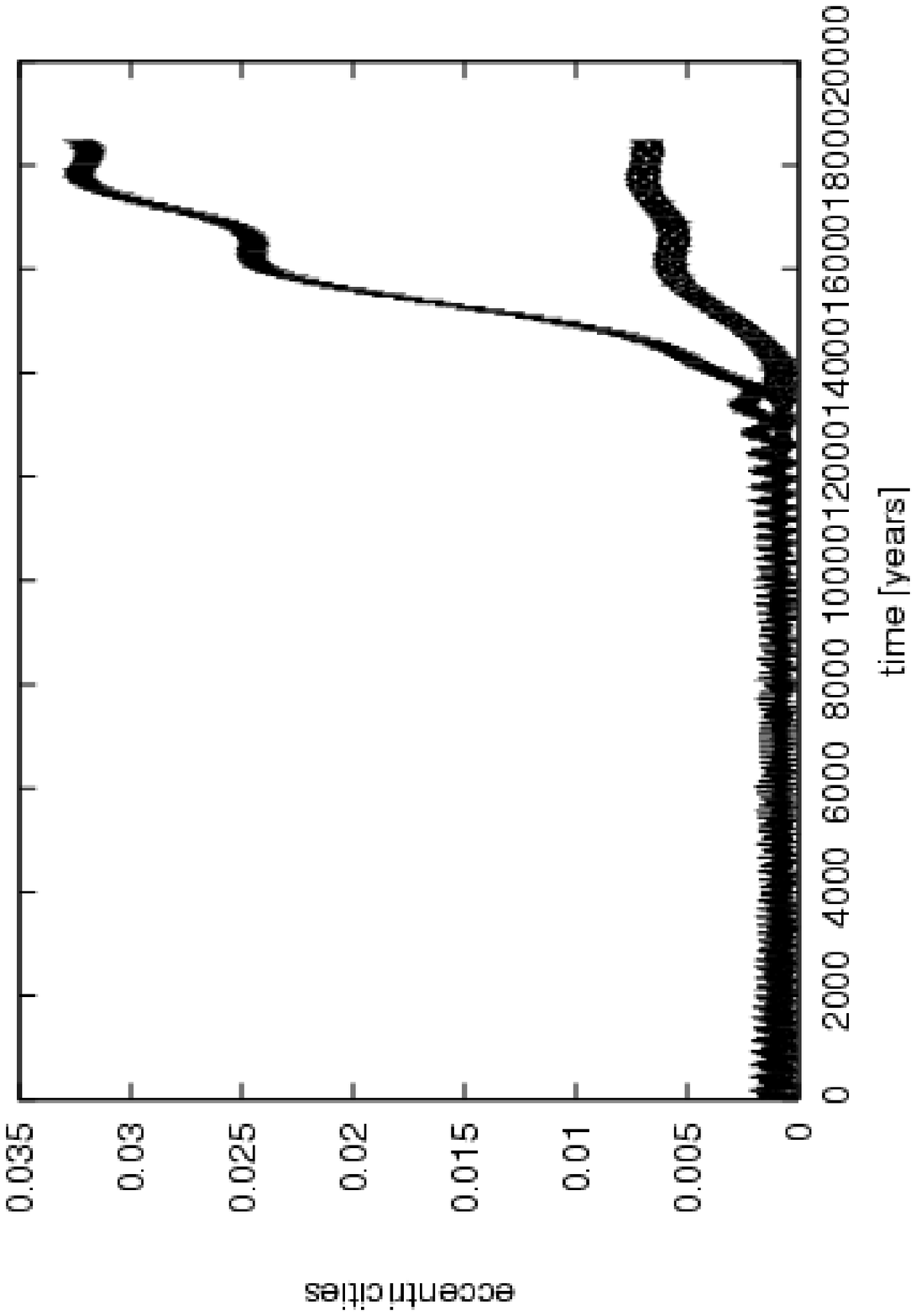}
\includegraphics{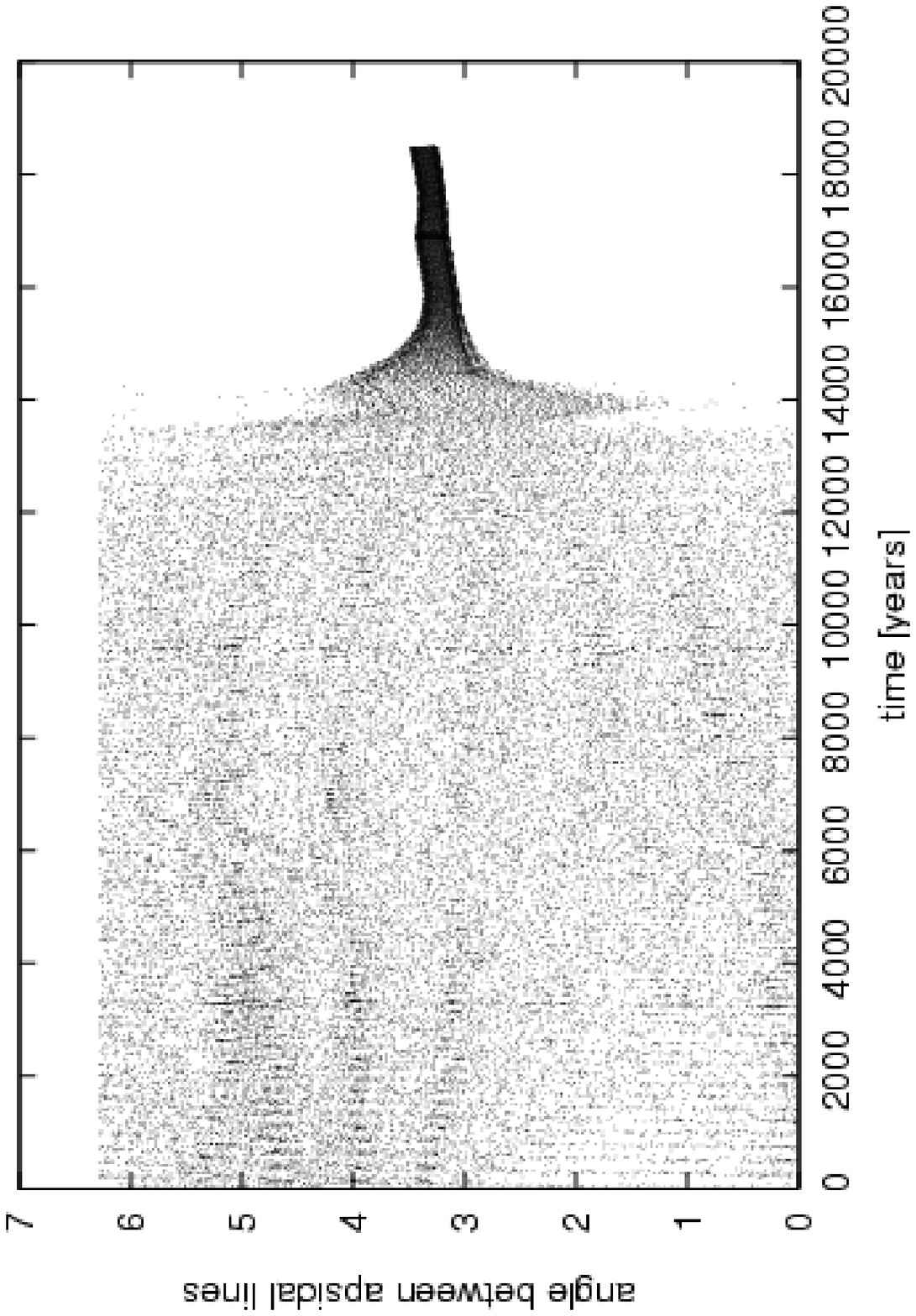}
\includegraphics{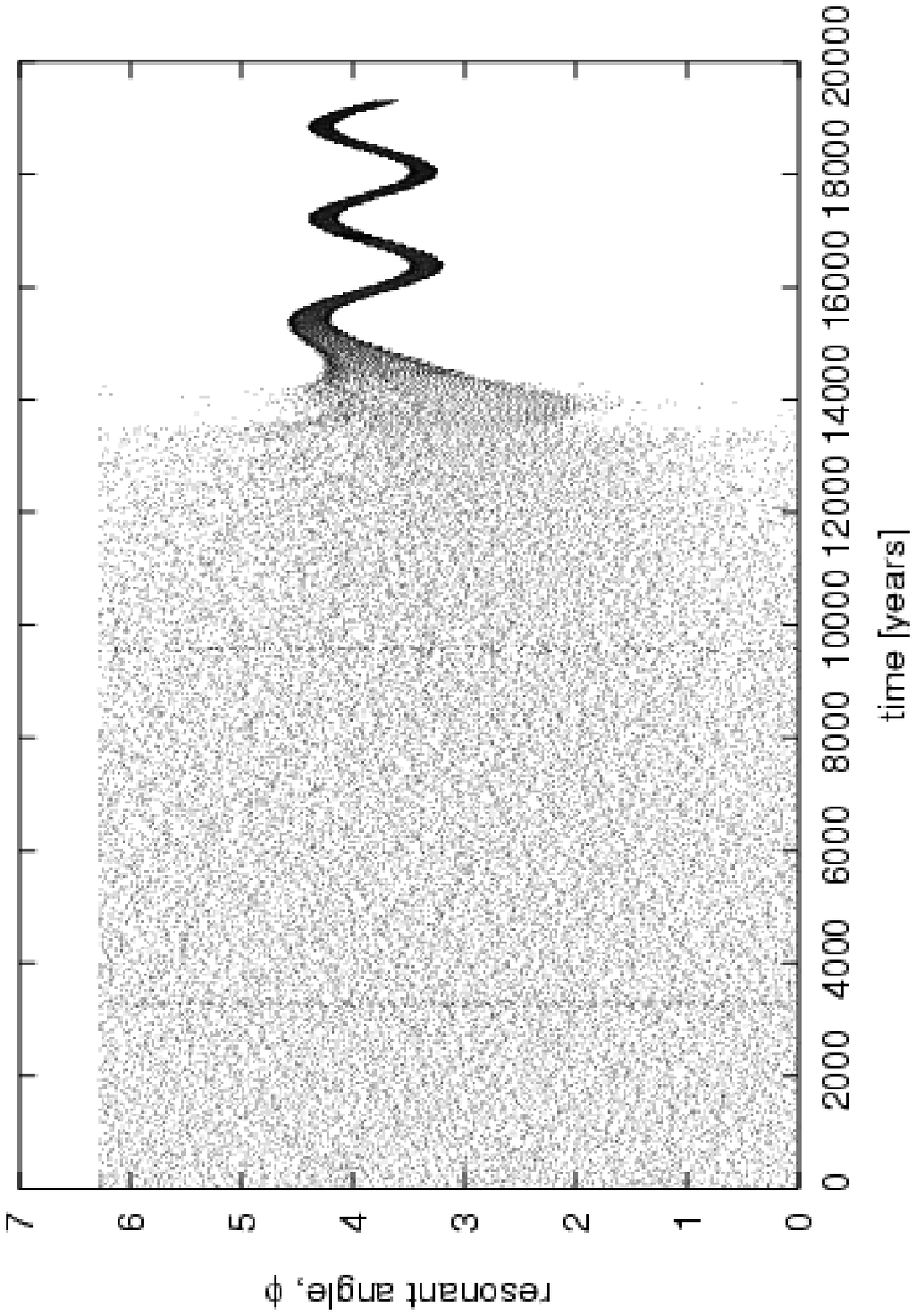}
\includegraphics{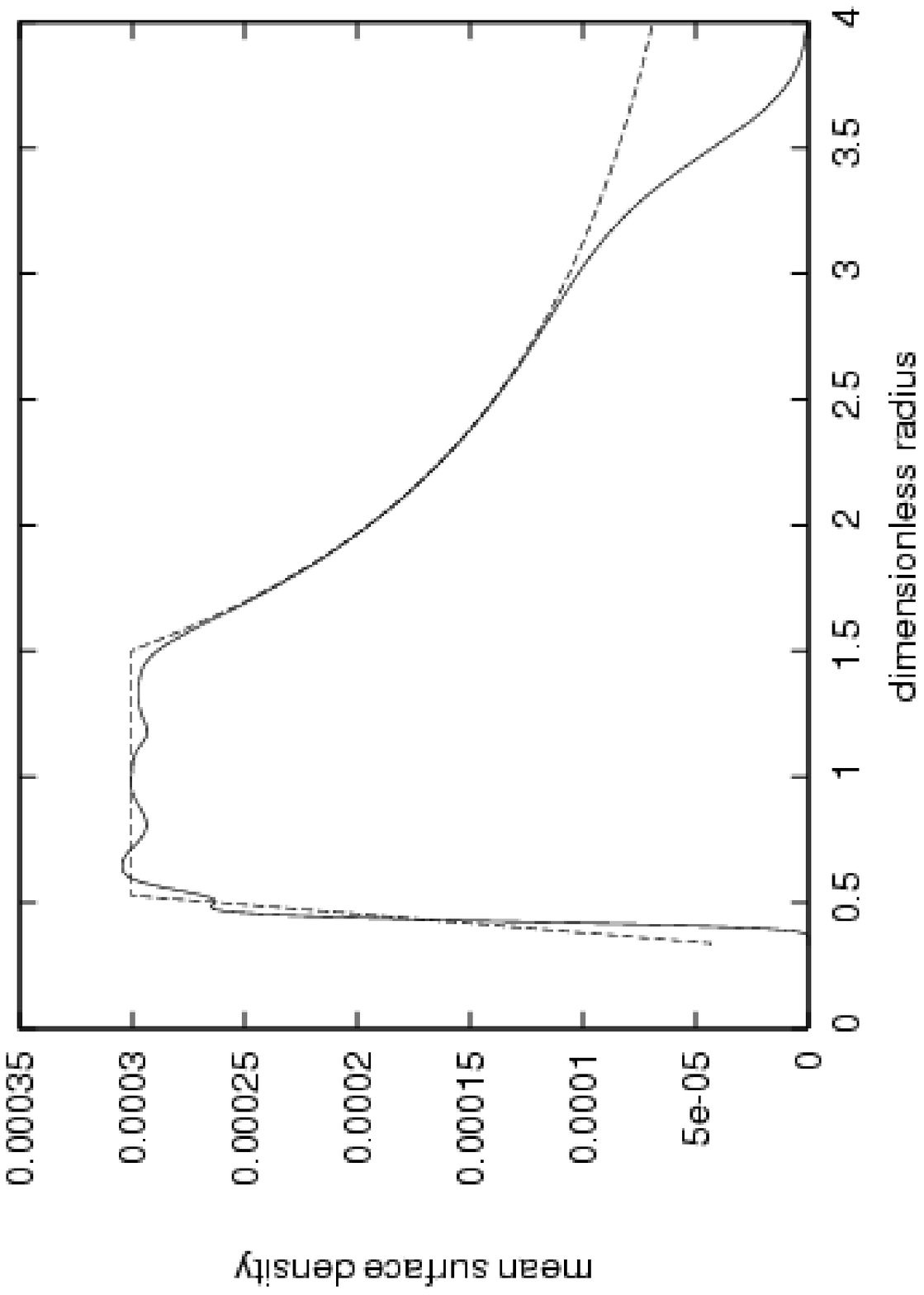}
\includegraphics{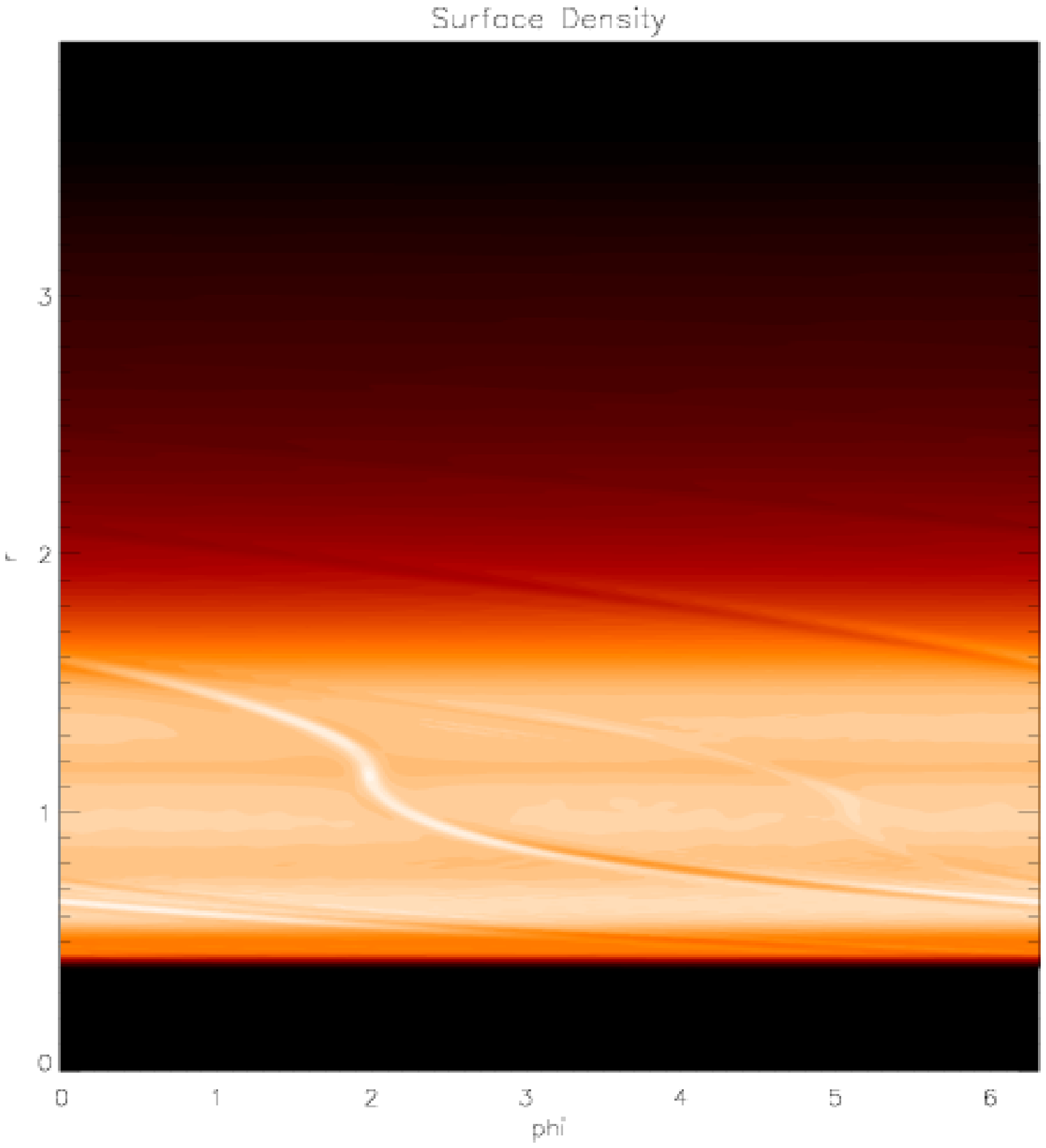}
\caption{\label{fig6}{As for Figure \ref{fig3}
 but   for the case when the planets
are embedded in a disc with $\Sigma_0 =\Sigma_{0.5}$.
}}
\end{minipage}
\end{figure*}

\noindent 
When $\Sigma_0 = \Sigma_2,$ leading to a slower rate of migration,
as before, the inner planet did not migrate significantly 
until a commensurability was achieved and maintained. A 7:6 
resonance was reached after 10000 years and 
subsequently maintained. The passage through the 5:4 and 6:5
commensurabilities are clearly marked by a local increase of 
inner planet eccentricity and also in behaviour of the angle 
between apsidal lines. In fact after the passage through 5:4 resonance
the eccentricities do not completely  decay potentially
making the outcome of  subsequent resonance passages probabilistic
\citep{klg93}.
 In this case the planets relative migration 
is slower and the planets stay longer in the vicinity of these 
resonances. At the end of the evolution the eccentricities ratio
is $e_{1}/e_{2}$ =0.14. The eccentricity of  the inner planet 
is slightly higher than it was in the case with 
$\Sigma_0 = \Sigma _4$. However, the eccentricities
continue to grow and the equilibrium is not quite
 established at the end of the simulation. The amplitude
of oscillations in both the  angle between the apsidal lines 
and the  resonant angle $\phi,$
is smaller than in the previous case.
For the two simulations that started with even lower disc surface 
densities  such that $ \Sigma_0 =  \Sigma _1$ and 
$ \Sigma_0 =  \Sigma _{0.5},$
the evolution  is shown in Figures \ref{fig5}-\ref{fig6}.

\noindent 
When $ \Sigma_0 = \Sigma_1,$ the evolution of two planets leads to 
locking in 5:4 resonance.  This happens at around 8000 years. Before 
that the inner planet migrated very slowly. After becoming trapped 
in the commensurability the inner planet semi-major axis changes 
in a characteristic oscillatory manner. These oscillations are 
clearly visible in all other   related plots (eg. Figure  \ref{fig5}). 
The eccentricity ratio at the end of the simulations is $e_{1}/e_{2} 
=0.18 $. The eccentricities are  still growing.
Similar behaviour is seen in angle between the apsidal lines, which 
starts to librate at time 8000 years around 180$^{\circ}$ and at time 
18000 years oscillates around 212$^{\circ}$. The resonant angle changes 
have very  large amplitude and seem to grow with time, which might 
indicate that the planets will   not remain in this resonance.
 In this context note that equation (\ref{MIG}) 
indicates that $t_{migi}n_i$ increases inwards for a uniform 
surface density. This should favour resonance locking at smaller radii.

\noindent 
Also the  planets in a disc with  $ \Sigma_0 = \Sigma_{0.5}$
end  in a  5:4  commensurability. It occurs after approximately
 14000 years. Because of slow migration rates
in low surface density discs, the subsequent
evolution   could not be continued for
long enough to be able to make a definitive statement
about   the final outcome. However,
the oscillations in the resonance angles are   much  smaller
and more closely centred around $\pi$
than is  the case when  $\Sigma_0 = \Sigma_1.$
This could indicate greater  stability of the 5:4 resonance 
in this case.

\subsection{Comparison with the analytic model}

It is of interest to compare the eccentricities obtained above 
when the planets are trapped in a commensurability with what 
is expected when resonant effects and disc tides are in balance.
In the simple analytic model given in the appendix, when
the eccentricities are small these satisfy equation (\ref{ejcons0}).
In principle orbital circularization acting on both planets
should be included. However, for the simulations presented
here, $e_2 >> e_1$ and even though the circularization time may be 
smaller for the possibly  more massive outer planet, the effect 
can be neglected in comparison to that associated with the 
inner planet. Accordingly we set $e_1 =0.$ Then equation 
(\ref{ejcons0}) gives the simple relation
\be  
{e_2^2\over t_{c2}}\left({m_2n_1a_1
\over m_1n_2a_2} +f\right)
=\left({1\over t_{mig1}}-{1\over t_{mig2}}\right){f\over 3},
\label{ejcons01}
\ee
where we recall that   $ f =m_2a_1/((p+1)(m_2a_1+m_1a_2)).$
We further simplify matters by assuming that $p$ is large.
Then we may set $a_1=a_2, n_1=n _2$ and
neglect $f$ on the left hand side of (\ref{ejcons01}).
In the same spirit, we replace it by $m_2/((p+1)(m_2+m_1))$ 
on the right hand side. Then we obtain
\be  
{e_2^2\over t_{c2}}
=\left({1\over t_{mig1}}-{1\over t_{mig2}}\right)
{m_1\over 3 (p+1)(m_2+m_1)},
\label{ejcons02}
\ee
Interestingly equation (\ref{ejcons02}) shows that it is the 
rate of convergent migration of the two planets that determines 
the eccentricities which are small when this is small in agreement 
with our results.
Further, other things being equal, the eccentricities decrease 
with increasing $p.$
We also note that use of equations (\ref{MIG}) and (\ref{CIRC}) 
gives $t_c = (\tau_{r}/W_c) (H/ r)^2.$
Thus we obtain
after recalling that in general  the e folding rate for mean 
motion is a factor of $1.5$ greater than that for radius, or 
$ t_{mig} = 2\tau_{r}/3$
\be  
e_2^2
= \left({m_1 \over m_2}\left({a_1\over a_2 }\right)^{1/2} - 1\right)
{m_1\over 0.578 (p+1)(m_2+m_1)}
 (H/ r)^2
\label{ejcons03}.
\ee
We apply this to the case illustrated in Figure \ref{fig3}
for the two planets with masses, $m_{1} = 4 M_{\oplus}$ and
$m_{2} = M_{\oplus}$ 
embedded in a disc with initial surface
density scaling  $\Sigma_0 =\Sigma _4$ and obtain $e_2 =0.037$ 
in reasonable agreement with the simulations.

\subsection{Dependence on the initial separation of the  planets}

\begin{figure}
\vskip 8cm
\includegraphics{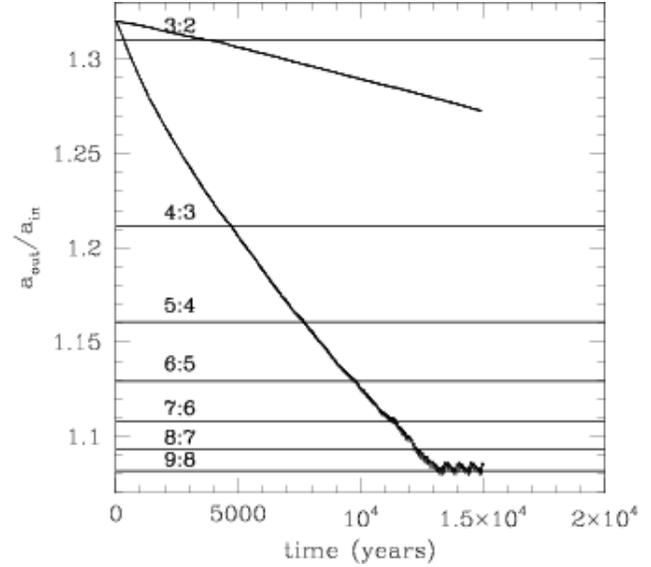}
\caption{\label{fig7}{The semi-major axis ratio  when  
$m_1 =4 M_{\oplus}$, $m_2 = 1M_{\oplus}$   with 
 $\Sigma_0 = \Sigma_{0.5}$ - upper curve and $\Sigma_0 = \Sigma_{4}$ 
- lower curve.
        }}
\end{figure}

Here  we  investigate the dependence of attained 
commensurabilities on the initial radial separation of 
the planets. To do this  we have performed  simulations with 
planets of the same mass as above  but  starting  with
orbital separations a little larger than that required for strict  
3:2 and 4:3 commensurabilities.  In the former case  the two planets   
were initiated with  $r_{1}=1.32$ and $r_{2}=1.00.$ 
Two initial surface density scalings were considered, namely
$\Sigma_0 = \Sigma_{0.5}$ and 
$\Sigma_0 = \Sigma_{4}$.
The evolution of the ratio of the  
semi-major axis ratios for both  cases is   plotted in  Figure \ref{fig7}.
For $ \Sigma_0 = \Sigma_{4}$ the end state  has the planets
in a 9:8 resonance. The evolution for
$ \Sigma_0 = \Sigma_{0.5}$ is   shown for comparison.
The result in the $ \Sigma_0 = \Sigma_{4}$  case  is a clear 
indication that the system gets into the chaotic regime as 
a value of $p$ of the attained resonance is increased when the initial 
semi-major axis ratio  was larger.
Stability would have implied that the same resonance should 
have been attained. Unfortunately the migration rate in the 
$ \Sigma_0 = \Sigma_{0.5}$
case was too slow for us to be able to attain a commensurability
with the available computational resources and confirm or 
otherwise the stability of the  4:3  commensurability.    

\begin{figure*}
\begin{minipage}{175mm}
\vspace{220mm}
\includegraphics{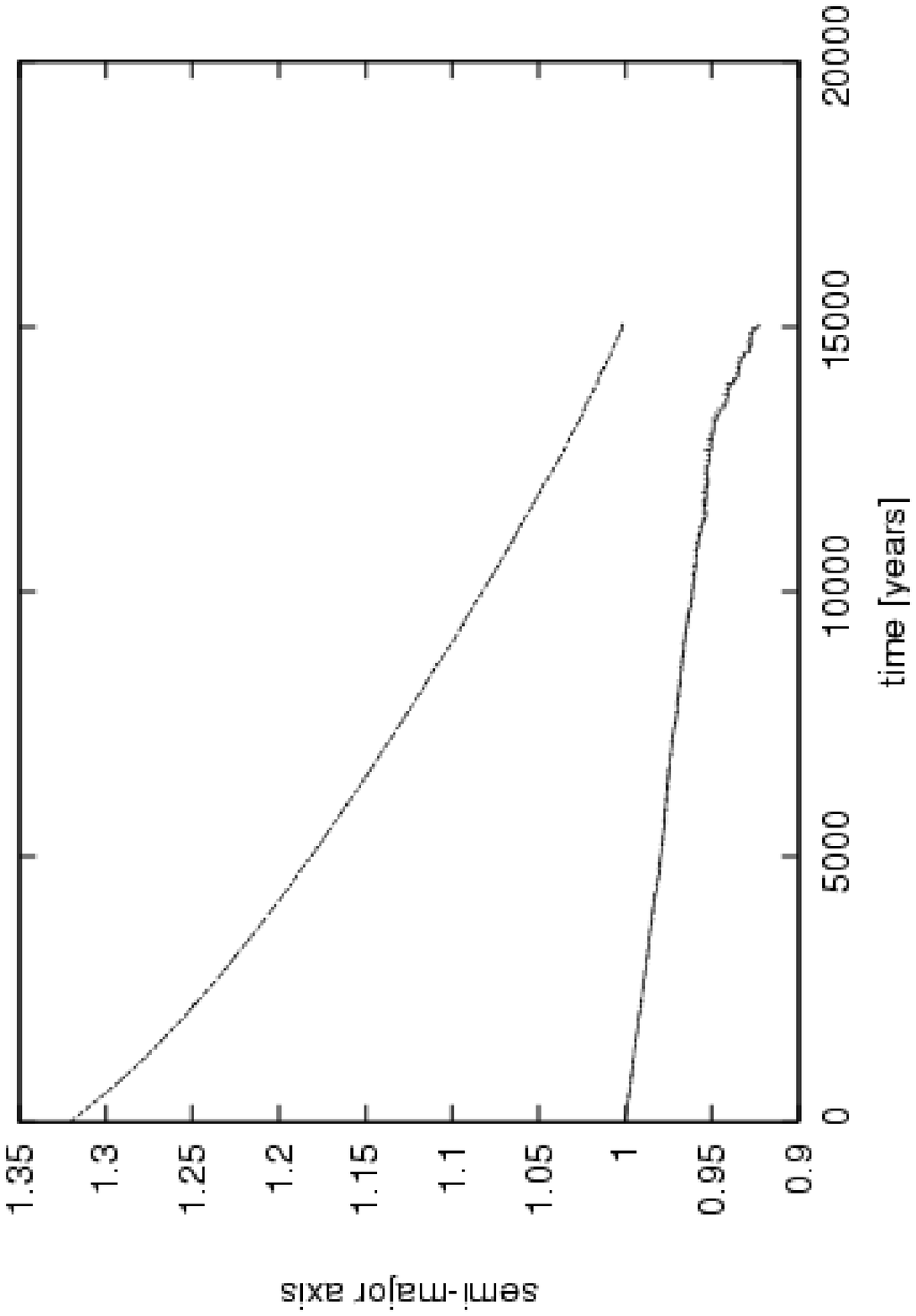}
\includegraphics{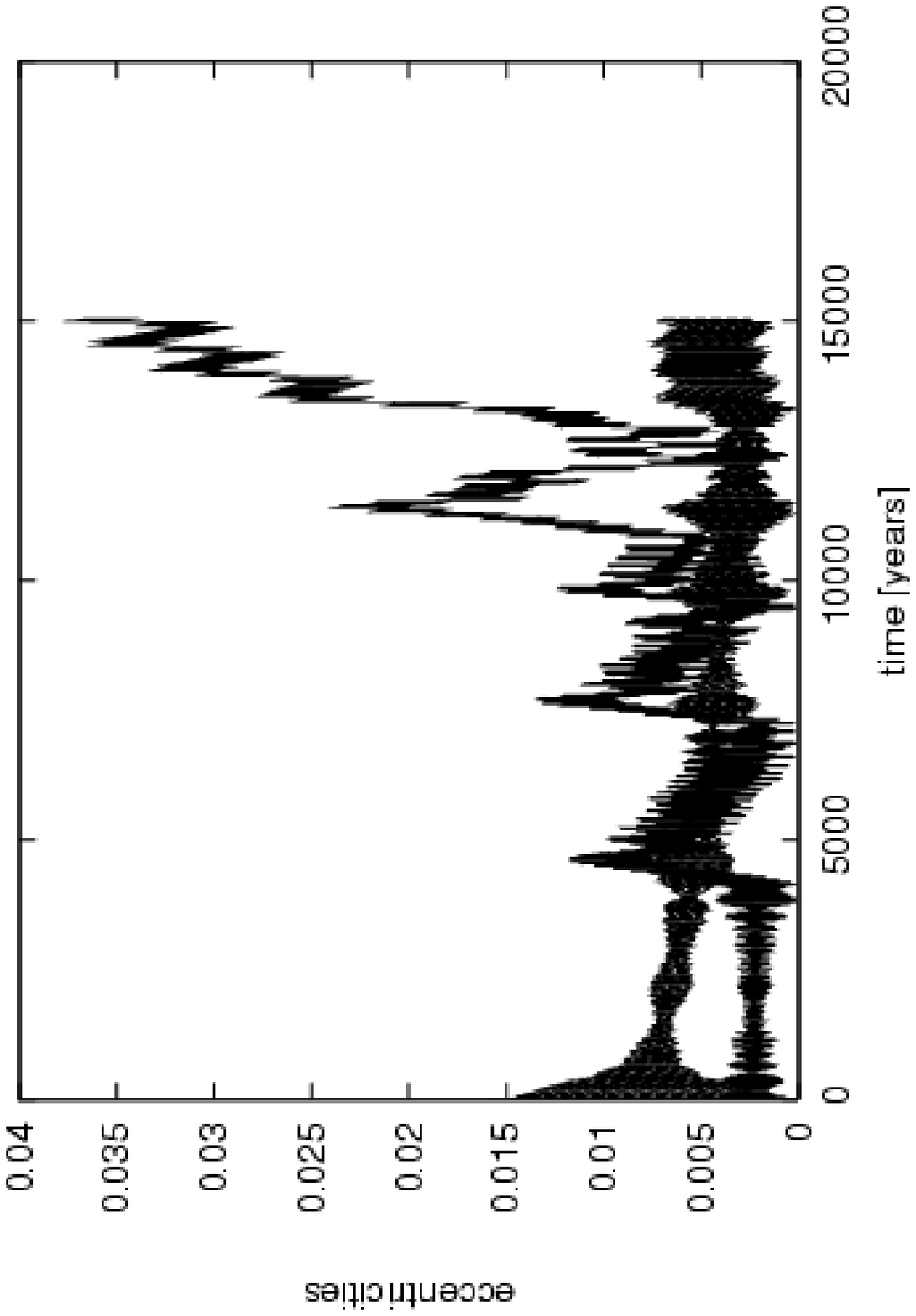}
\includegraphics{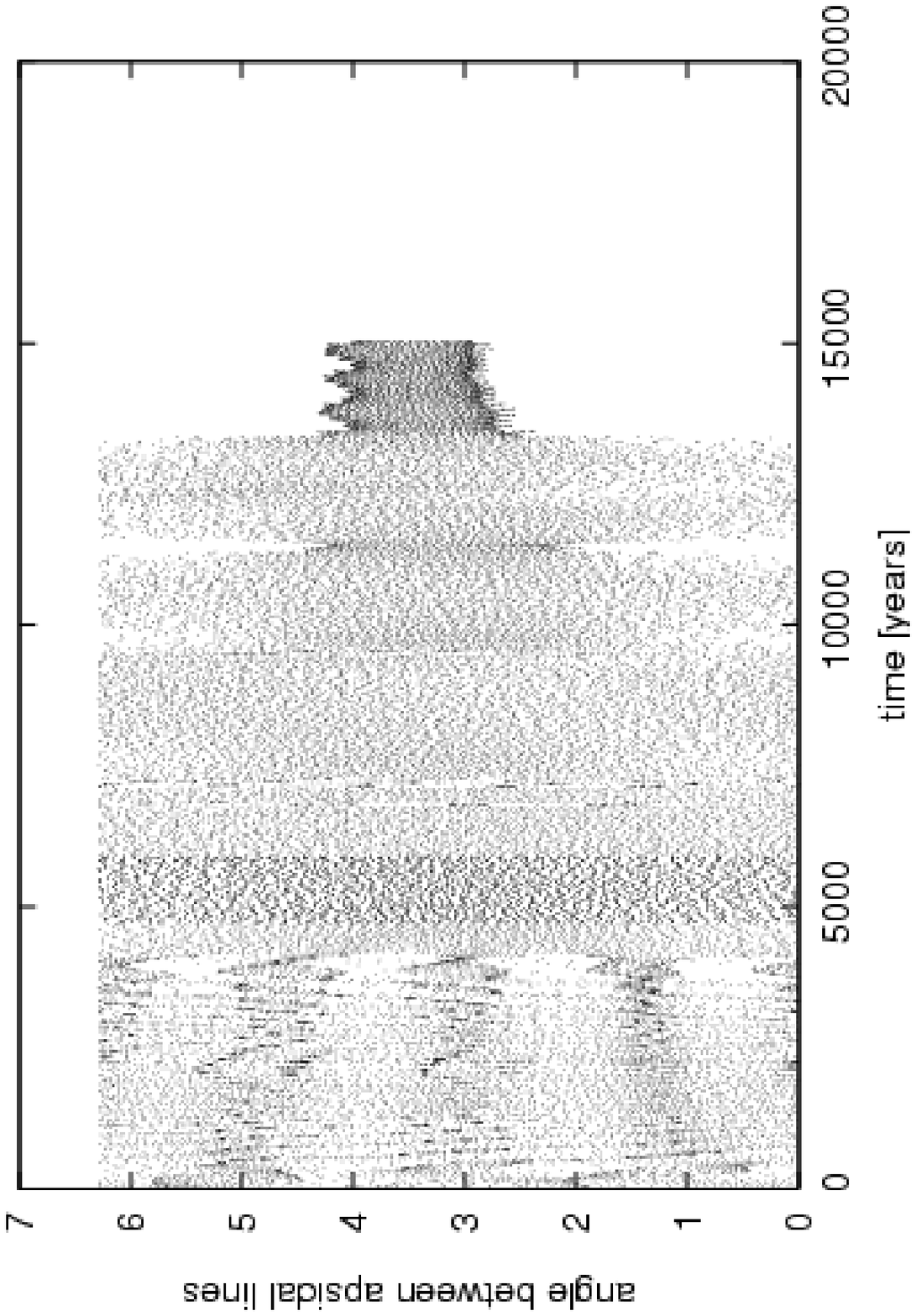}
\includegraphics{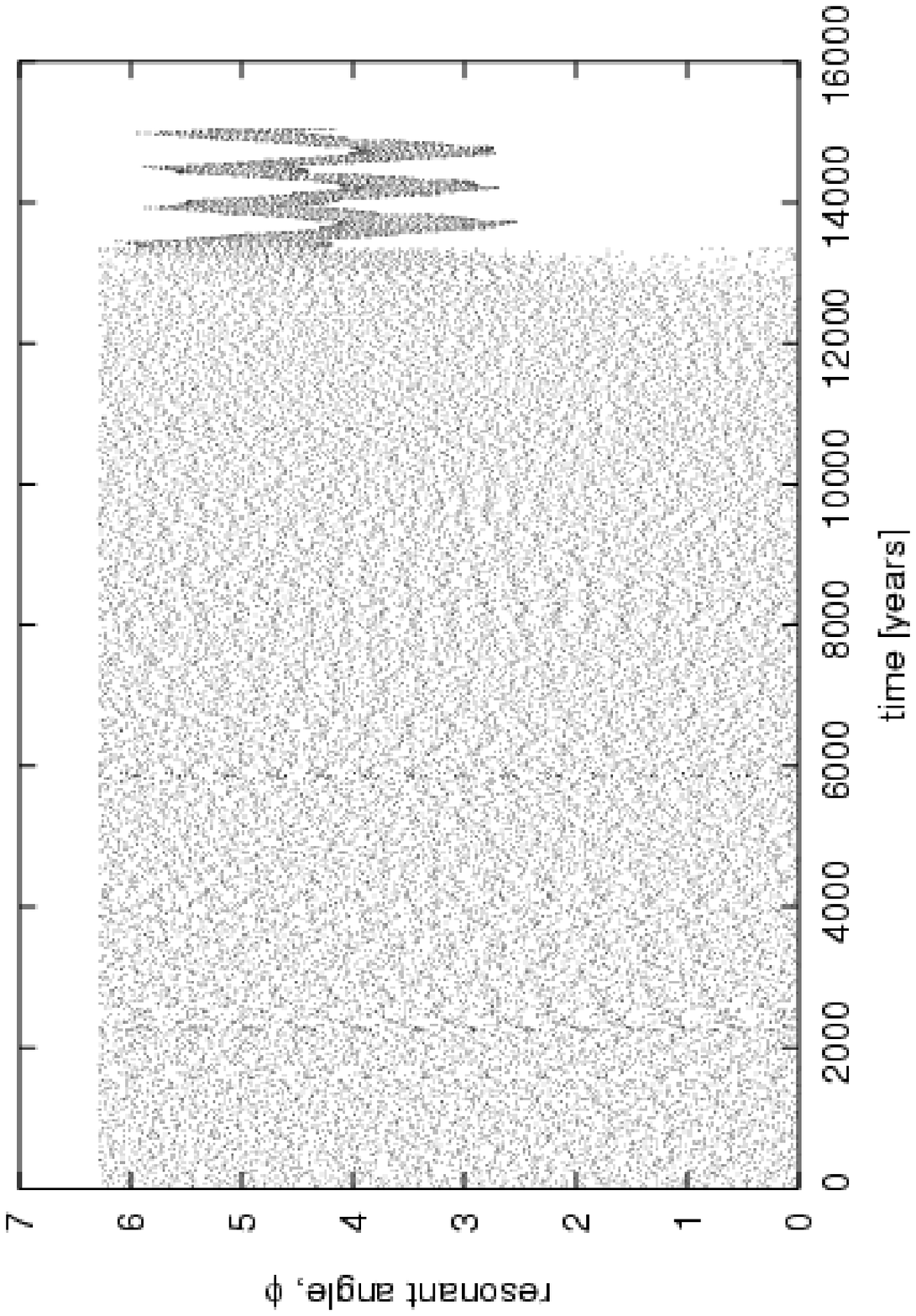}
\includegraphics{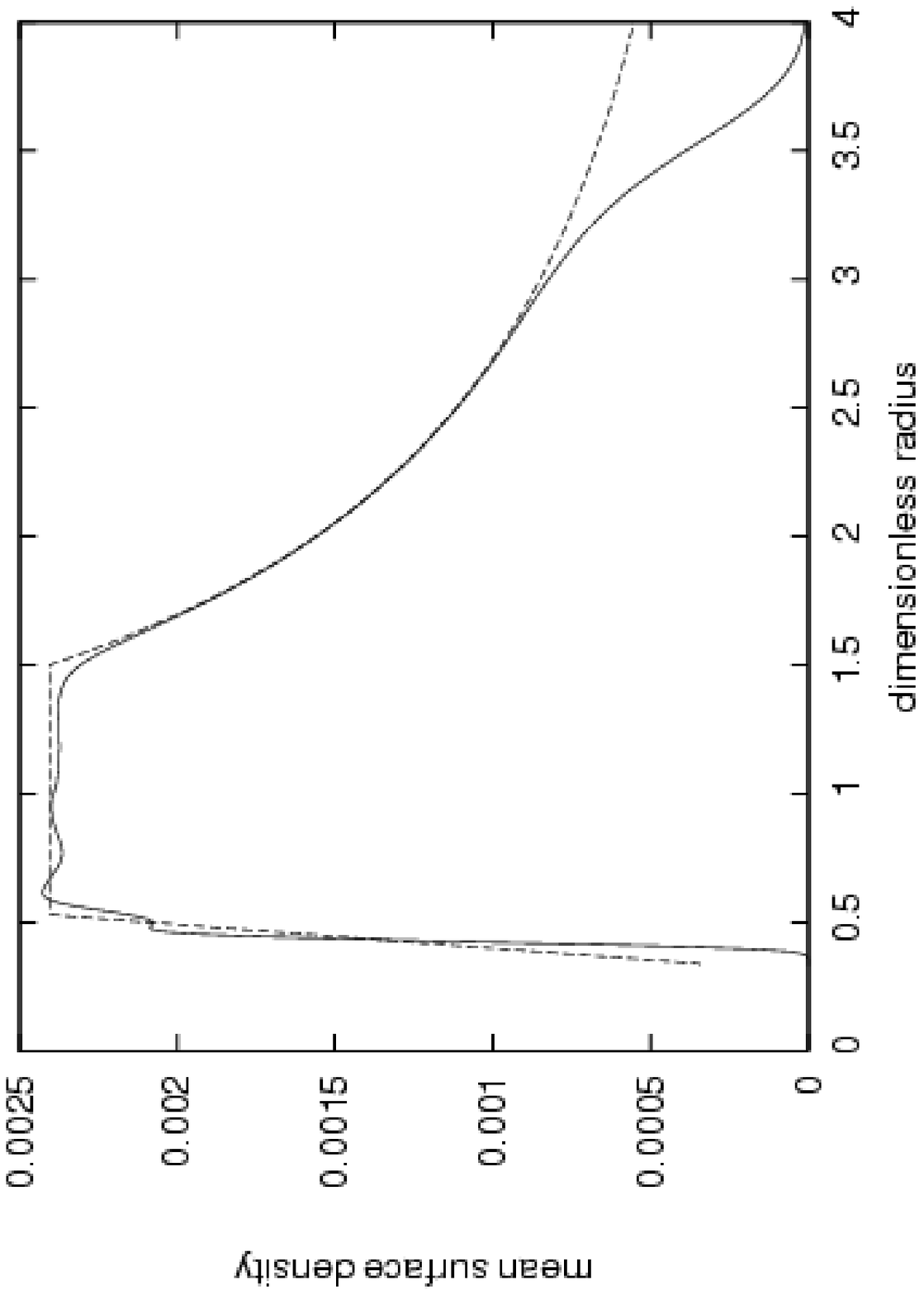}
\includegraphics{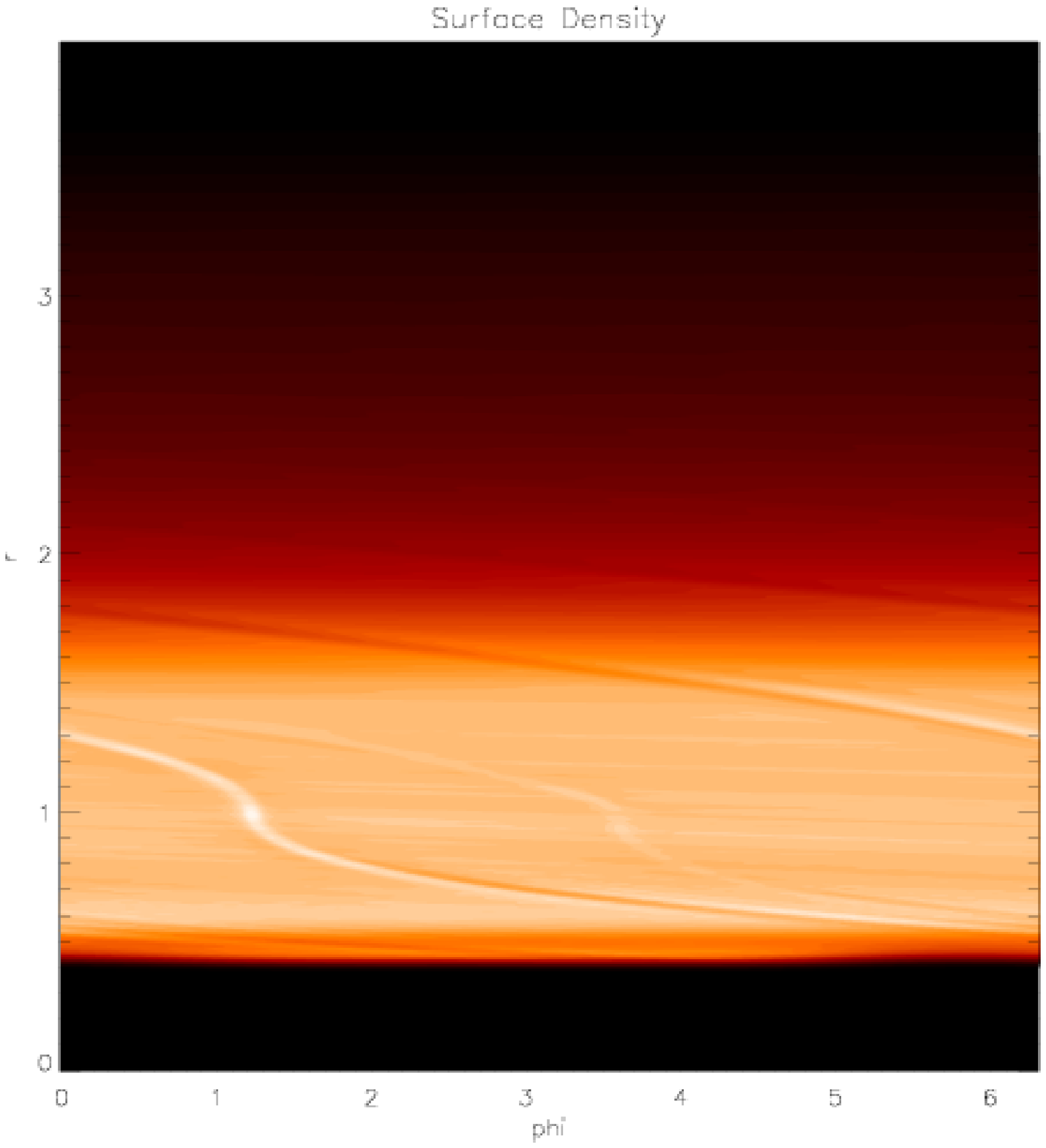}
\caption{\label{fig8}{As for Figure \ref{fig3} but here the 
planets initially  have $a_1/a_2 = 1.32.$ This system ended up in a 9:8
commensurability. Note the  eccentricity  increases produced by prior
resonance passages which do not completely decay between them. 
}}
\end{minipage}
\end{figure*}

\noindent 
When the  ratio of the initial semi-major axes is changed 
from  1.2 to 1.32 the two planets embedded in a disc with 
$\Sigma_0 = \Sigma_4$ become
trapped in a different resonance (compare Figure \ref{fig3}
and Figure \ref{fig8}). This is again consistent with
the presence  of  stochastic behaviour as  a different migration
history leads to different end states and it suggests that having  
crossed different resonances before attaining a current one may 
influence the final outcome.
 
\noindent 
A similar experiment has been performed but now  placing the  
planets close to  4:3 resonance. In this case
we have chosen  discs with  lower surface density,  such that
$\Sigma_0 = \Sigma_{0.5}$ and  $\Sigma_0 = \Sigma_{1}.$
In the former case the system enters into 4:3 resonance while 
in the latter the system passes through the 4:3 resonance  and 
becomes trapped in  the 6:5 resonance. 
\begin{figure}
\vskip 8cm
\includegraphics{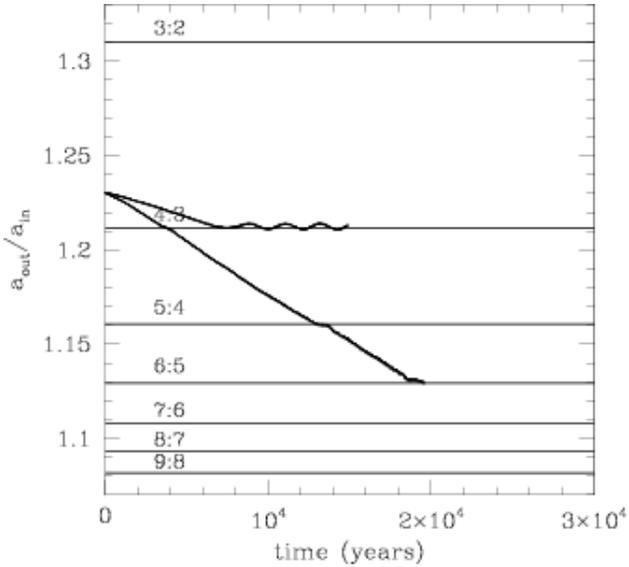}
\caption{\label{fig9}{The semi-major axis ratio  for 
$m_1 = 4 M_{\oplus}$ and  $m_2 =  1 M_{\oplus}$  with
surface density scaling parameter  $ \Sigma_0 = \Sigma_{0.5}$ - 
upper curve and $\Sigma_0 = \Sigma_{1}$ - lower
curve.}}
\end{figure}

\noindent  
When $\Sigma_0 = \Sigma_{0.5}$ the planets enter  4:3 commensurability
at a  time around 6000 years. The eccentricities of both planets  grow.
The resonant angle oscillates around 260$^{\circ}$ with large amplitude,
higher than it was in the case when the initial planet separation 
was  smaller, namely 1.2 (Figure \ref{fig6}).

\begin{figure*}
\begin{minipage}{175mm}
\vspace{220mm}
\includegraphics{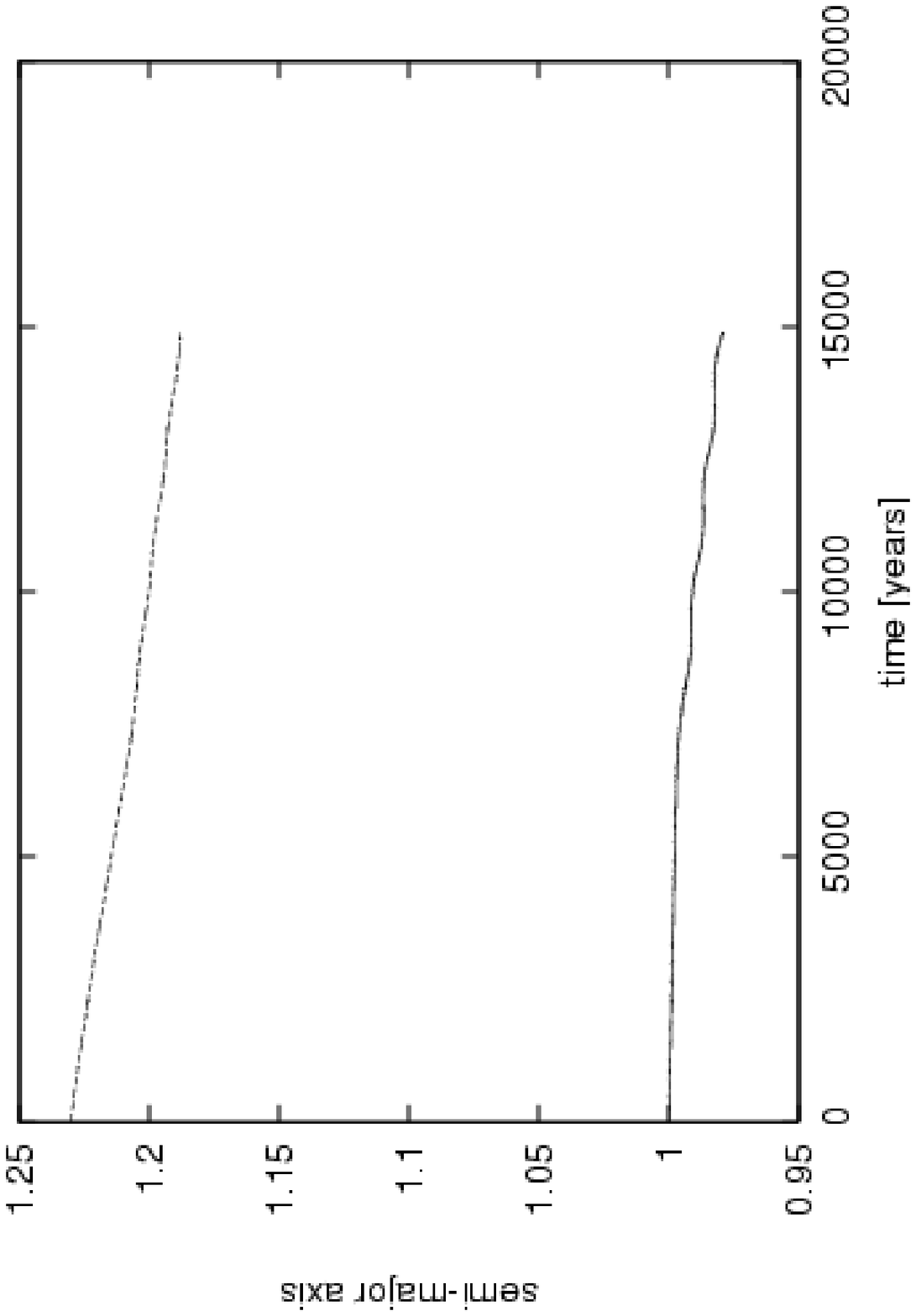}
\includegraphics{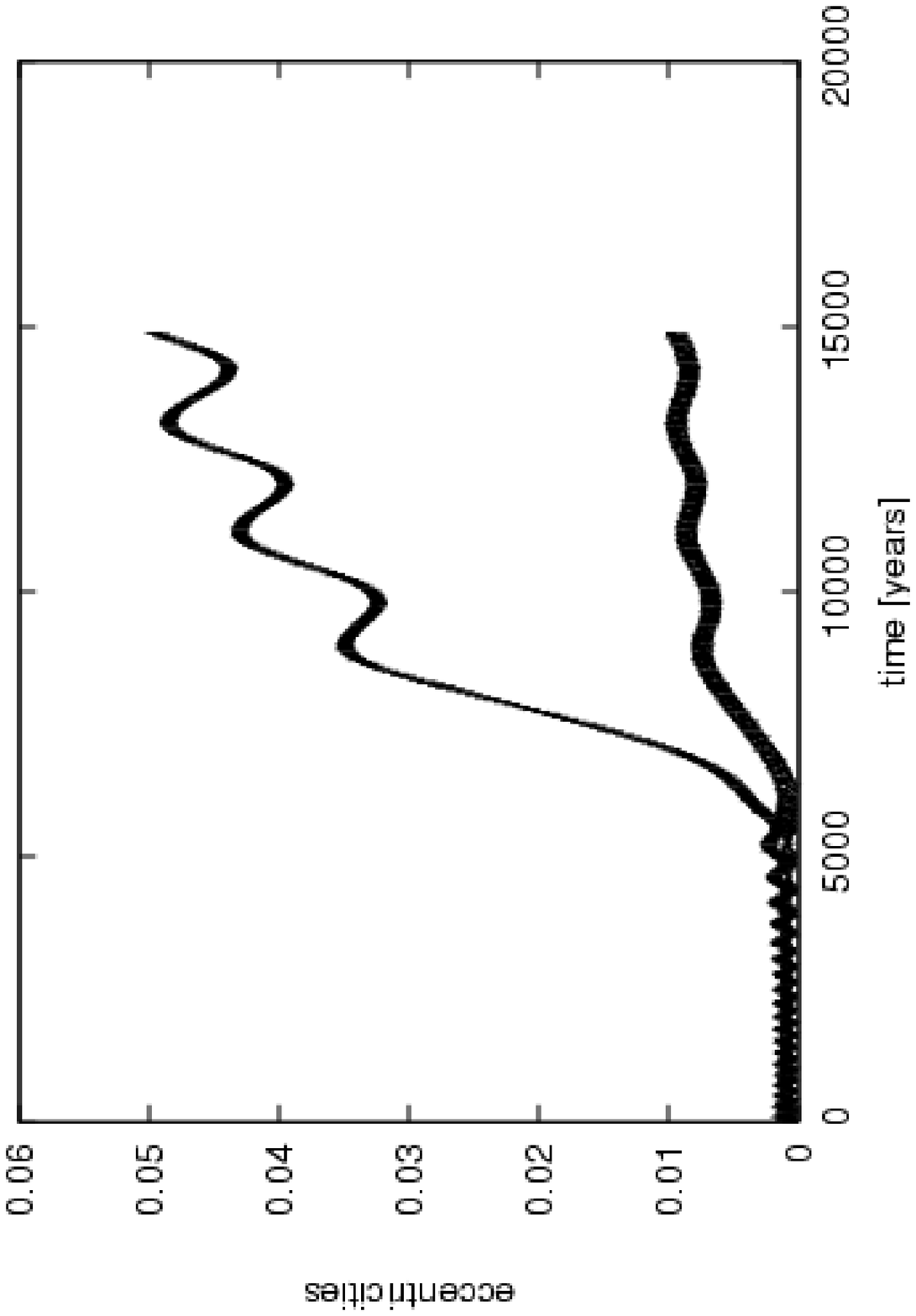}
\includegraphics{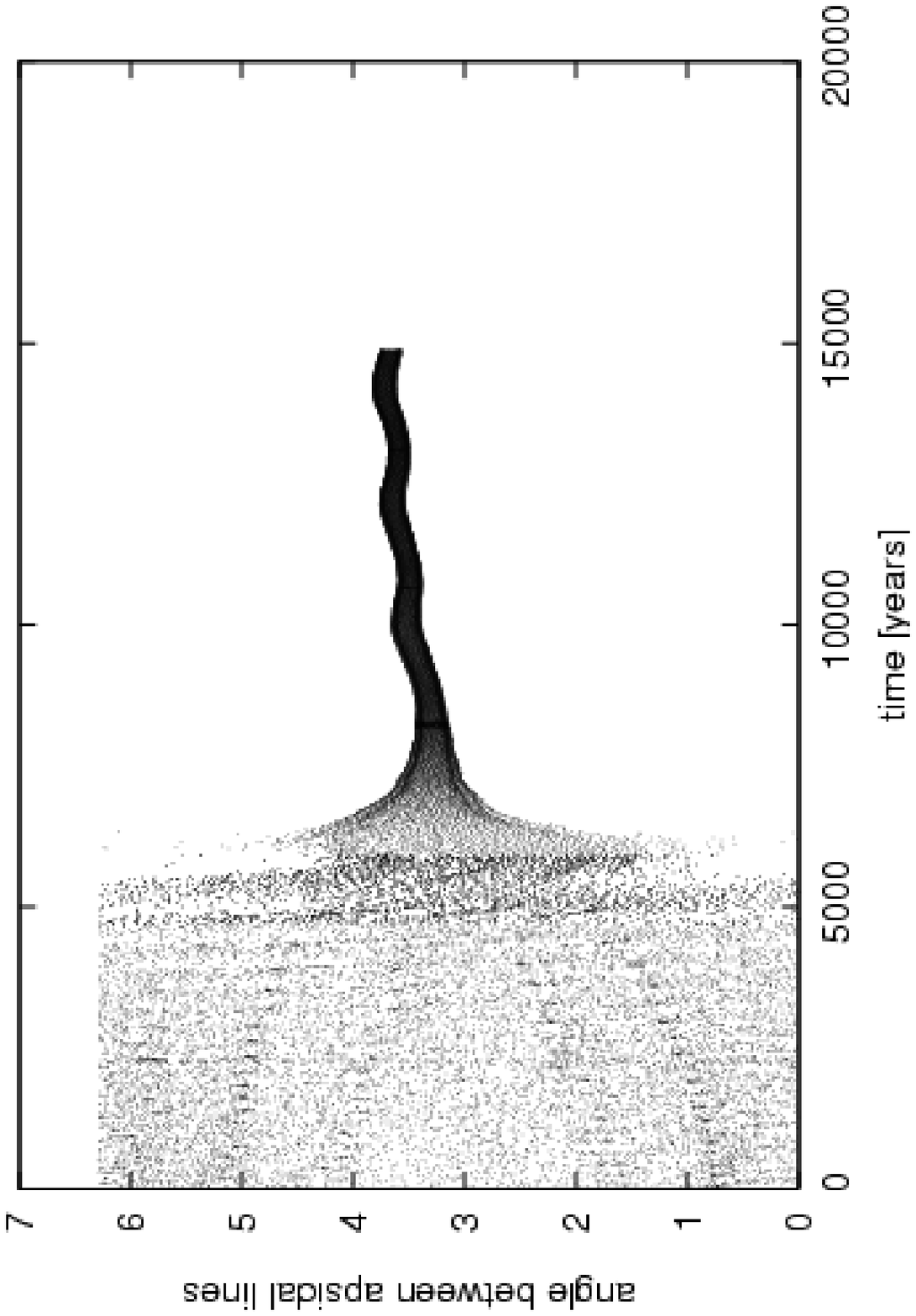}
\includegraphics{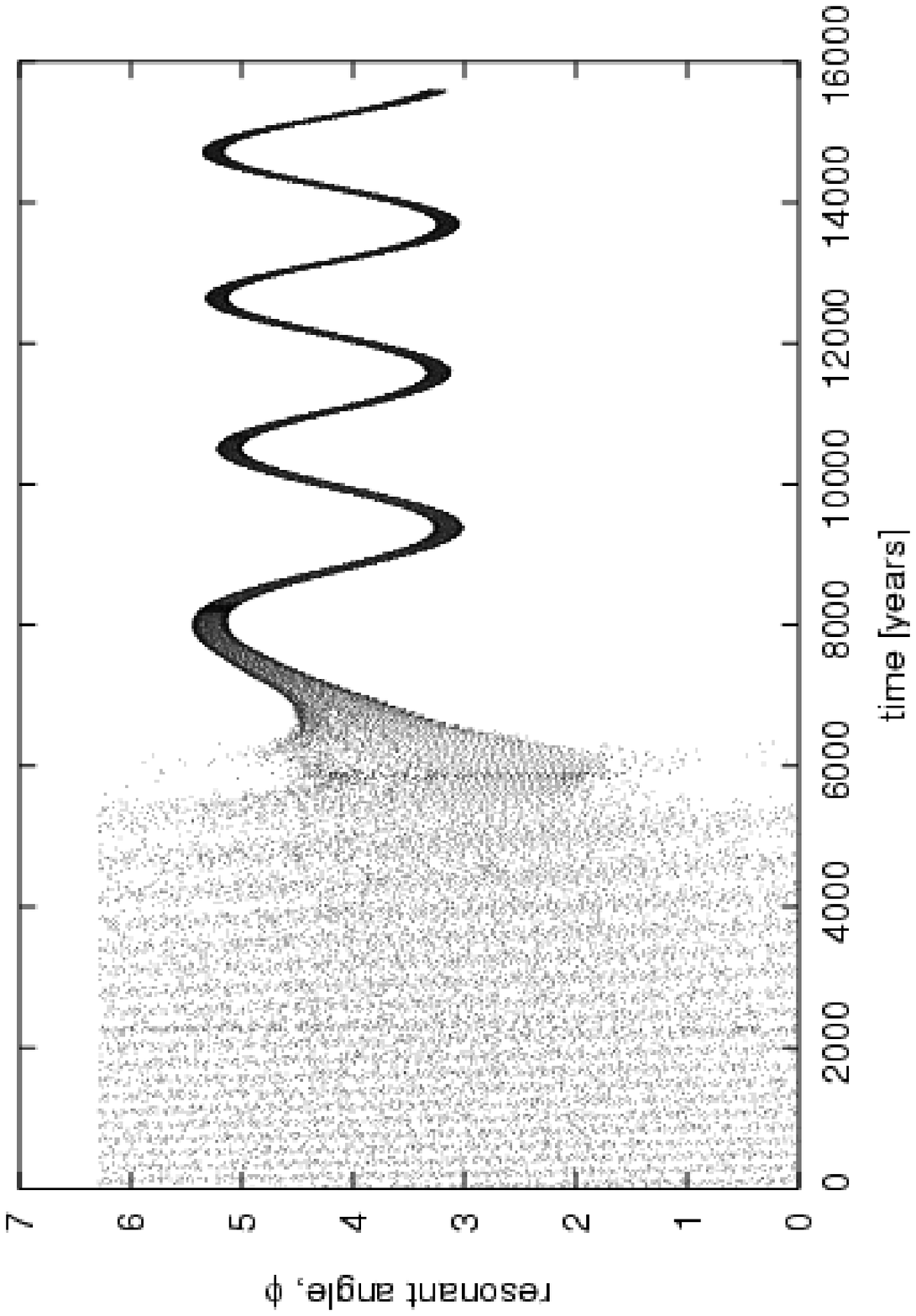}
\includegraphics{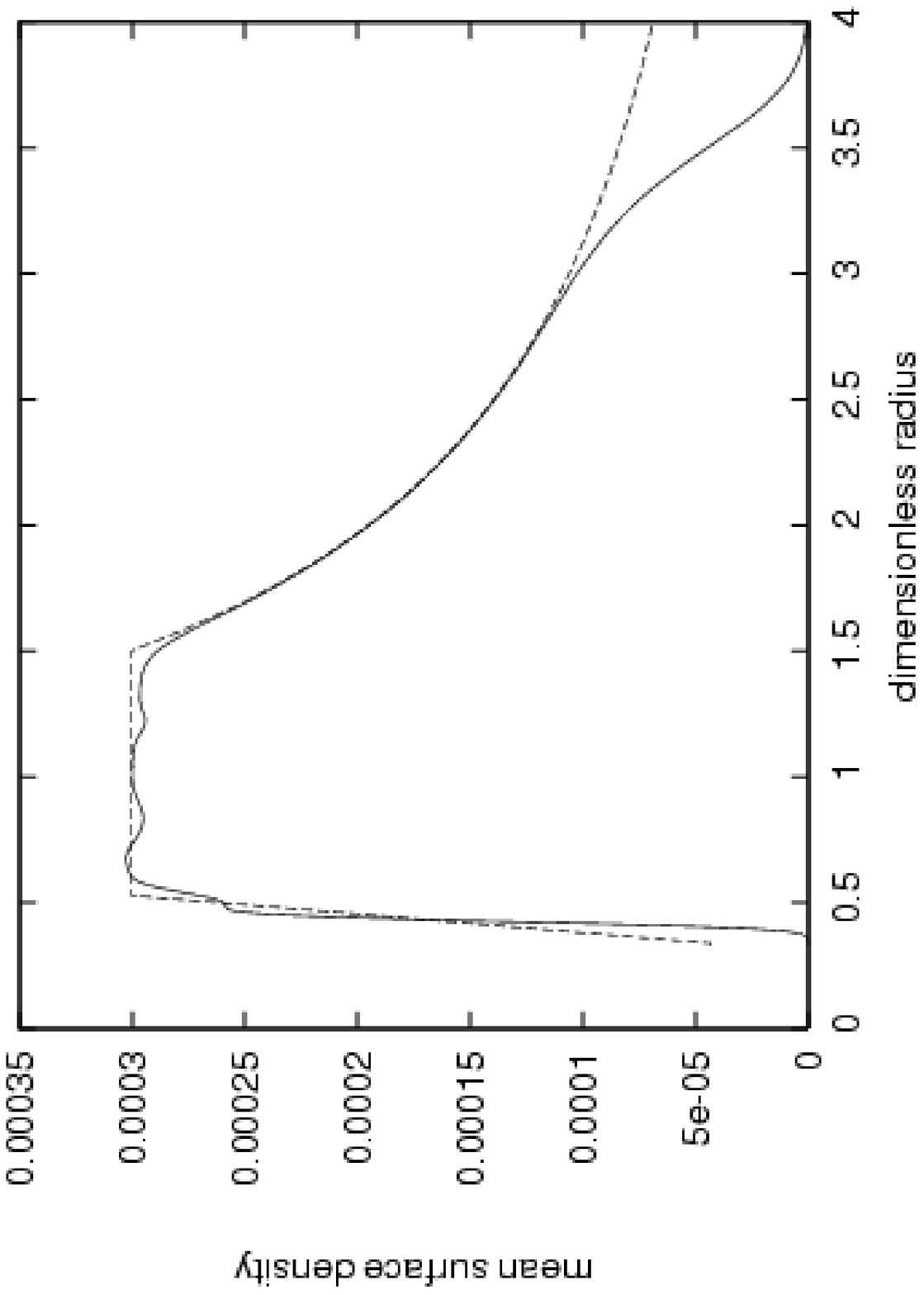}
\includegraphics{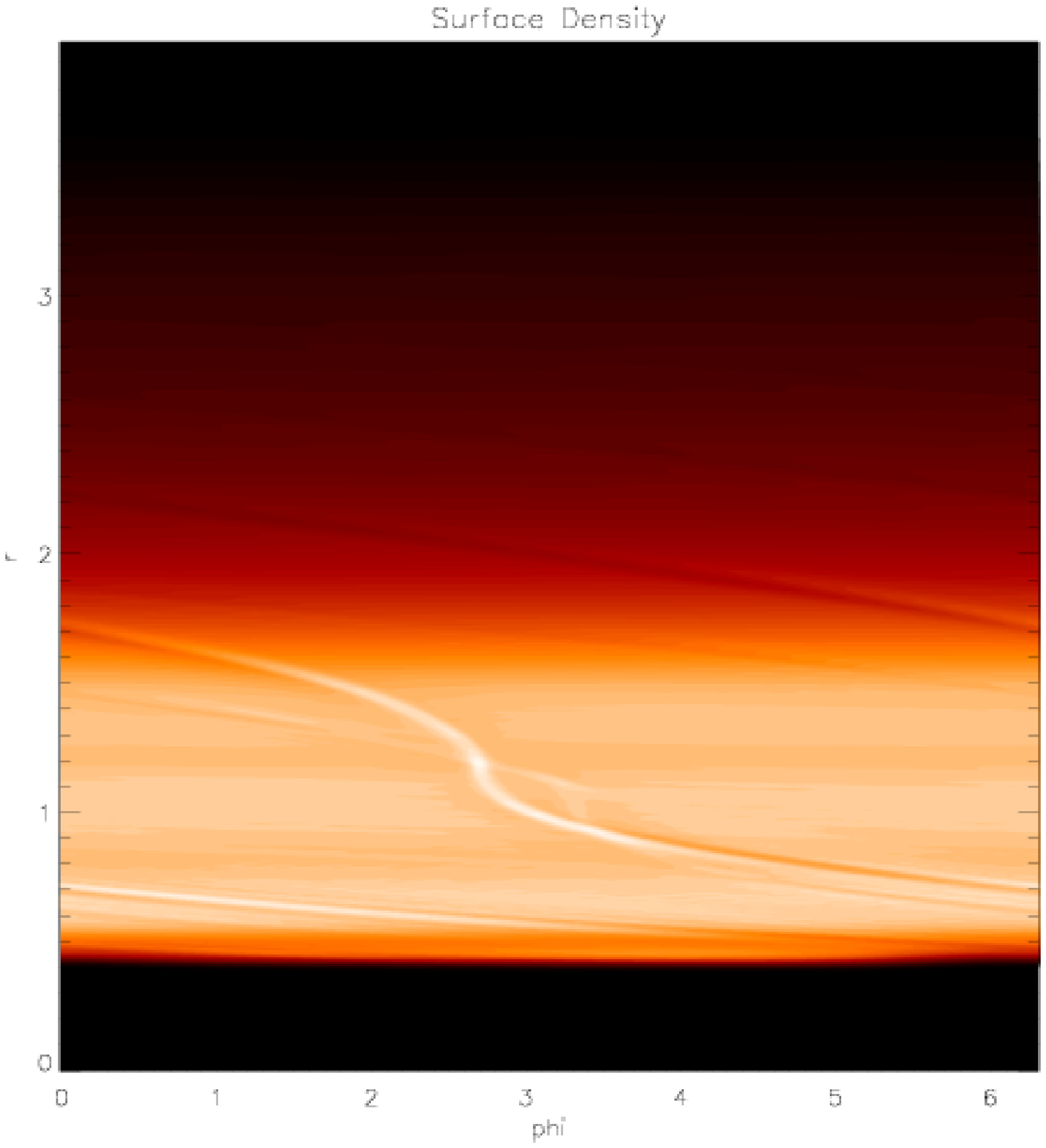}
\caption{\label{fig10}{As for Figure \ref{fig6} but here the planets
initially have $a_1/a_2 = 1.23.$
}}
\end{minipage}
\end{figure*}

\begin{figure*}
\begin{minipage}{175mm}
\vspace{220mm}
\includegraphics{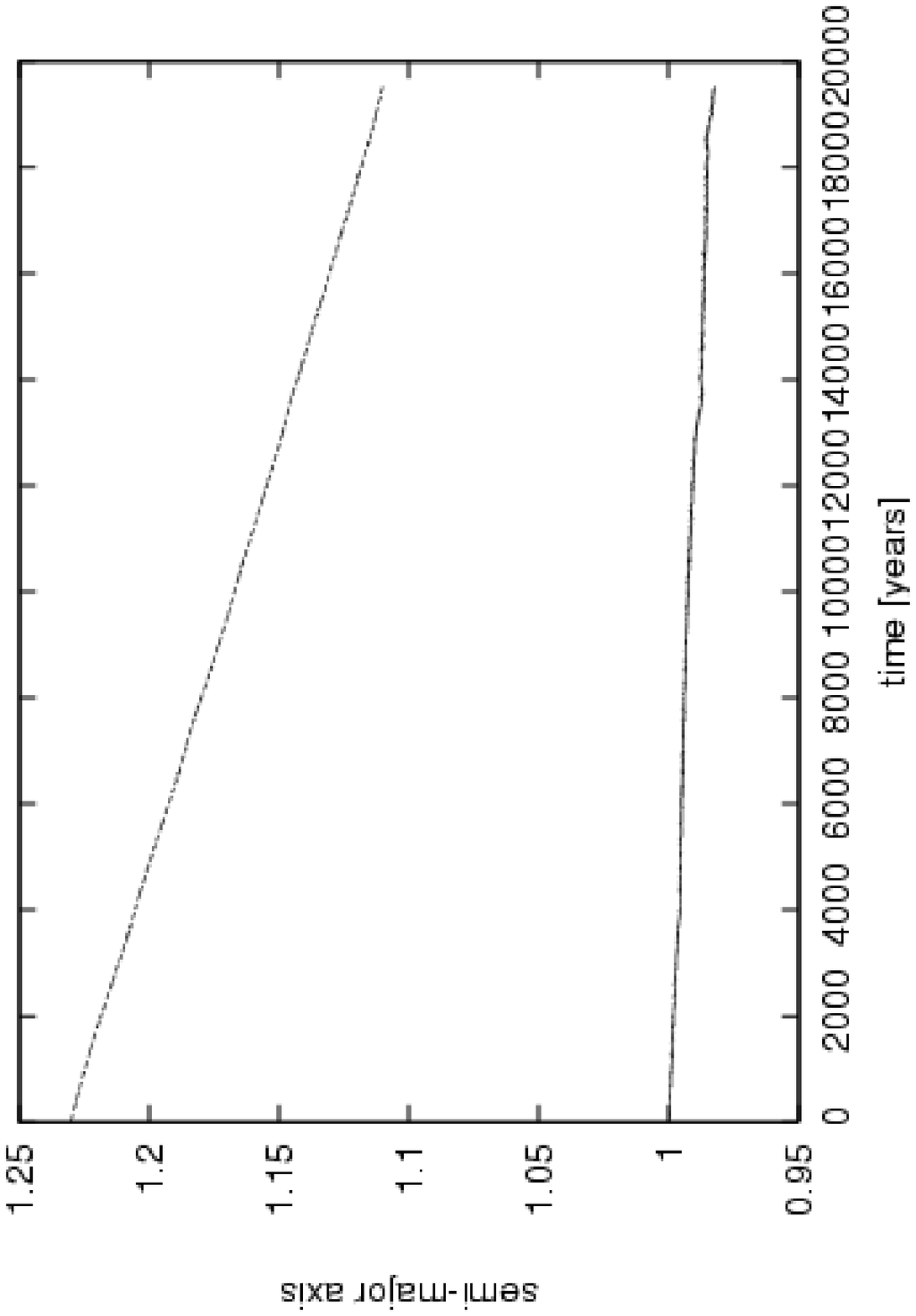}
\includegraphics{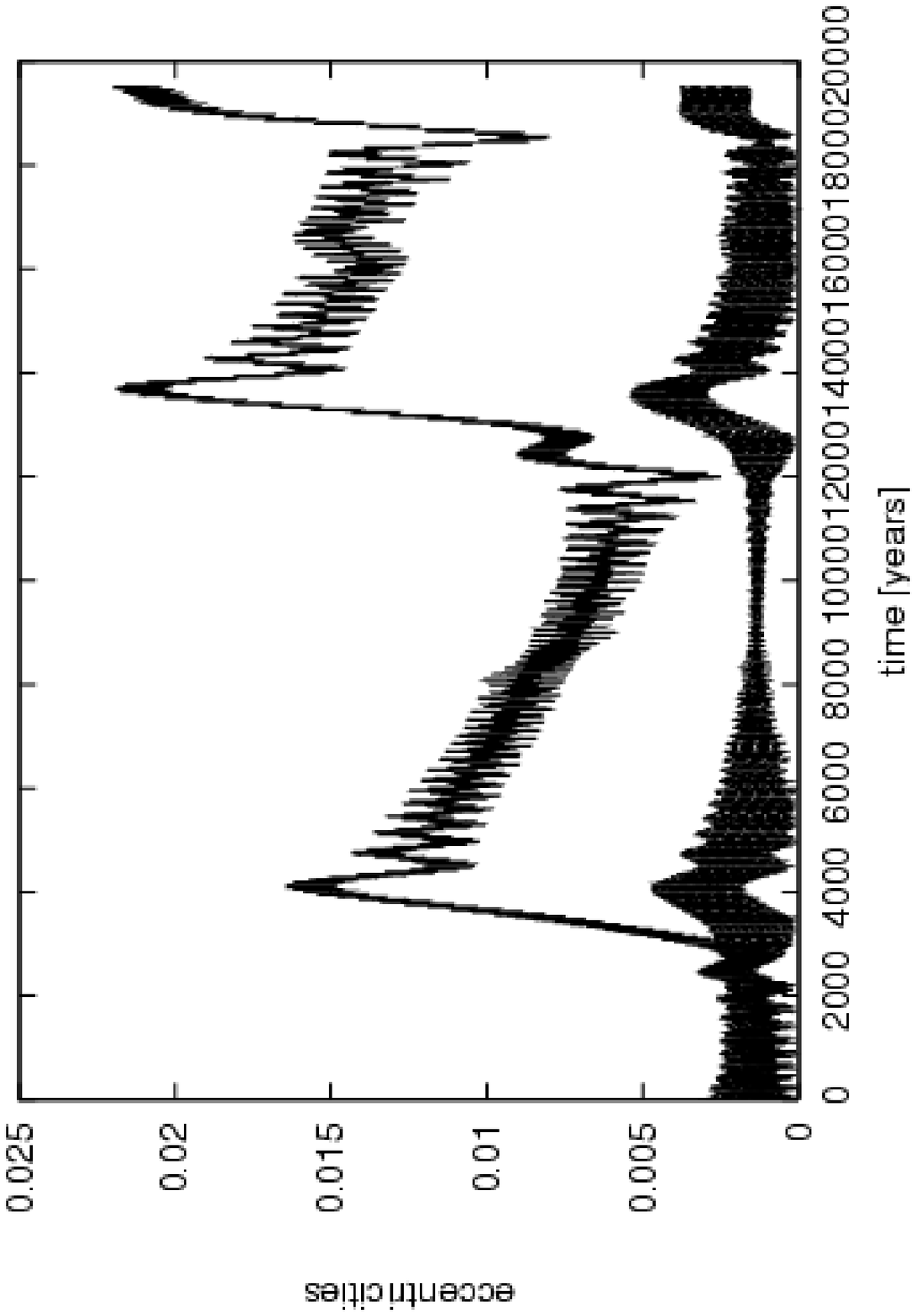}
\includegraphics{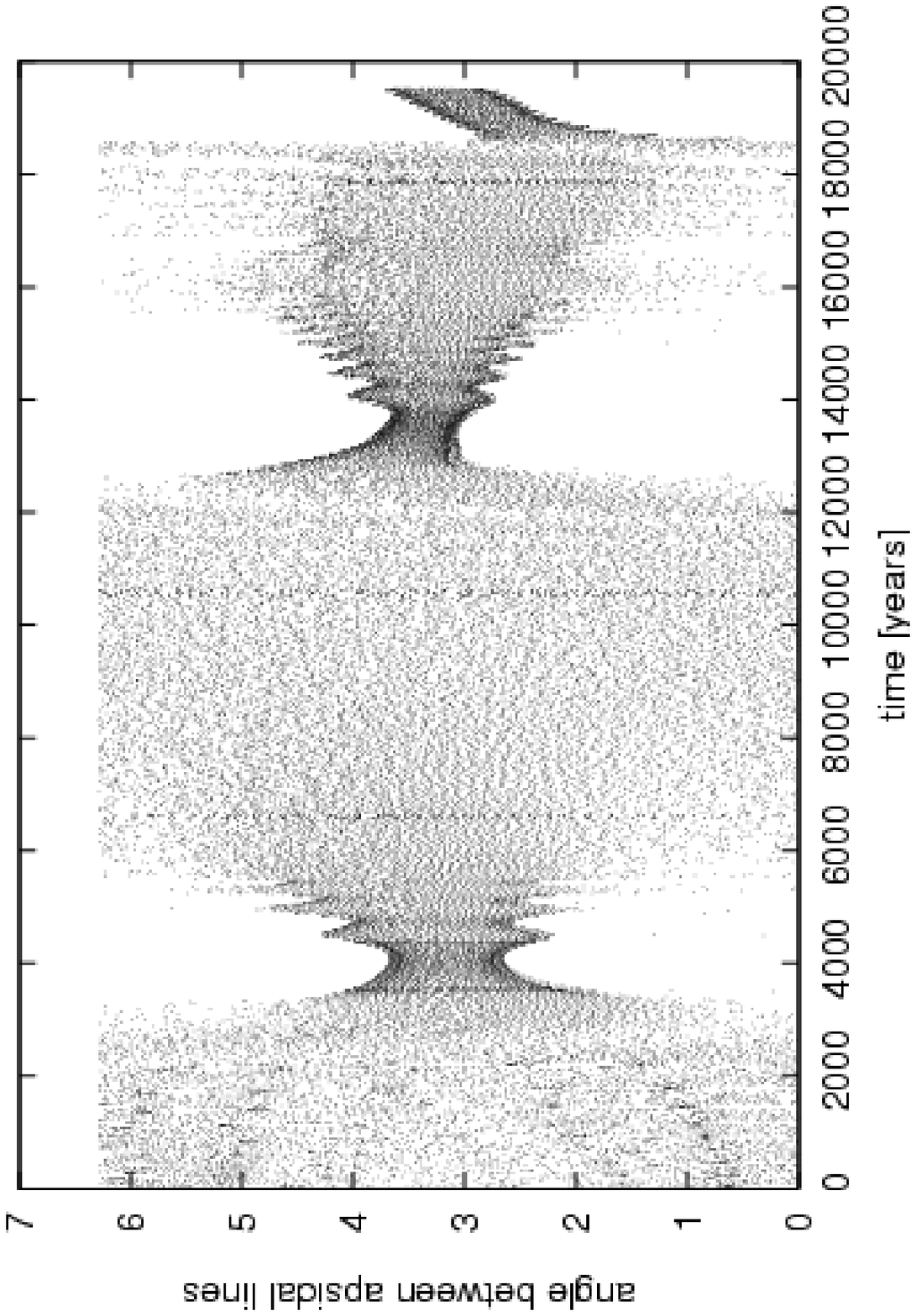}
\includegraphics{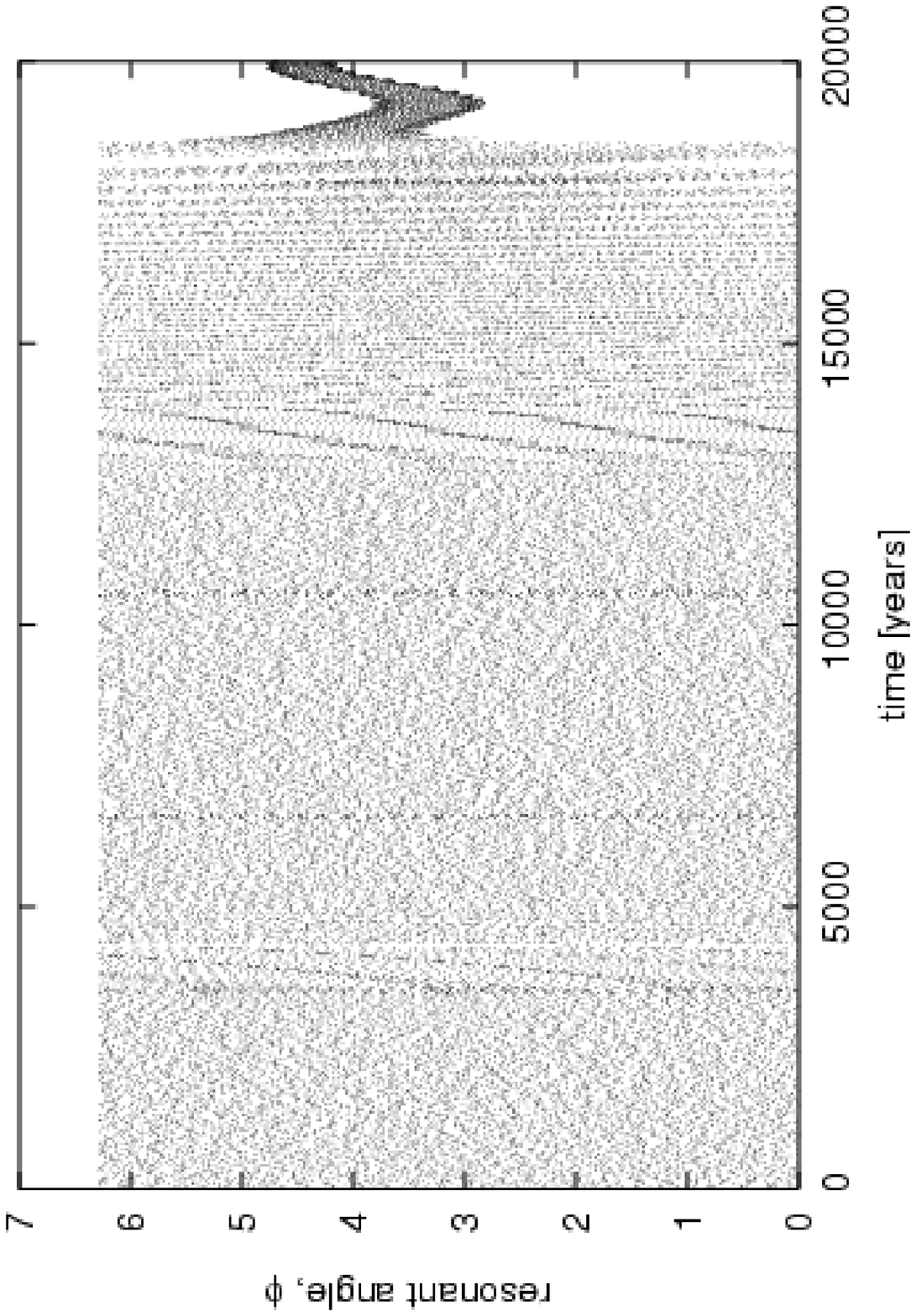}
\includegraphics{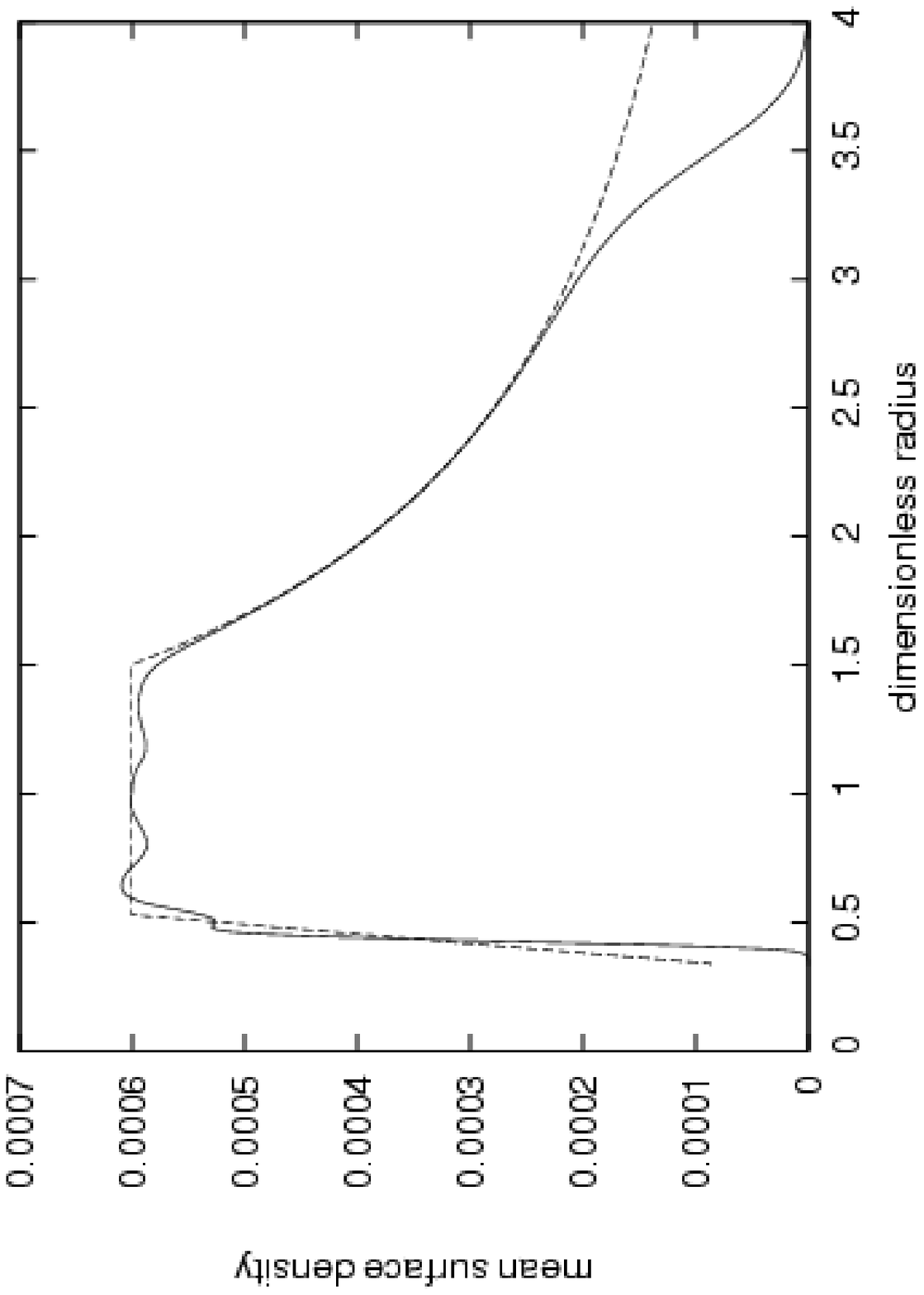}
\includegraphics{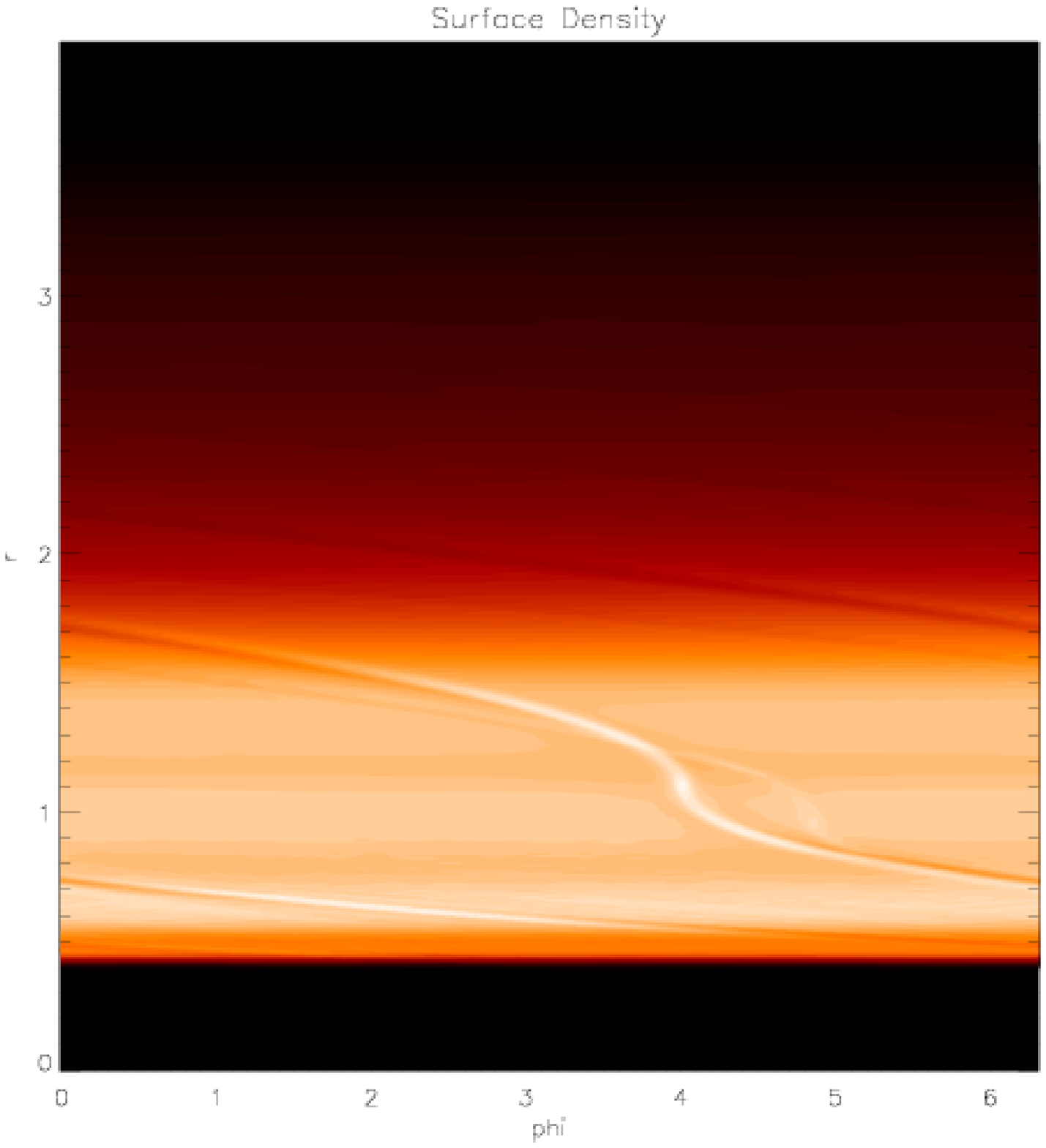}
\caption{\label{fig11}{As for Figure \ref{fig5} but   when 
the planets are initiated with $a_1/a_2 =  1.23.$
Note that the system fails to become trapped in
the 4:3 and 5:4 commensurabilities
but nonetheless shows associated
strong resonant interaction for long periods of time
around $t = 4000$ and $t = 14000$ years respectively.
}}
\end{minipage}
\end{figure*}

\noindent 
For the disc with the surface density $\Sigma_0 = \Sigma _1$ we 
can follow in detail the passage of the planets through the  4:3 
resonance  and the temporary trapping in the  5:4
resonance. In Figure \ref{fig11} at time 4000 years there is 
a characteristic  rapid  increase in the eccentricities of both 
planets   followed by a slower decrease. The angle
between the  apsidal lines also shows the  expected behaviour 
during resonance crossing. This evolution should be compared with 
that   shown in Figure \ref{fig5},
for the same pair of planets embedded in the same disc but  starting 
with  a smaller initial planet separation. The effect of a prior 
resonance passage before approaching and  becoming relatively stably 
trapped in the 5:4 resonance is not present in this case. So it is 
likely that the mutual interaction of planets during resonance 
crossing influences their subsequent evolution and whether higher 
$p$ resonances are attained or not and that these planets are either 
close to or in the chaotic regime. This behaviour may be related to
the fact that eccentricities excited during resonance passages
do not completely decay between them. 

\subsection{Slower relative migration - planets with equal masses}

\begin{figure}
\vskip 8cm
\includegraphics{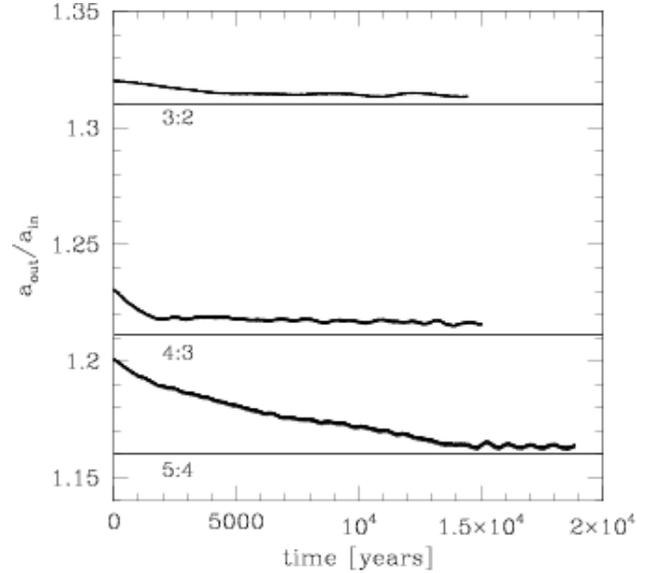}
\caption{\label{fig12}{The semi-major axis ratio when
$m_1 = m_2 = 4M_{\oplus}$ with 
$\Sigma_0 = \Sigma_{1}$ (uppermost curve) and with 
$\Sigma_0 = \Sigma_{4}$ (two lower curves) }}
\end{figure}

\begin{figure*}
\begin{minipage}{175mm}
\vspace{220mm}
\includegraphics{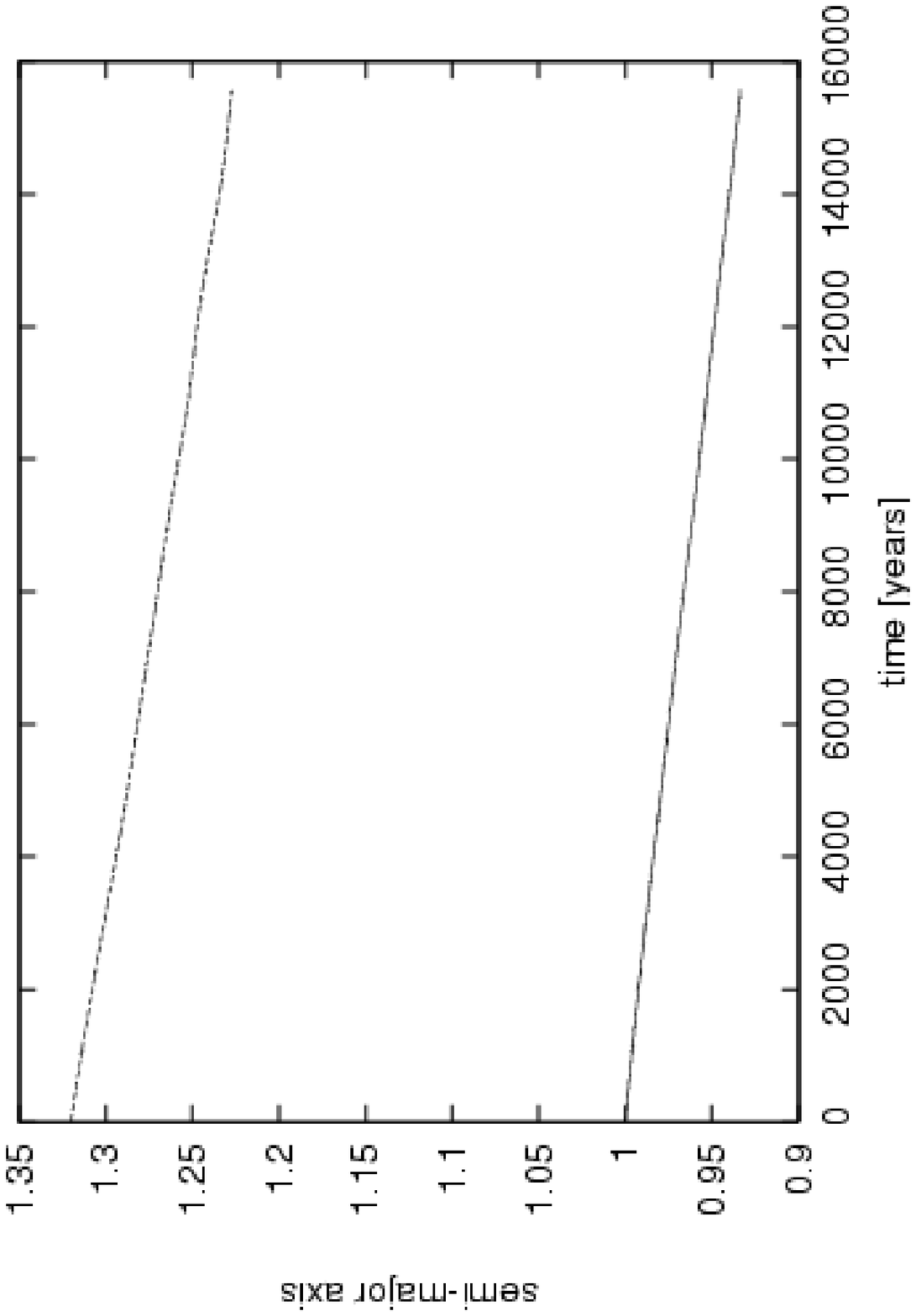}
\includegraphics{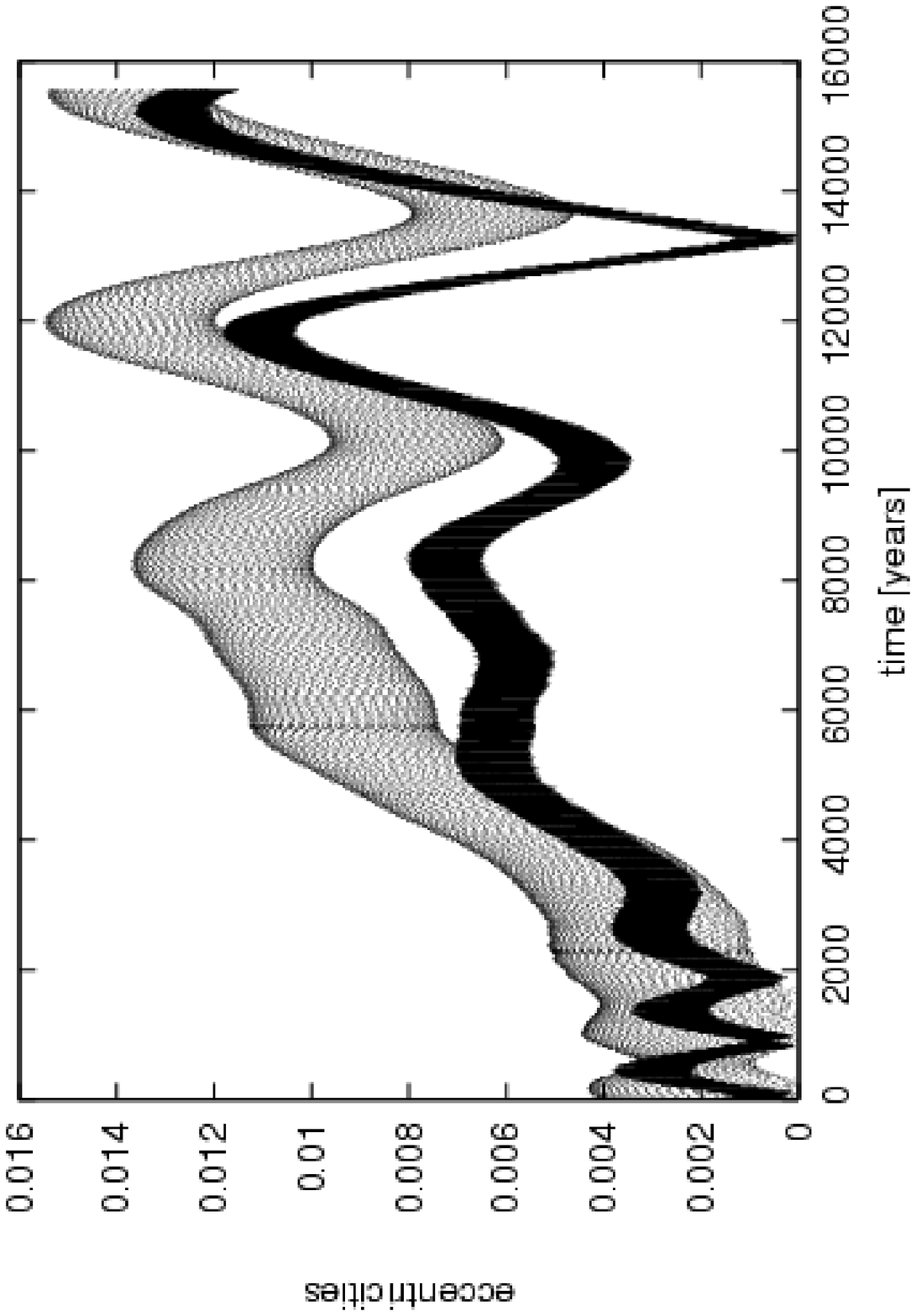}
\includegraphics{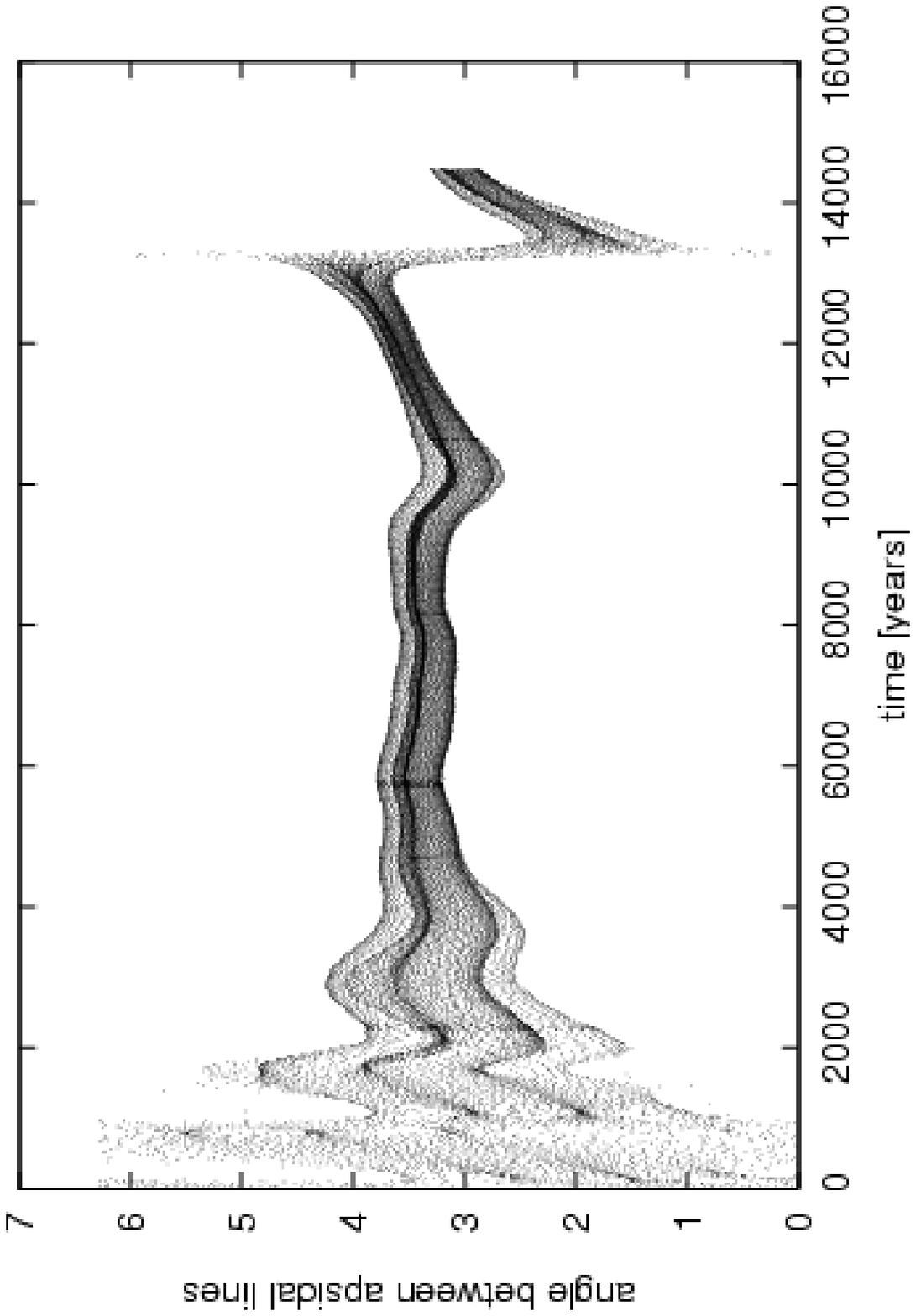}
\includegraphics{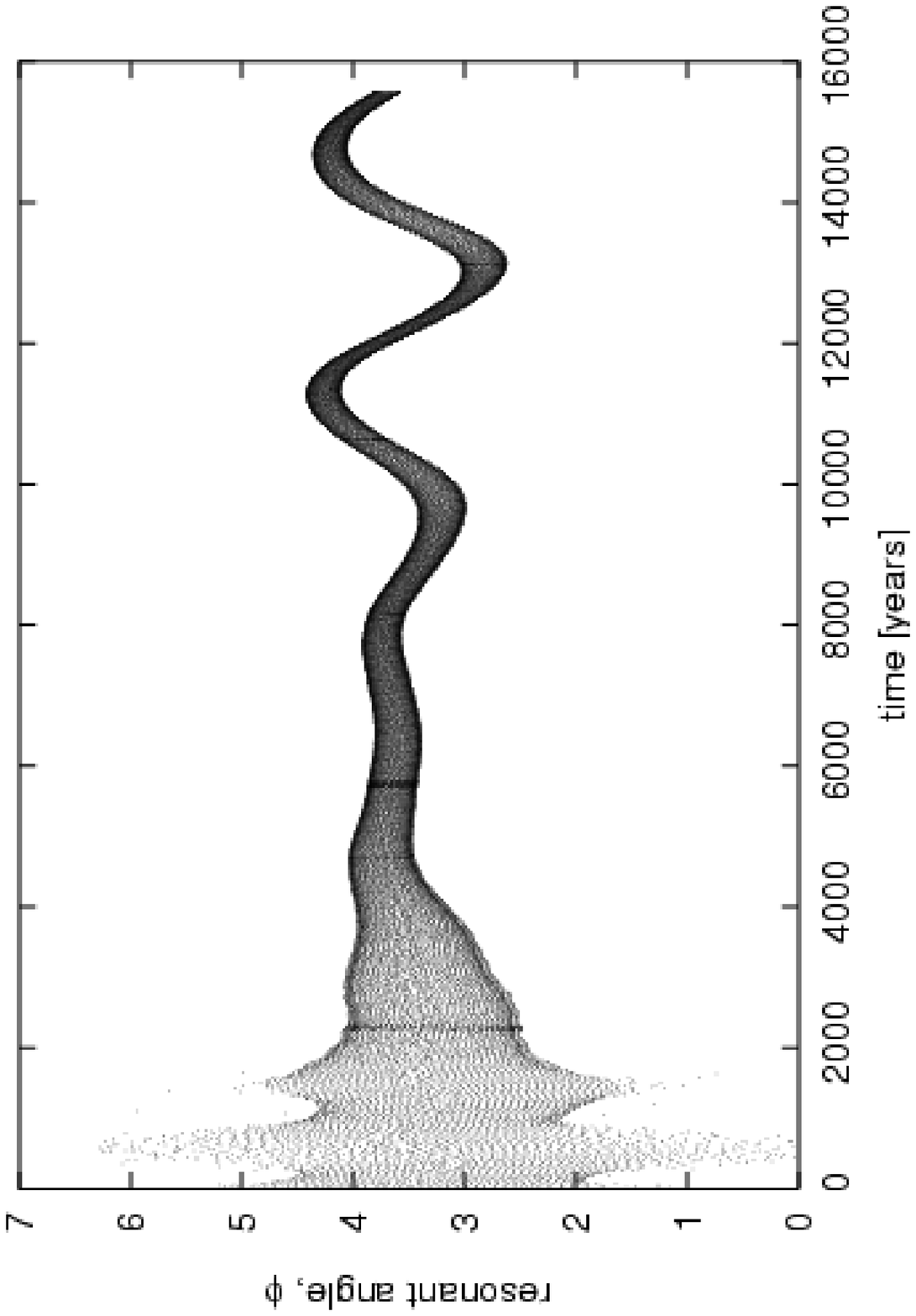}
\includegraphics{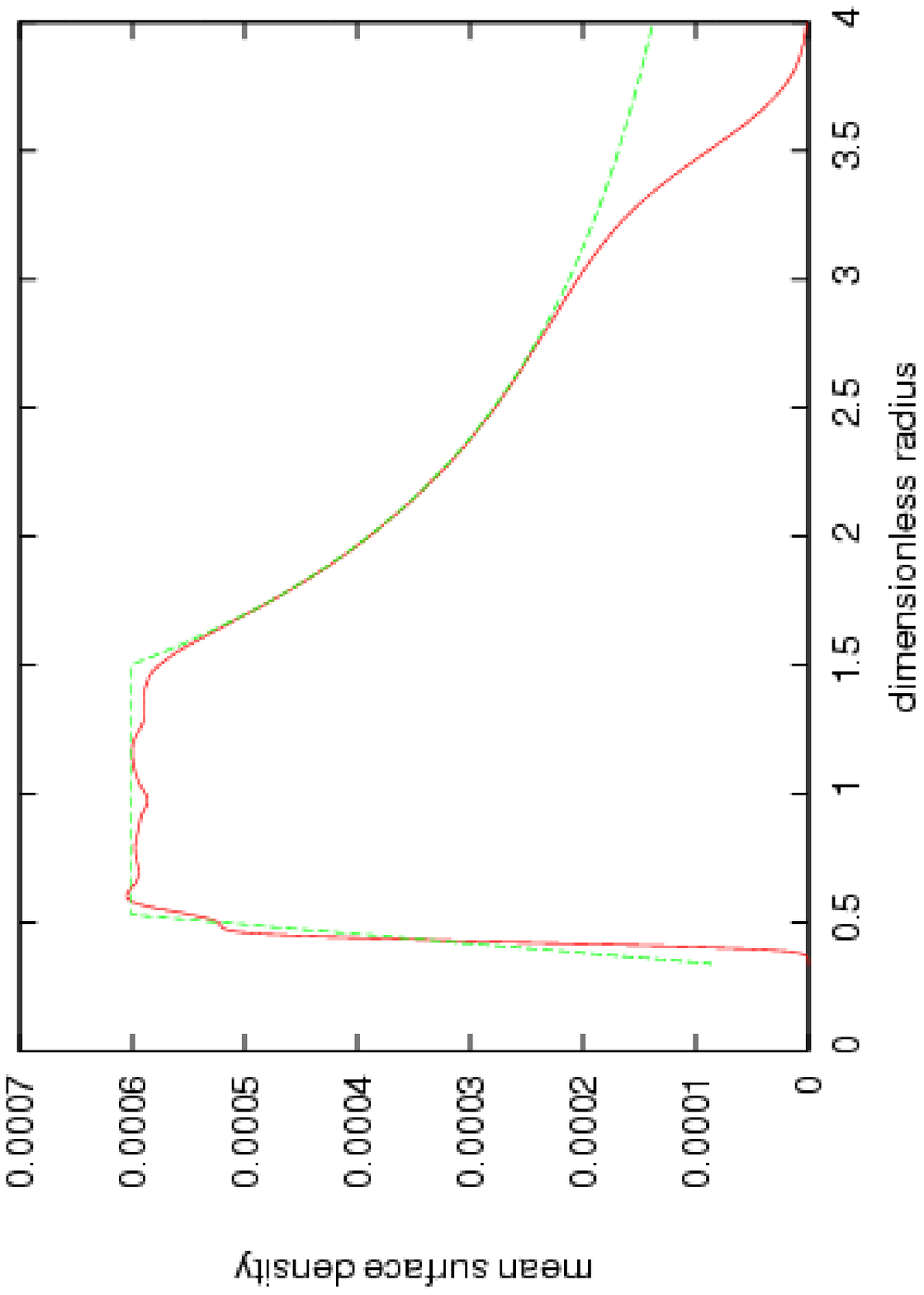}
\includegraphics{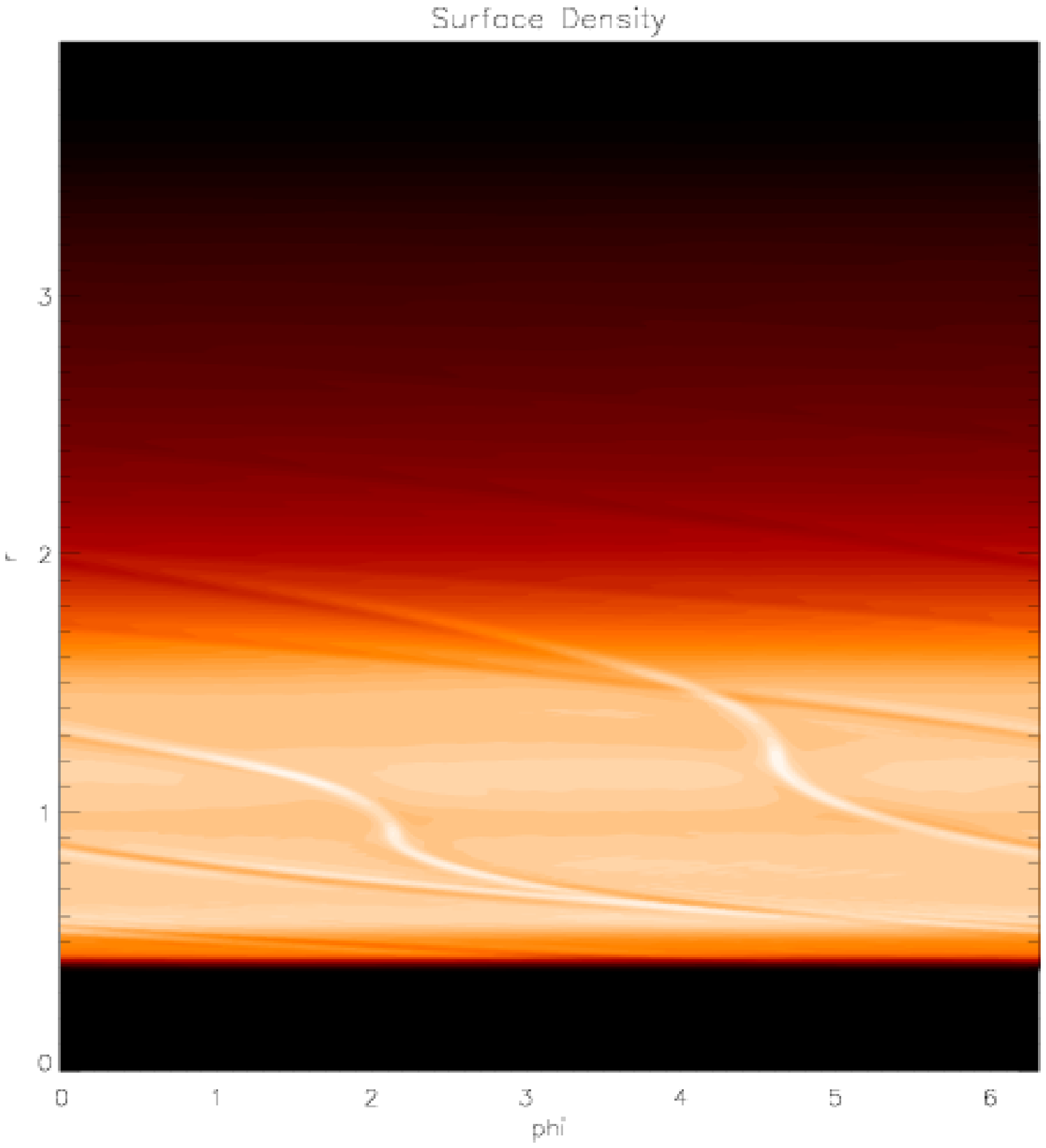}
\caption{\label{fig13}{
The evolution of semi-major axes, eccentricities, angle 
between apsidal lines and resonant angle for two planets 
with  equal mass, $m_{1} = m_{2}  = 4 M_{\oplus},$ 
migrating towards a central star and embedded in a disc 
with $\Sigma_0 =\Sigma _1.$ (four upper panels).
The mean surface density profile of the disc near the end 
of the simulations (solid line),  together with
the initial surface density profile (dashed
line) and  a surface density contour plot
near the end of the simulation  are given in the   
lowest left and  right  panels respectively.
}}
\end{minipage}
\end{figure*}

\begin{figure*}
\begin{minipage}{175mm}
\vspace{220mm}
\includegraphics{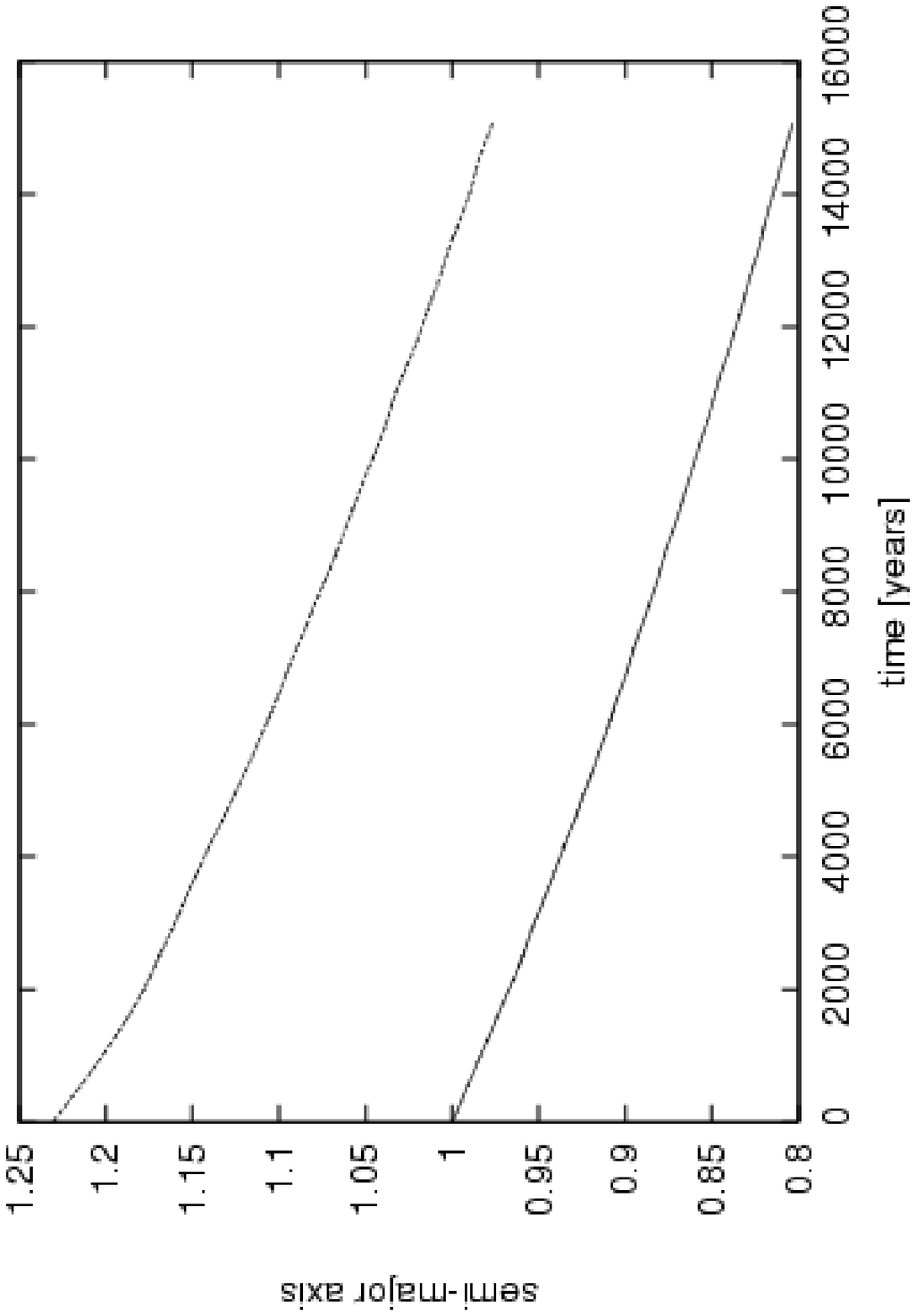}
\includegraphics{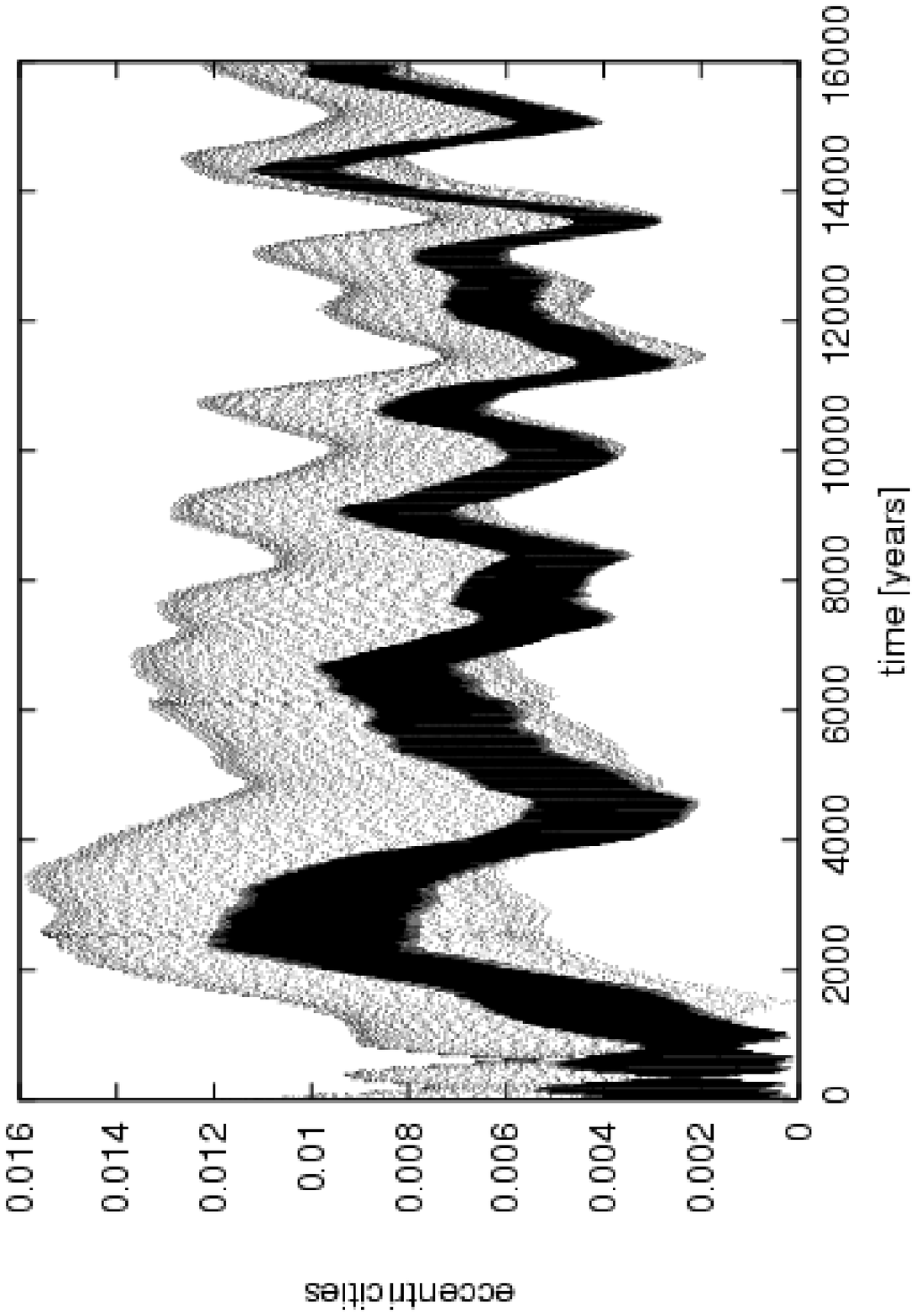}
\includegraphics{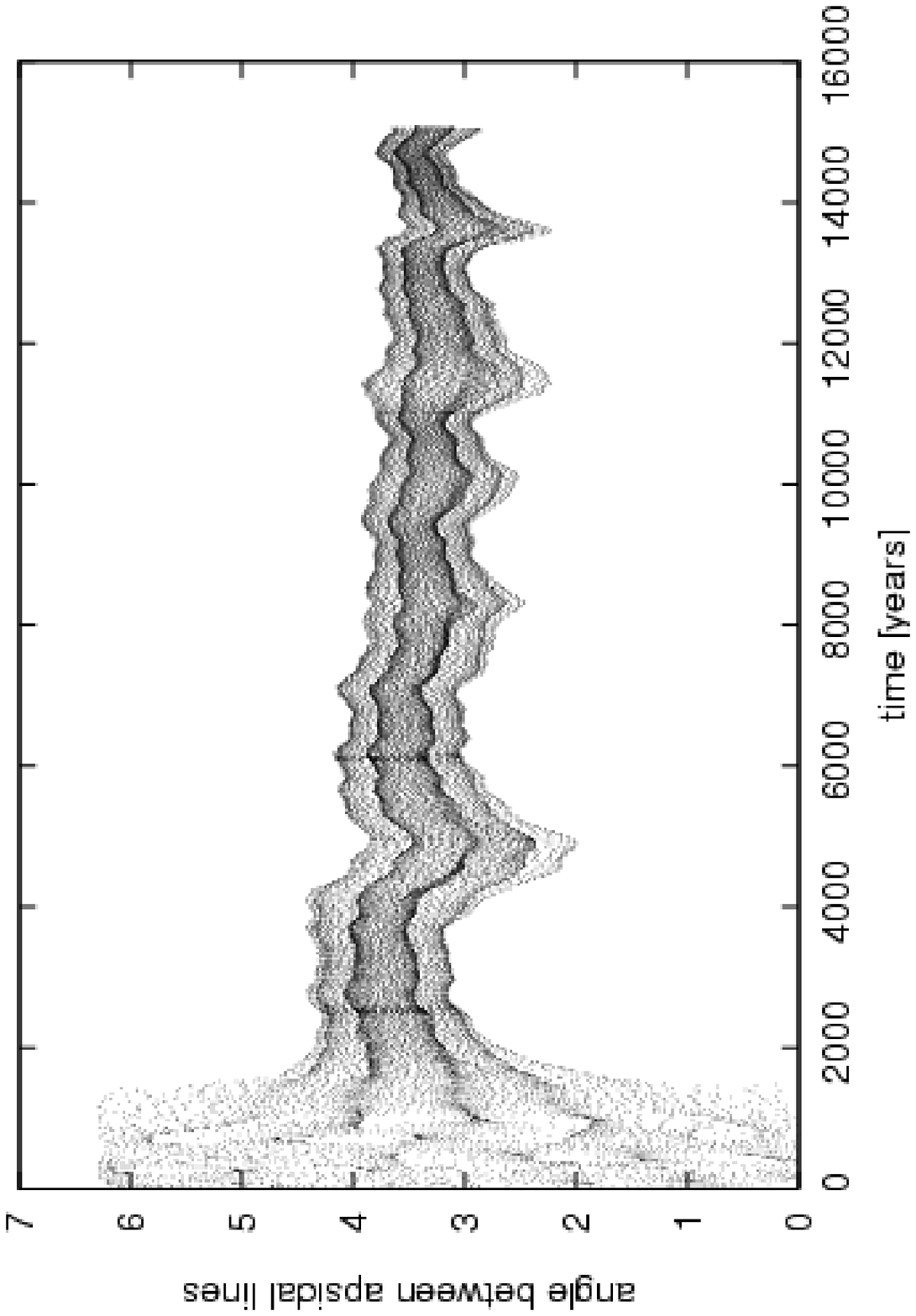}
\includegraphics{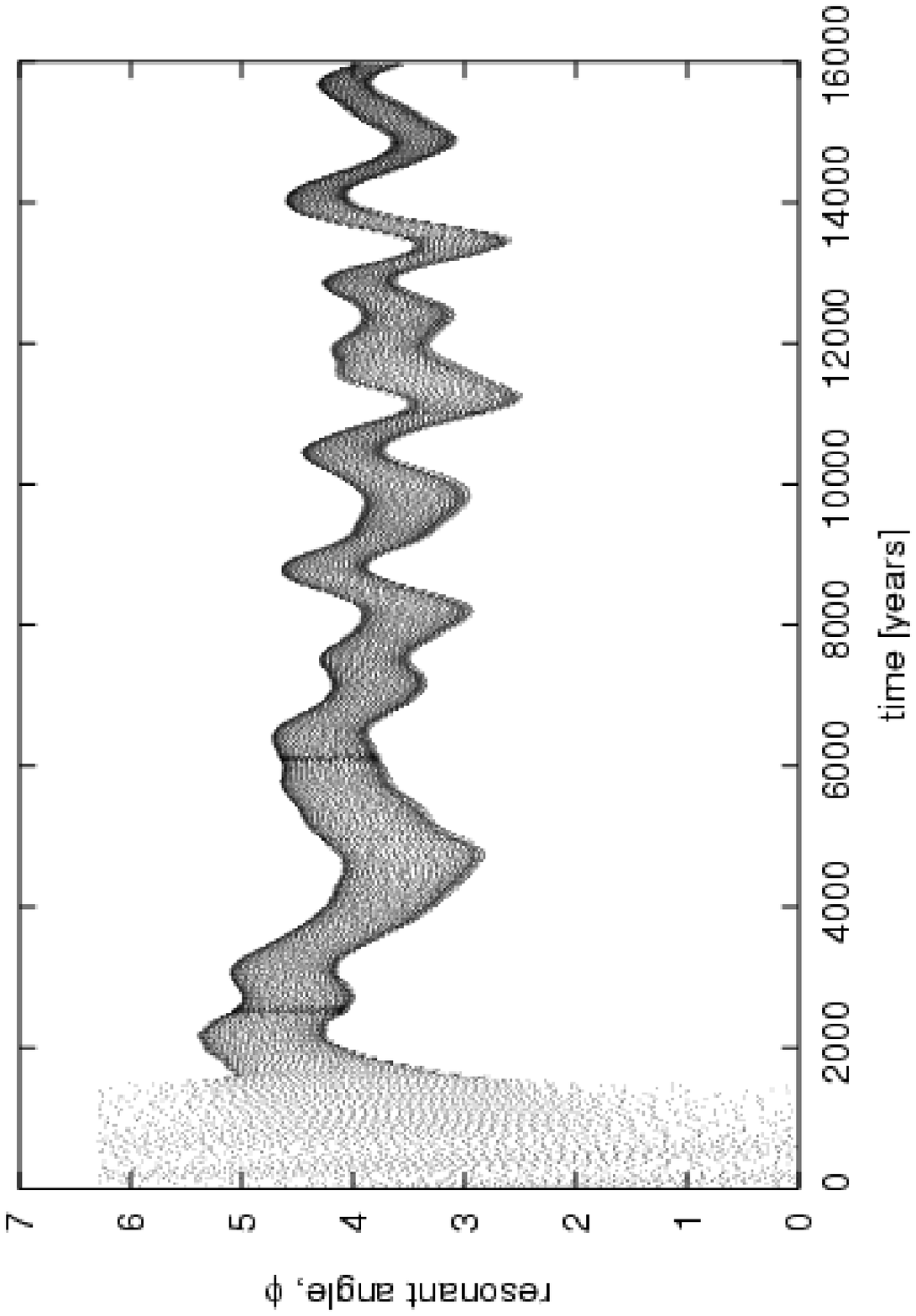}
\includegraphics{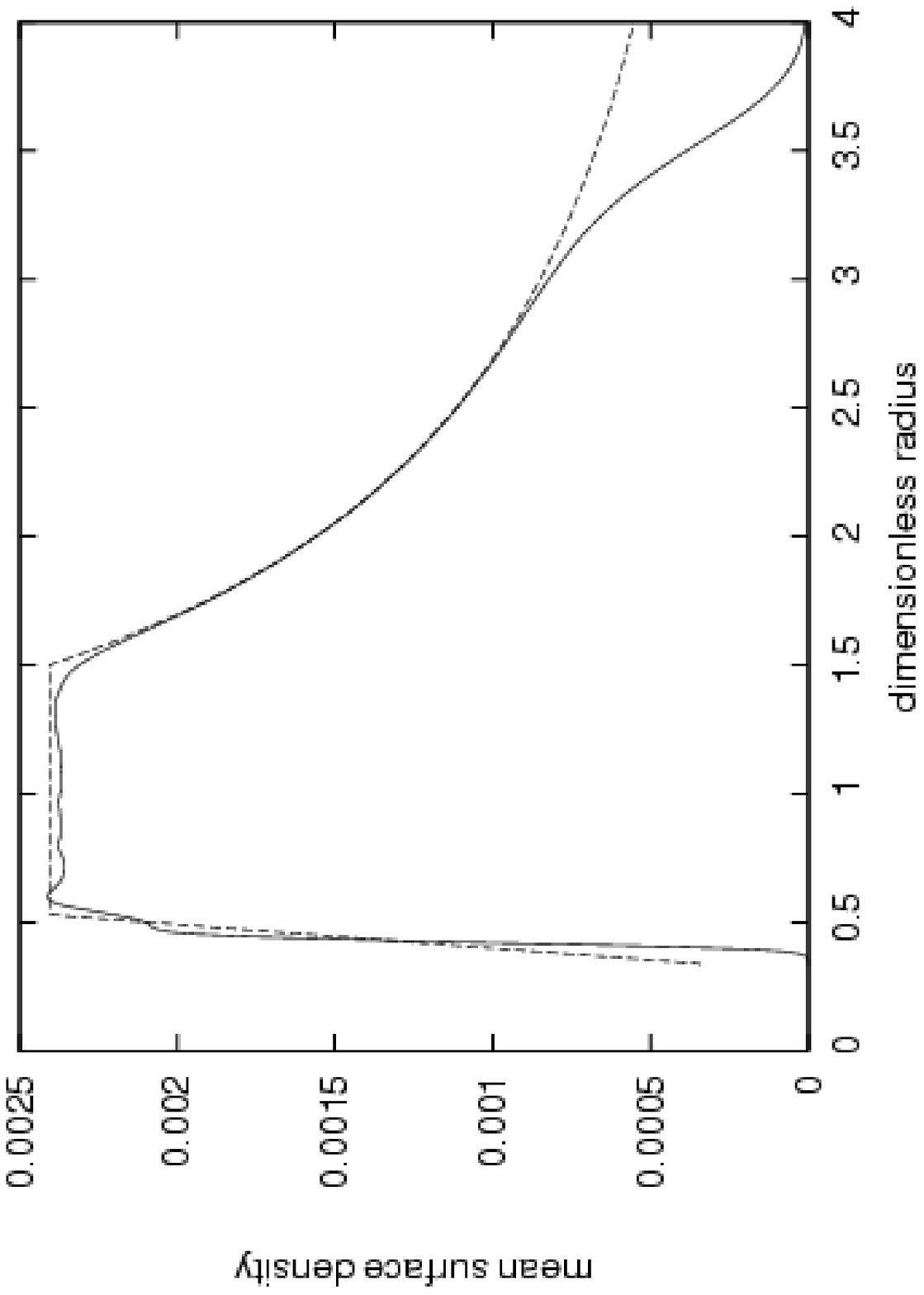}
\includegraphics{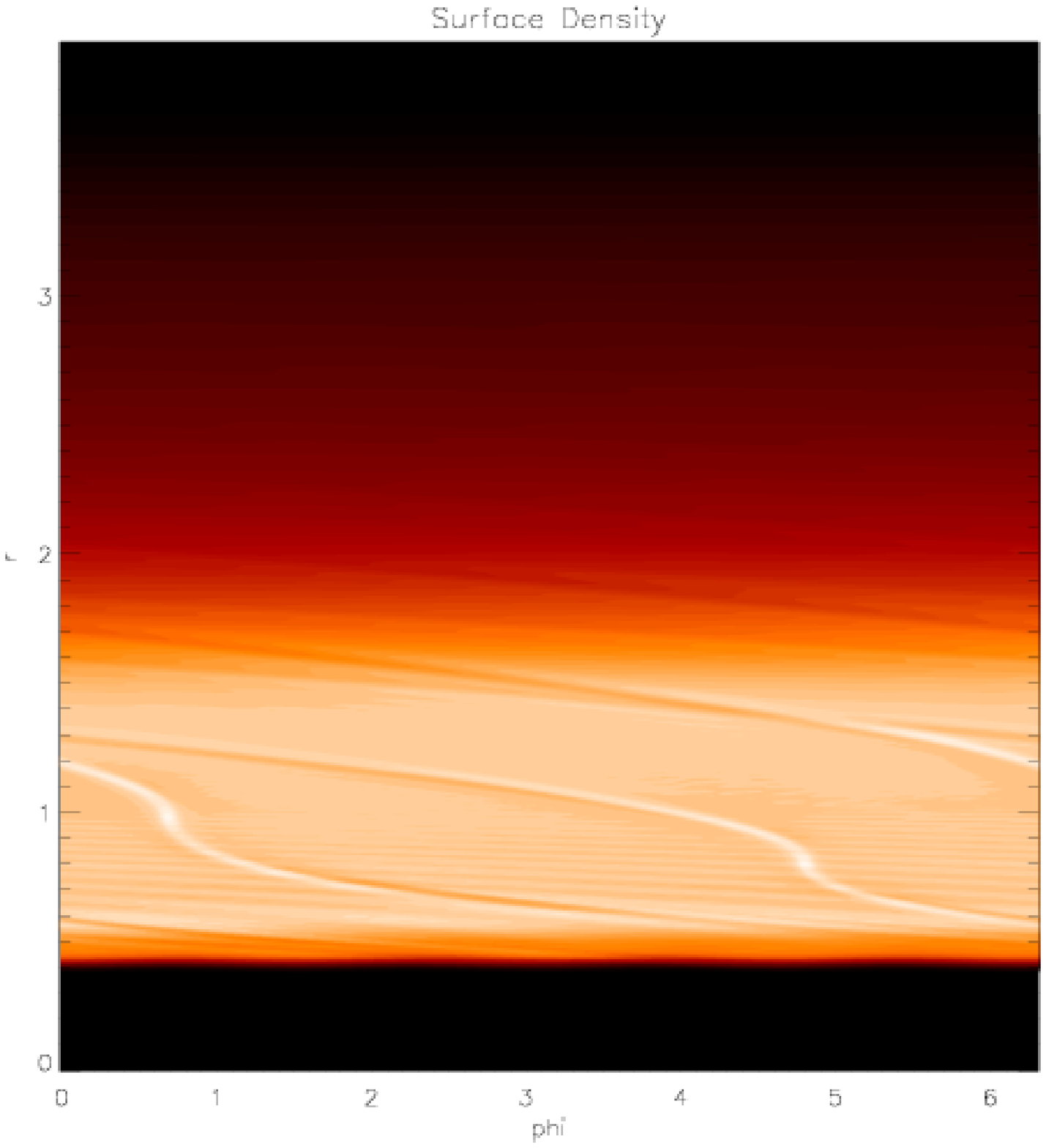}
\caption{\label{fig14}{The same as for Figure \ref{fig13} 
but the planets initially   had $a_1=1.23$ and $a_2 = 1$
and they are embedded in  a disc with  $\Sigma_0 =\Sigma _4$
}}
\end{minipage}
\end{figure*}

\begin{figure*}
\begin{minipage}{175mm}
\vspace{220mm}
\includegraphics{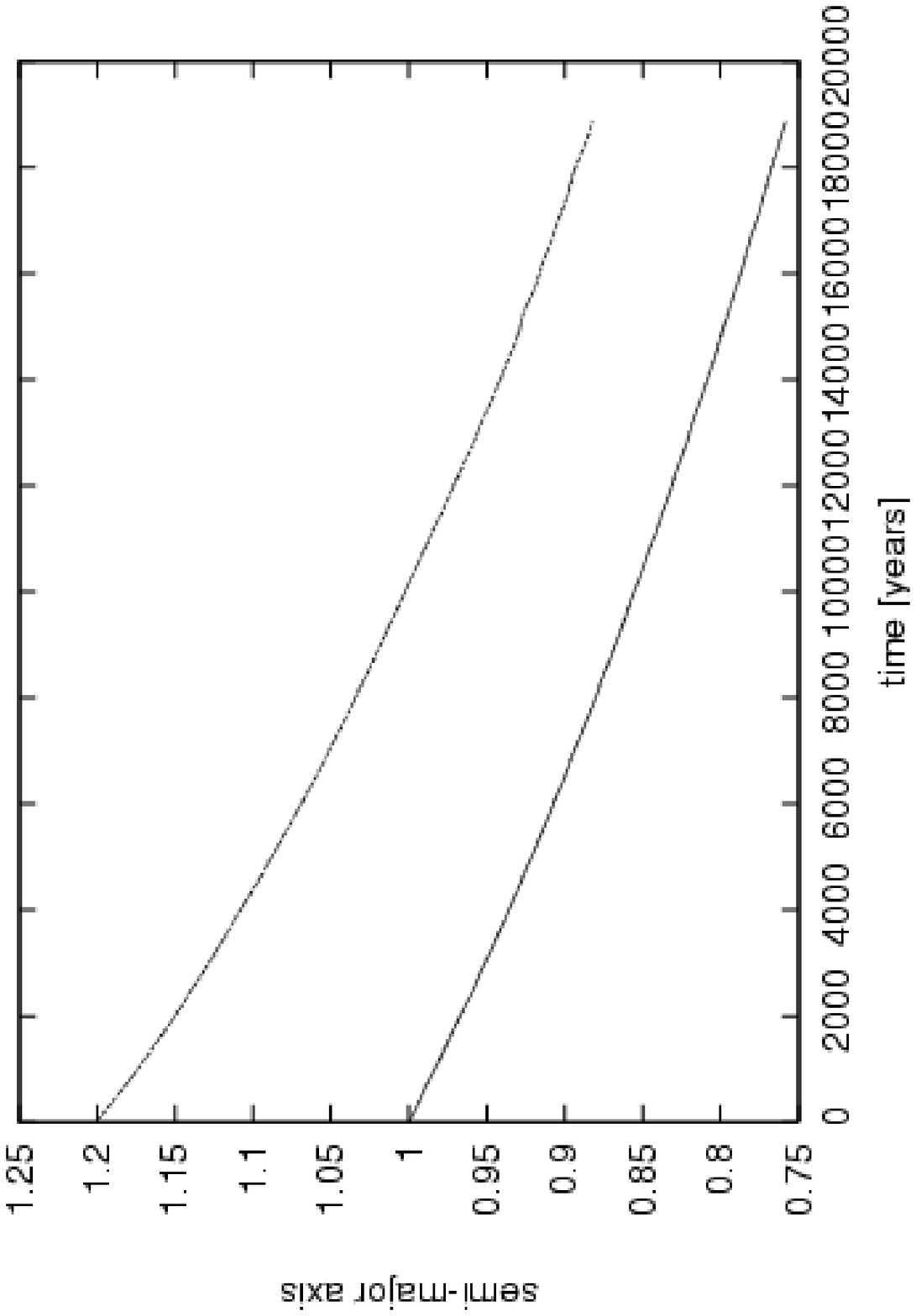}
\includegraphics{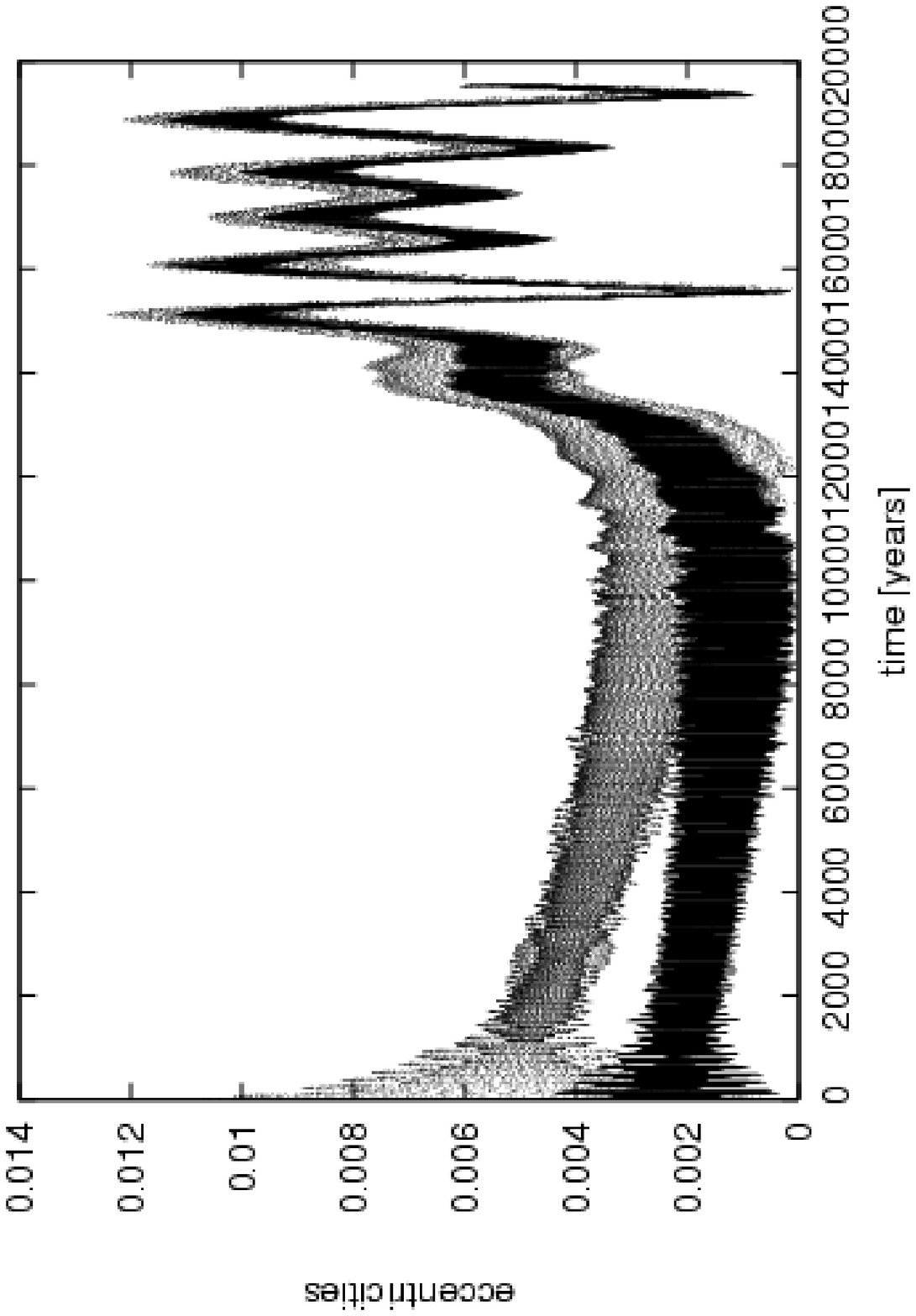}
\includegraphics{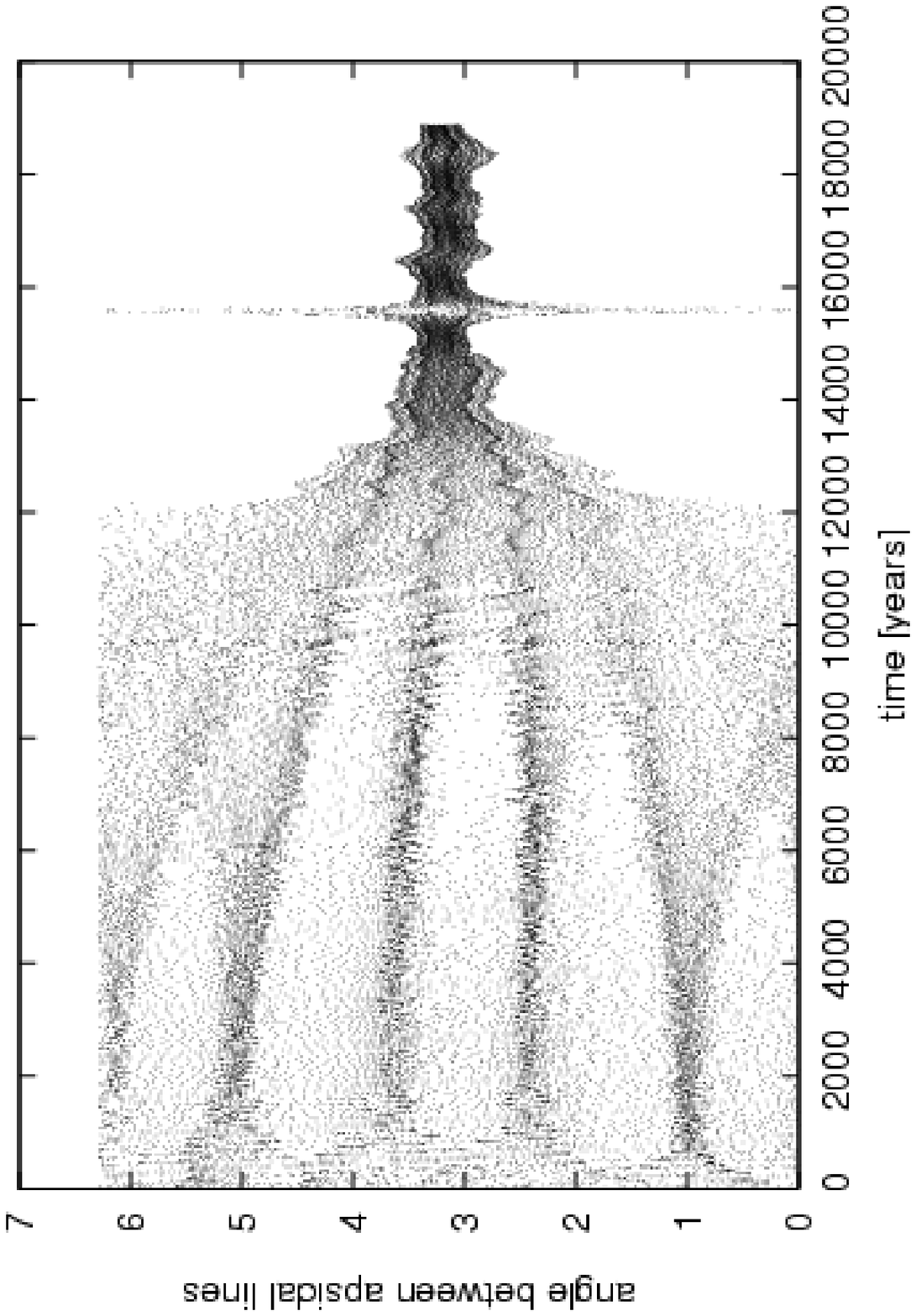}
\includegraphics{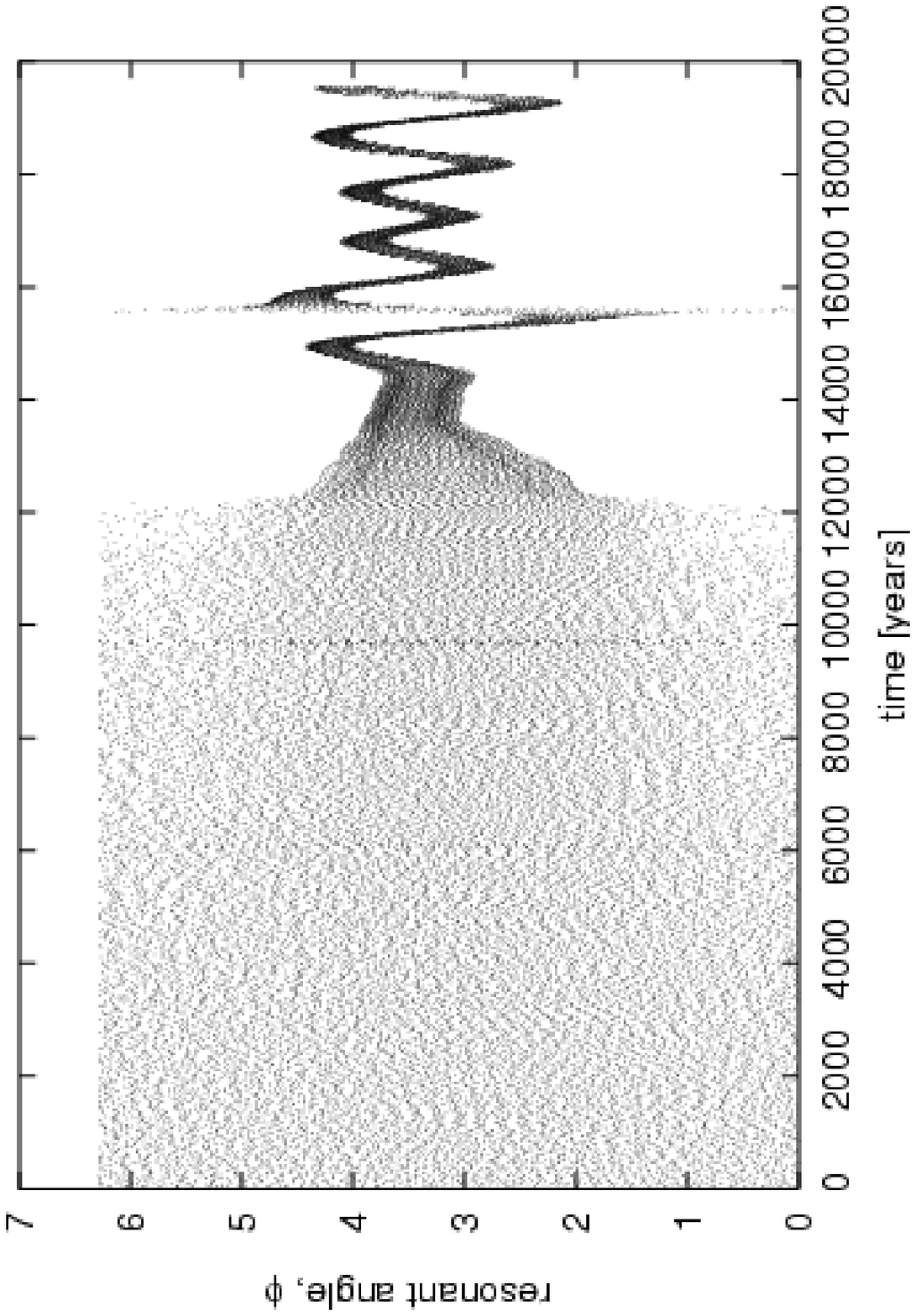}
\includegraphics{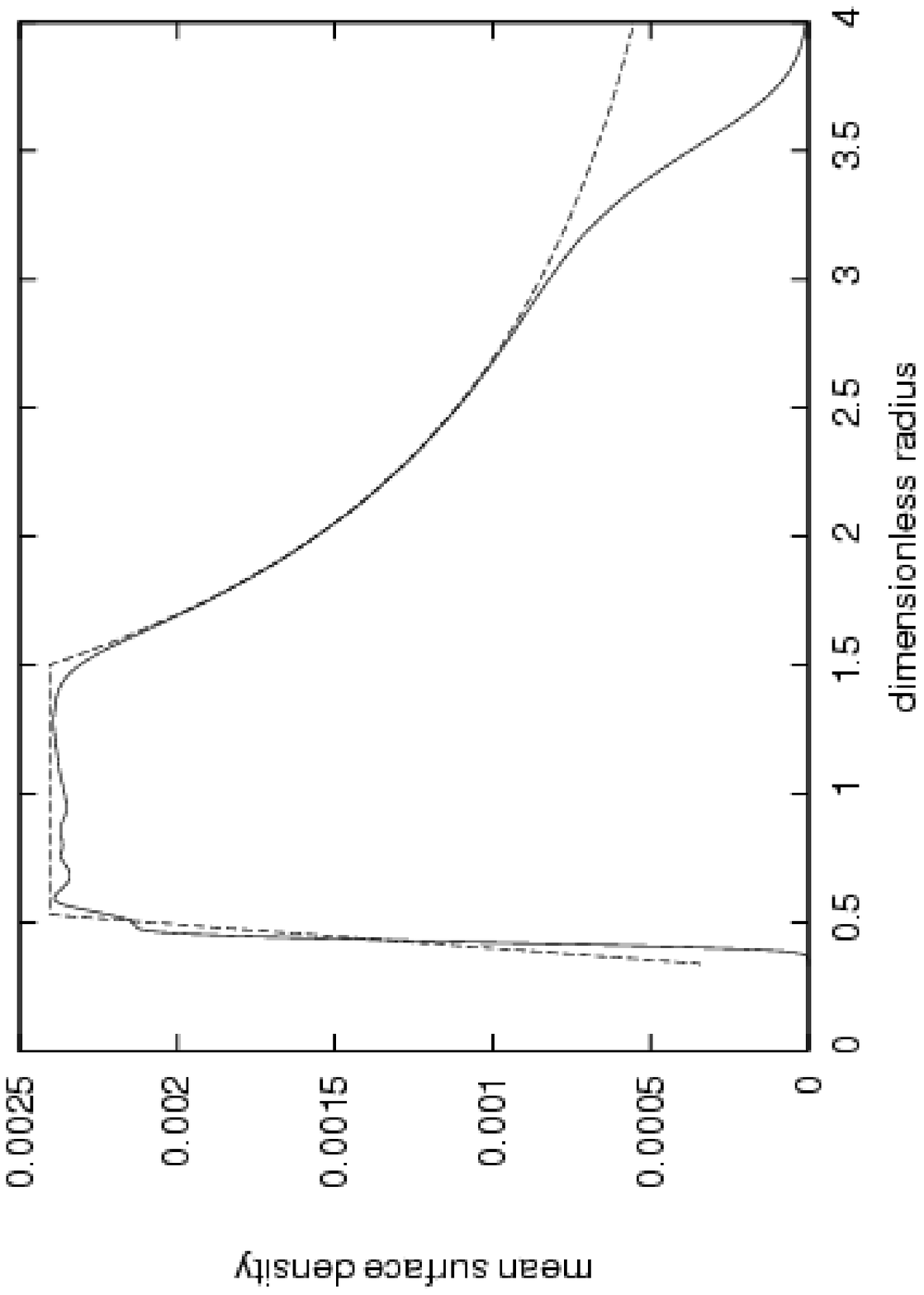}
\includegraphics{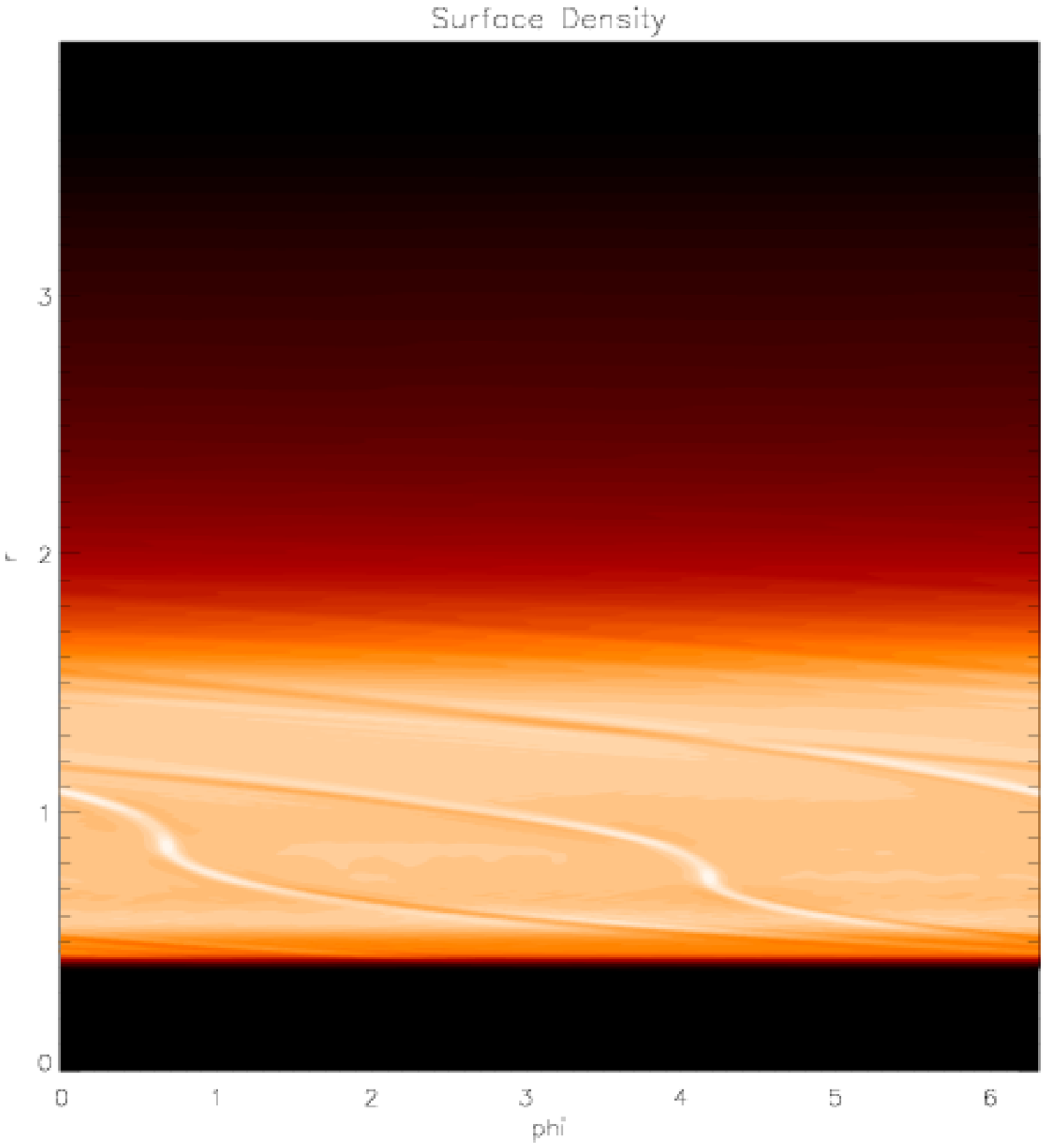}
\caption{\label{fig15}{The same as for Figure \ref{fig13} but 
the planets initial separation ratio  is 1.2 and they are embedded 
in  the disc with $\Sigma =\Sigma _4$
}}
\end{minipage}
\end{figure*}

As described above, the resonant interaction must balance
the tendency towards  relative migration of the two planets. 
This is smaller when the planets have the same mass. Thus
lower $p$ commensurabilities and less tendency
to be driven into  a chaotic state  might be expected in this 
case when compared to the situation where the inner planet
has significantly smaller mass.  In order to investigate this,
two planets of  mass $4M_{\oplus}$ were initiated  close to 
3:2 resonance  with $a_{1}=1.32$ and $a_{2}=1$ in a disc
with  $\Sigma_0 = \Sigma_{1}$.
The  evolution of 
the  semi-major axis ratio for the planet  orbits  is shown 
in Figure  \ref{fig12}.
We also performed  simulations for the same masses  starting
with $a_{1}=1.23$ and $a_{2}=1.00$ in a disc
with  $\Sigma_0 = \Sigma_{4}$ and $a_{1}=1.2$ and $a_{2}=1$ 
in a disc with $\Sigma_0 = \Sigma_{4}.$
The planets   become trapped in the   nearest available 
resonance which is a good indication of stability.
The   evolution of the equal mass pair of planets is shown 
in Figures  \ref{fig13}-\ref{fig15}.
The planets  attaining 3:2 resonance after around 4000 years 
show the following behaviour. Their eccentricities increased  
until at  8000 years 
$e_{1} =0.013$ and $e_{2} =0.007.$ They  then start to 
oscillate.
At around time 13000 the inner planet eccentricity dropped  
to zero for short period of time. 
The resonant angle oscillates around 200$^{\circ}$.

\begin{figure}
\vskip 8cm
\includegraphics{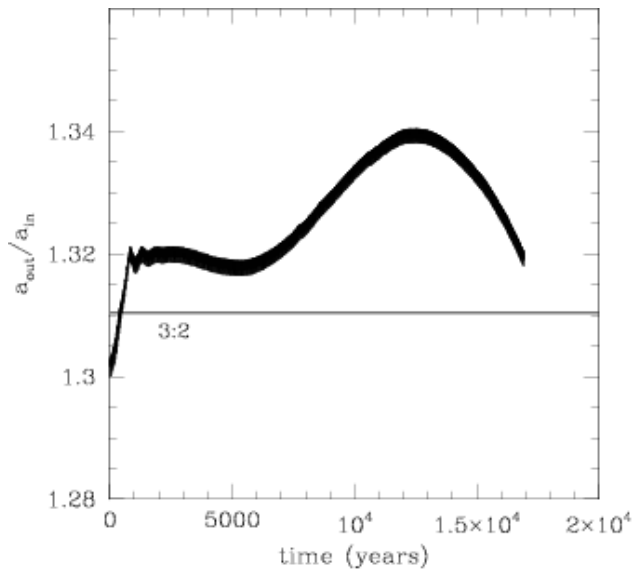}
\caption{\label{fig16}{The ratio of semi-major axes for two 
equal mass  planets of $30M_{\oplus}$, and disc surface 
density  scaling $\Sigma_0 =\Sigma_{1}$.}}
\end{figure}

\begin{figure*}
\begin{minipage}{175mm}
\vspace{220mm}
\includegraphics{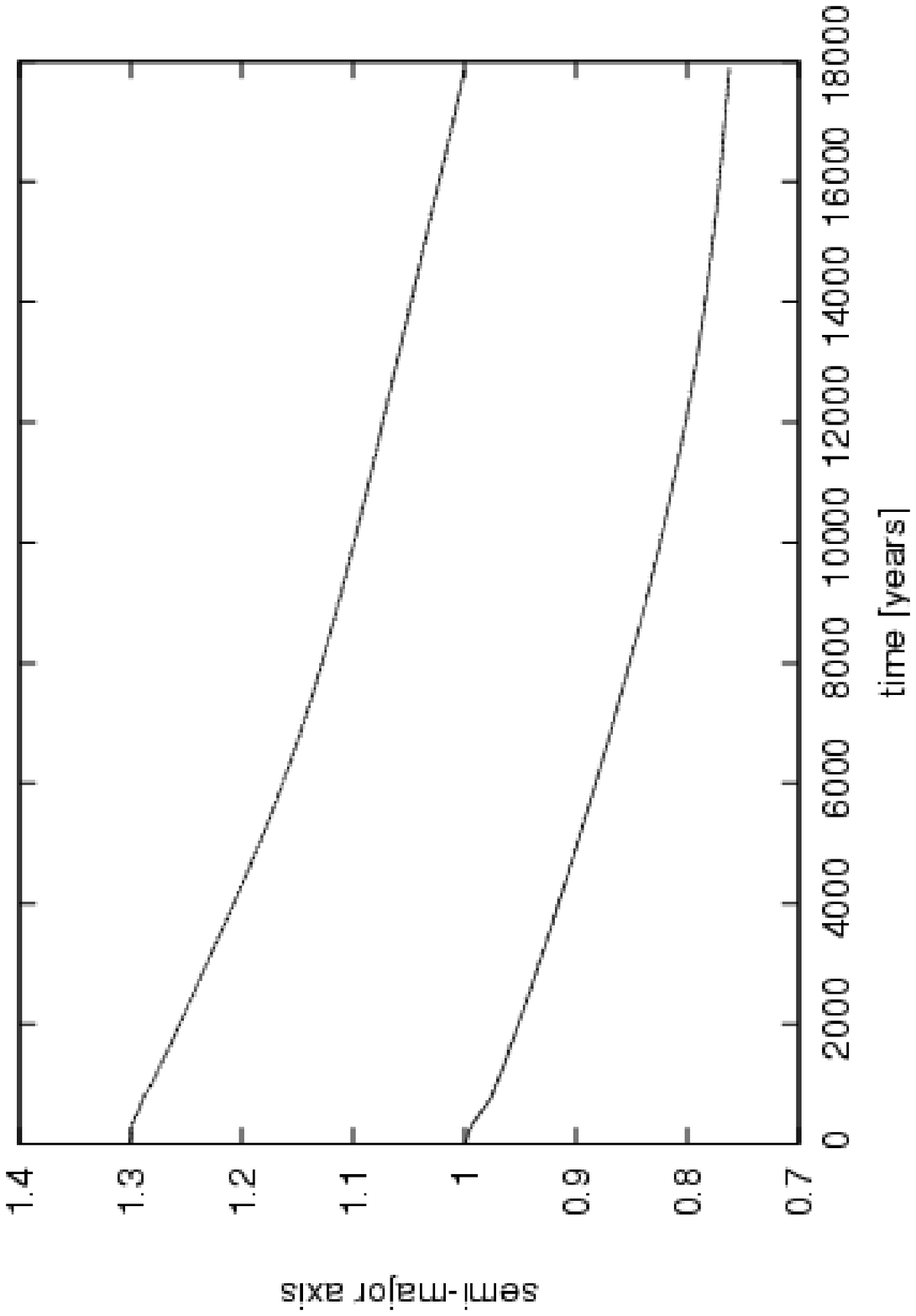}
\includegraphics{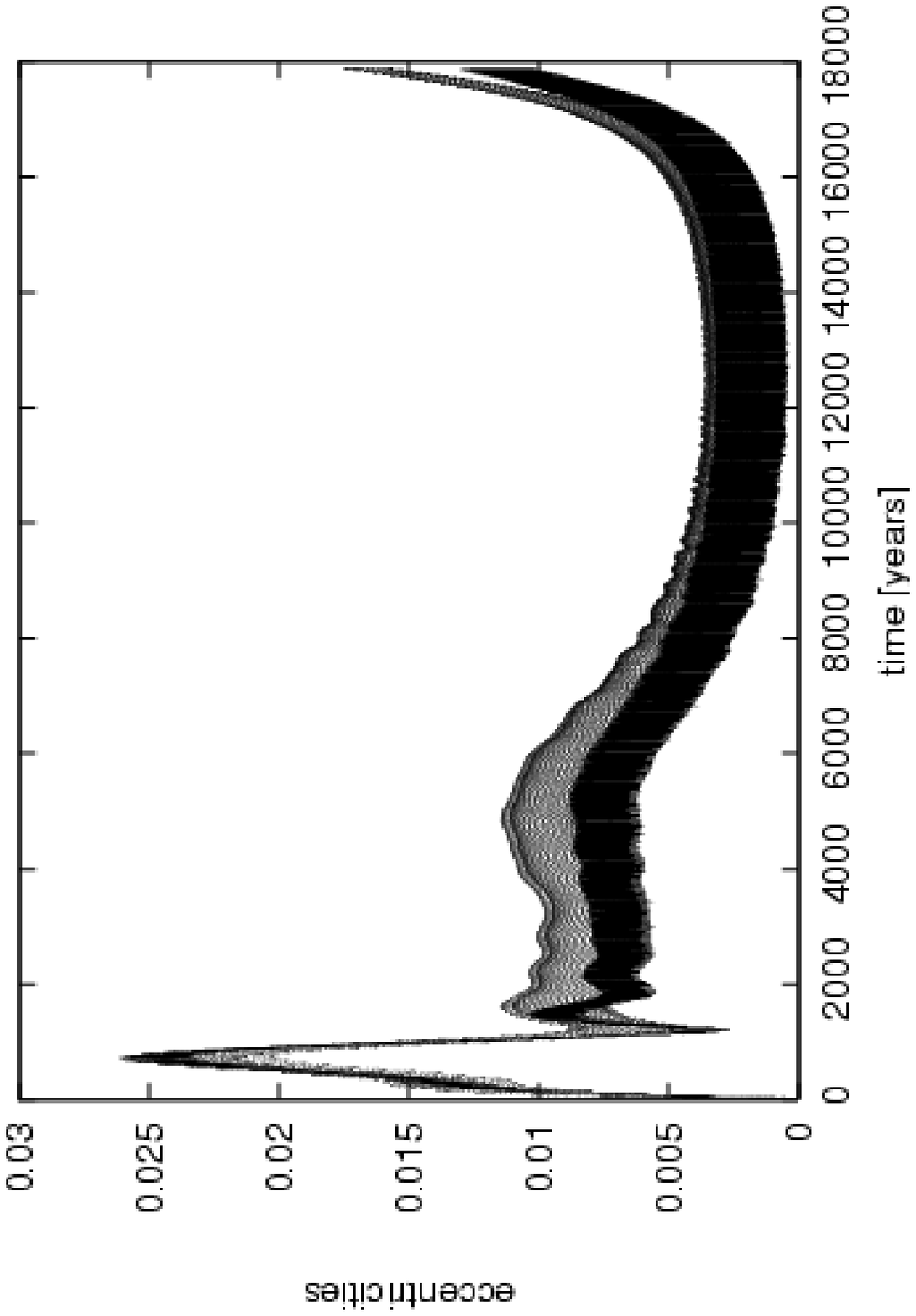}
\includegraphics{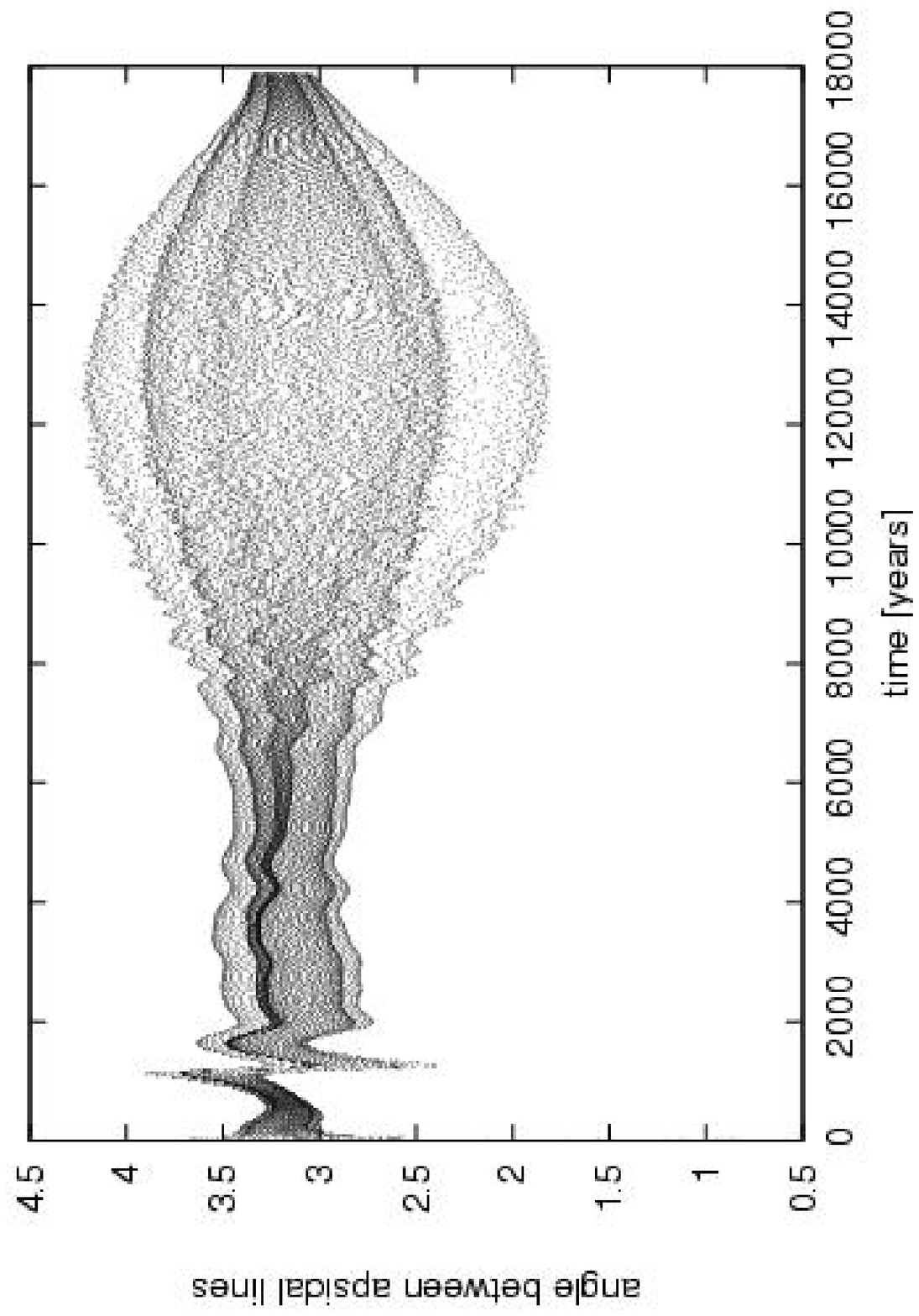}
\includegraphics{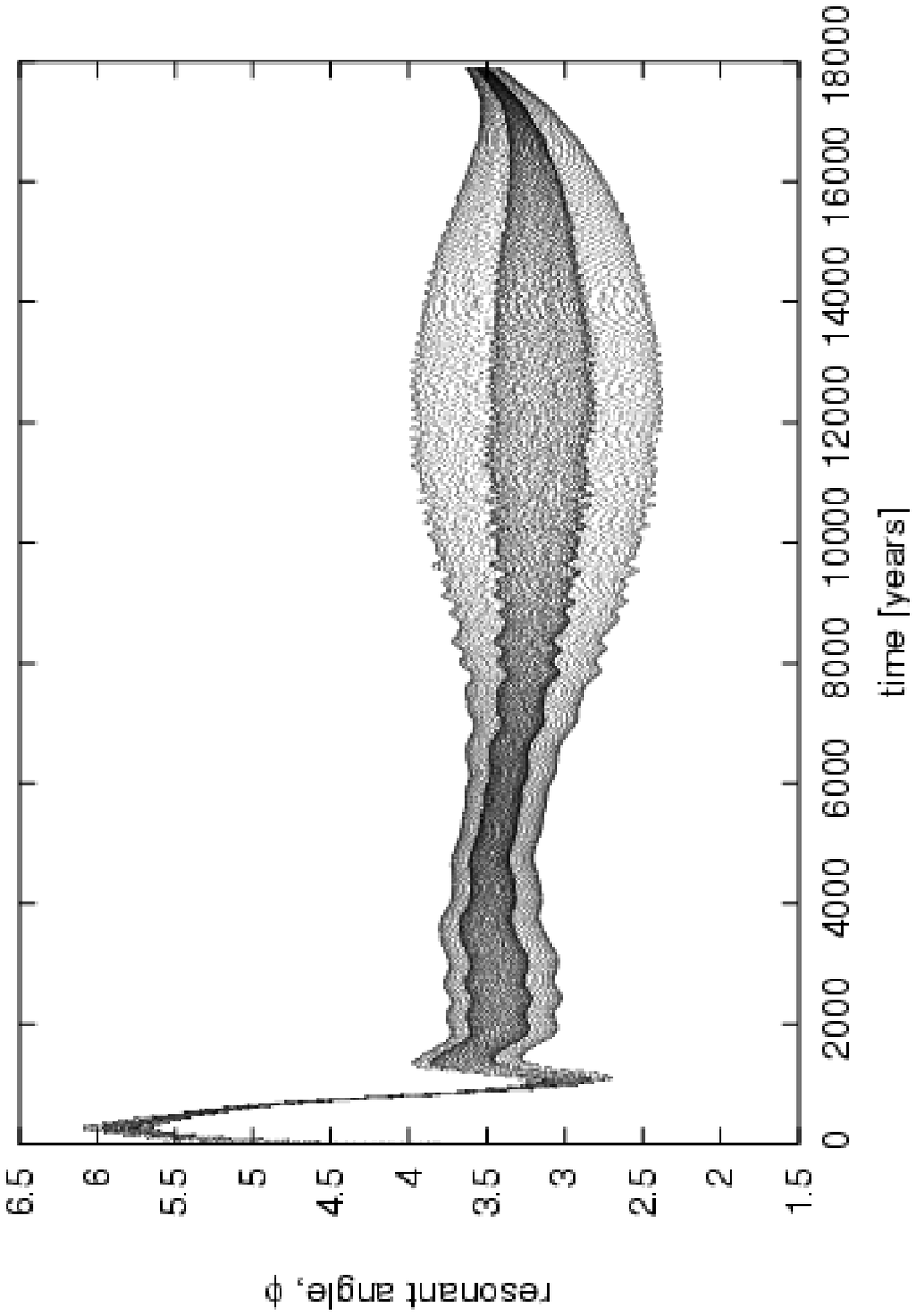}
\includegraphics{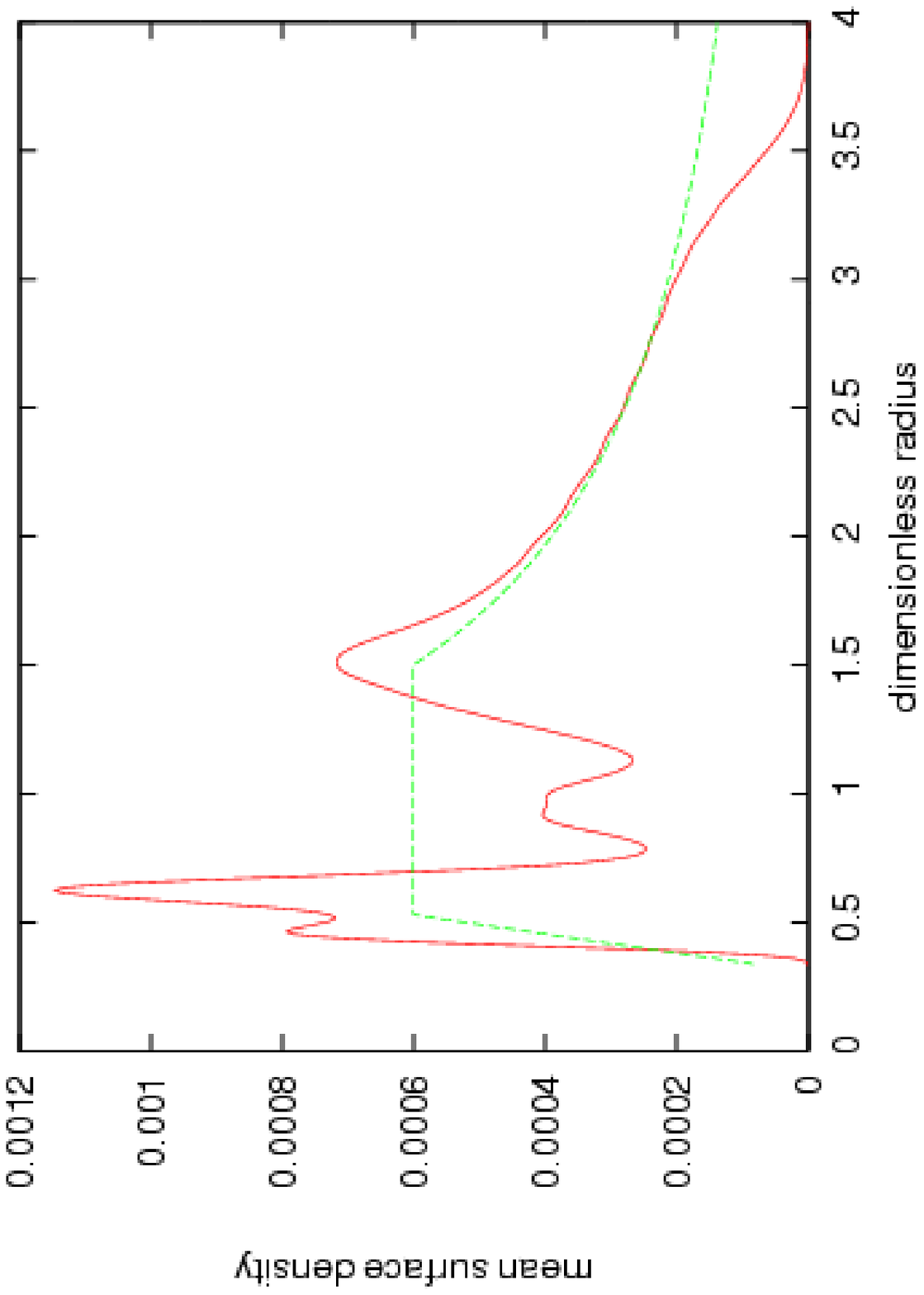}
\includegraphics{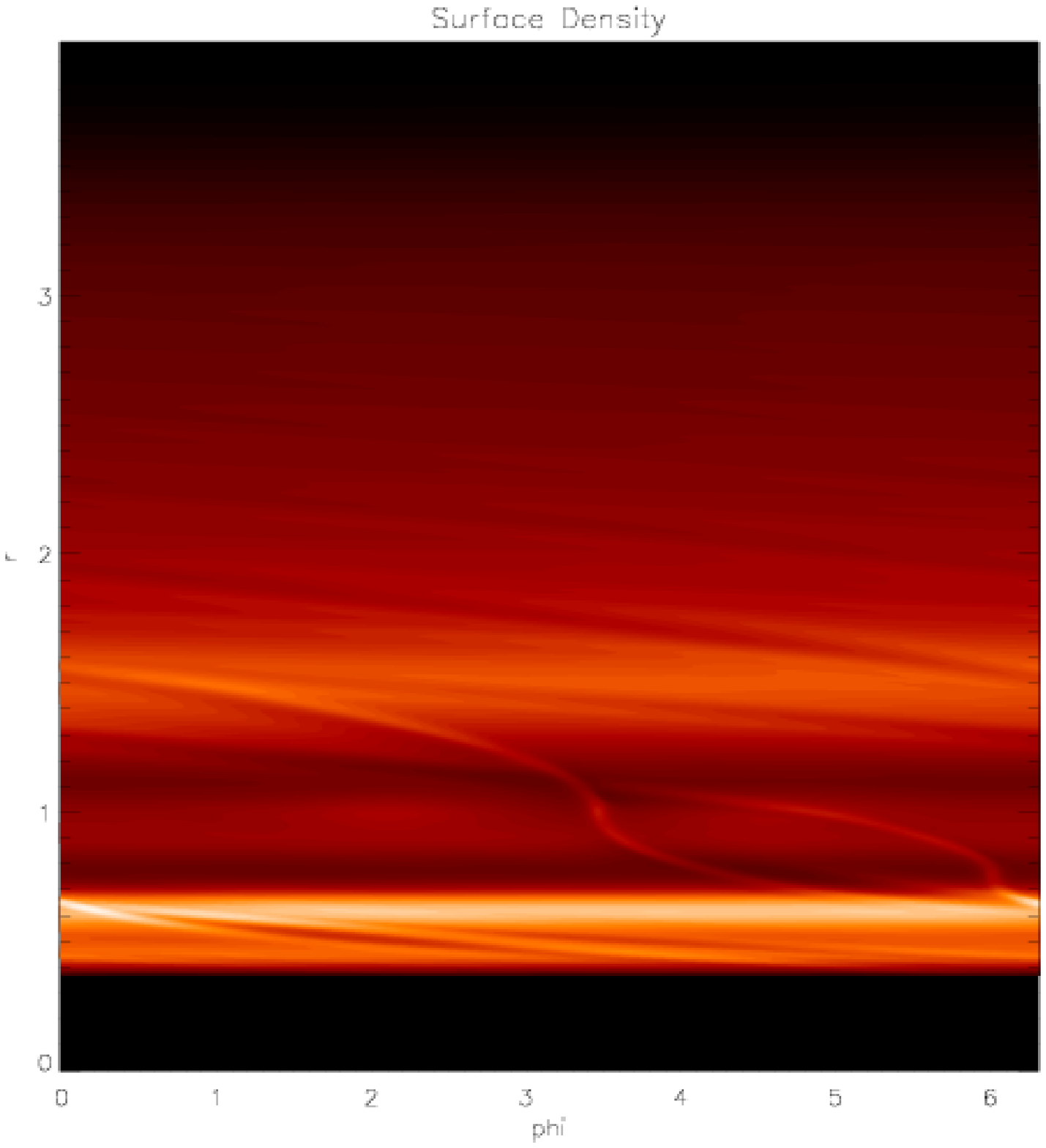}
\caption{\label{fig17}{
The evolution of semi-major axes, eccentricities, angle 
between apsidal lines and resonant angle
for two planets with masses, $m_{1} = 30 M_{\oplus}$ and
$m_{2} = 30 M_{\oplus}$ migrating towards
a central star embedded in a disc with $\Sigma_0 =\Sigma _1$
(four upper panels).
The mean surface density profile of the disc near the end 
of the simulation (solid line) together 
 with the initial surface density profile (dashed
line) and  a surface density
contour plot near the end of the simulation
are given in  lowest left and right  panels respectively.
}}
\end{minipage}
\end{figure*}

\noindent 
The planets placed close to 4:3 resonance become locked 
in this resonance at around 2000 years. The eccentricities 
of both planets change in a similar way. They do not grow 
much, towards the end of the simulation they oscillate 
around 0.008. The angle between apsidal lines and the  
resonant angle oscillate around 180$^{\circ}$ and 230$^{\circ}$, 
respectively.

\noindent 
The last experiment of this series was as follows. The two 
planets are initially located just interior to 4:3 resonance. 
After 12000 years they arrive through convergent
migration at the 5:4 resonance and become trapped in it. 
The eccentricities of the two planets evolve together in an  
oscillatory way. At time around 16000 years they are very close 
to zero. Particular features at that time are seen 
in the evolution of the angle between apsidal lines as well 
as the resonant angle which are not properly defined for zero 
eccentricity.

We also  performed calculations with two planets 
of  30 $M_{\oplus}$ starting with  $a_{1}=1.3$ and $a_{2}=1.$
The planets were embedded in a disc with 
$\Sigma_0 = \Sigma _1$. They attained a stable 3:2 resonance.
The behaviour of the  ratio of semi-major axes  close to the 
resonance is shown in Figure \ref{fig16} and other aspects of 
the evolution are shown  in Figure \ref{fig17}.
The excursions around resonance are lager and noisier
in this case.
 Note too that  in 
this case the disc  planet interaction becomes non linear with 
significant perturbation to the  underlying disc surface density.
There is a tendency towards gap formation and the excavation 
of disc material into two mounds at the gap edges.
The inner mound occupies only a small radial  region and so
may not be well represented with the available 
numerical resolution in this case.

\section{N-body investigations - a survey}

In order to study  migrating planets  with a wider range
of masses and disc surface densities, we have performed 
a resonance survey using an  N-body code. The approach is 
the same as that used by \citet{spn01}
and \citet{np02} and it has also been used by \citet{lp02}
and \citet{kpb04}. The reader is referred to those papers for details.
The procedure is to model the effects of a disc by incorporating
orbital migration and eccentricity damping as described
by equations (\ref{MIG} - \ref{CIRC}) through the addition of appropriate
acceleration terms to the equations of motion.
\noindent 
We summarize the parameters and outcomes of some of the N body 
simulations in Table \ref{table1}.

\noindent 
We begin by discussing a calibration procedure which enables
a matching of the N body simulations with the hydrodynamic 
simulations. The advantage of the N body simulations is that
the time evolution of a system can be followed for a much 
longer period of time.
After discussing a few examples, we move on to discuss the 
results, shown in Table \ref{table1},
on the possible commensurabilities
to be expected for low mass planets  for  particular choices 
of their masses, disc parameters and migration and eccentricity 
damping rates.

\subsection{Comparison between hydrodynamic simulations and N-body
calculations}

The hydrodynamical calculations discussed in the previous Section
allow us to follow planets migrations for almost 2$\times$10$^4$
years. As a result of this we have been able to simulate planets 
becoming trapped in resonances. The behaviour and stability of 
the resonance trapping varies with the planet masses  and the 
surface density of the disc in which they are embedded.
The outcome of these simulations could be well matched to those 
of  N-body integrations where we  incorporate  the simple 
prescriptions for the migration and eccentricity damping given 
through equations  (\ref{MIG}-\ref{CIRC}).

\begin{table*}
\centering
\begin{minipage}{140mm}
\caption{ \label{table1} This table  lists some of N body 
calculations of interacting planets with imposed migration 
and eccentricity damping  rates, assumed to result from 
interaction with a disc, that we performed, together
with  their outcomes. During these calculations the migration
has been followed for a period of time in which the inner planet 
changes its semi-major axis  from $a_2 =1$ to $a_2 = 0.385 $ 
in dimensionless units. The outer planet is initially placed
in circular orbit at $r =  1.4 $ in dimensionless units.
Thus the situation corresponds with appropriate scaling to 
the inner planet being initiated at $5.2$AU and then migrating 
to $2$AU. 
Column 1 gives the planet mass ratio, columns 2 and 3 give 
the outer and inner planet mass in Earth masses  respectively. 
Columns 4-7 indicate the outcome of the interaction
as either attainment of a commensurability or a scattering 
for those cases where the migration was too rapid for that 
to occur. Where there is a scattering, the radius
where it occurred is indicated in dimensionless units in 
brackets. An asterisk indicates that the resonance showed 
signs of instability that would result in the transition 
to a higher $p$ resonance on further inward migration which 
could not be followed.}

\begin{tabular}{|r|r|r|c|c|c|c|}
\hline
        &    &       &           &            &     &   \\
m$_{1}$/m$_{2}$      &m$_{1}$ &
m$_{2}$  &
$ \Sigma_0 = \Sigma_{0.5}$ & $ \Sigma_0 = \Sigma_{1}$ & $\Sigma_0 =\Sigma_{2}$
& $ \Sigma_0 = \Sigma_{4}$ \\
        &    &       &           &            &     &   \\
\hline
100/3   &100/3   &1       &5:4*&scatter(0.77)&5:4         &scatter(0.87) \\
20      &20      &1       &4:3*&scatter(0.57)&6:5         &scatter(0.96) \\
10      &10      &1       &7:6*&8:7*   &scatter(0.87)&scatter(0.38) \\
10      &100/3   &10/3    &no resonance    &6:5         &scatter(0.67)&scatter(0.67) \\
25/3    &100/3   &4       &6:5     &scatter(0.67)&5:4*    &scatter(0.88) \\
6       &20      &10/3    &5:4*&6:5*    &6:5         &7:6          \\
5       &20      &4       &5:4     &7:6         &6:5         &7:6          \\
4       &4       &1       &6:5     &6:5*    &7:6         &10:9*    \\
10/3   &10/3   &1       &5:4     &6:5*    &8:7         &9:8*     \\
10/3   &100/3   &10      &4:3     &4:3         &5:4         &6:5          \\
3       &10      &10/3    &4:3     &5:4         &6:5         &7:6          \\
5/2     &10      &4       &4:3     &4:3         &5:4         &6:5          \\
2       &20      &10      &3:2     &3:2         &4:3         &5:4          \\
5/3     &100/3   &20      &3:2     &3:2         &4:3         &5:4          \\
6/5     &4       &10/3    &3:2     &3:2         &3:2         &4:3          \\
1       &1       &1       &3:2     &3:2         &3:2         &3:2          \\
1       &10/3   &10/3    &3:2     &3:2         &3:2         &3:2          \\
1       &4       &4       &3:2     &3:2         &3:2         &3:2          \\
1       &10      &10      &3:2     &3:2         &3:2         &3:2          \\
1       &20      &20      &3:2     &3:2         &3:2         &3:2          \\
1       &100/3   &100/3   &3:2     &3:2         &3:2         &3:2          \\
\hline
\end{tabular}
\end{minipage}
\end{table*}

\noindent 
Using these expressions in the N-body code we have extended
the hydrodynamic calculations for a longer period of time
and studied the long term stability of the
resonances we have found in that context.
As a first step we have adjusted the  numerical coefficients 
in equations (\ref{MIG}-\ref{CIRC}) in such a way that the 
hydrodynamic  and N-body approach
give the same qualitative evolution.

\begin{figure*}
\begin{minipage}{175mm}
\vspace{140mm}
\includegraphics{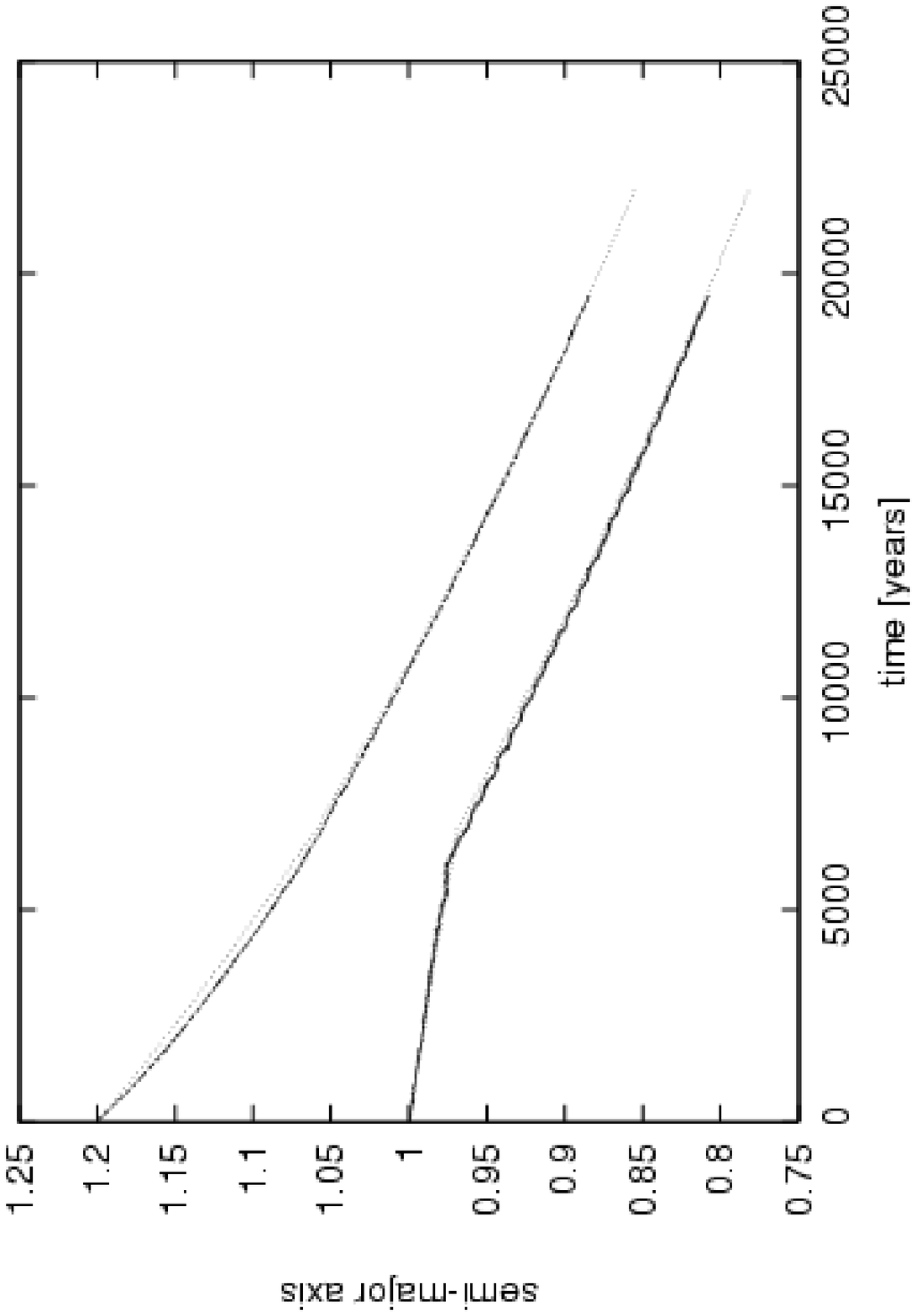}
\includegraphics{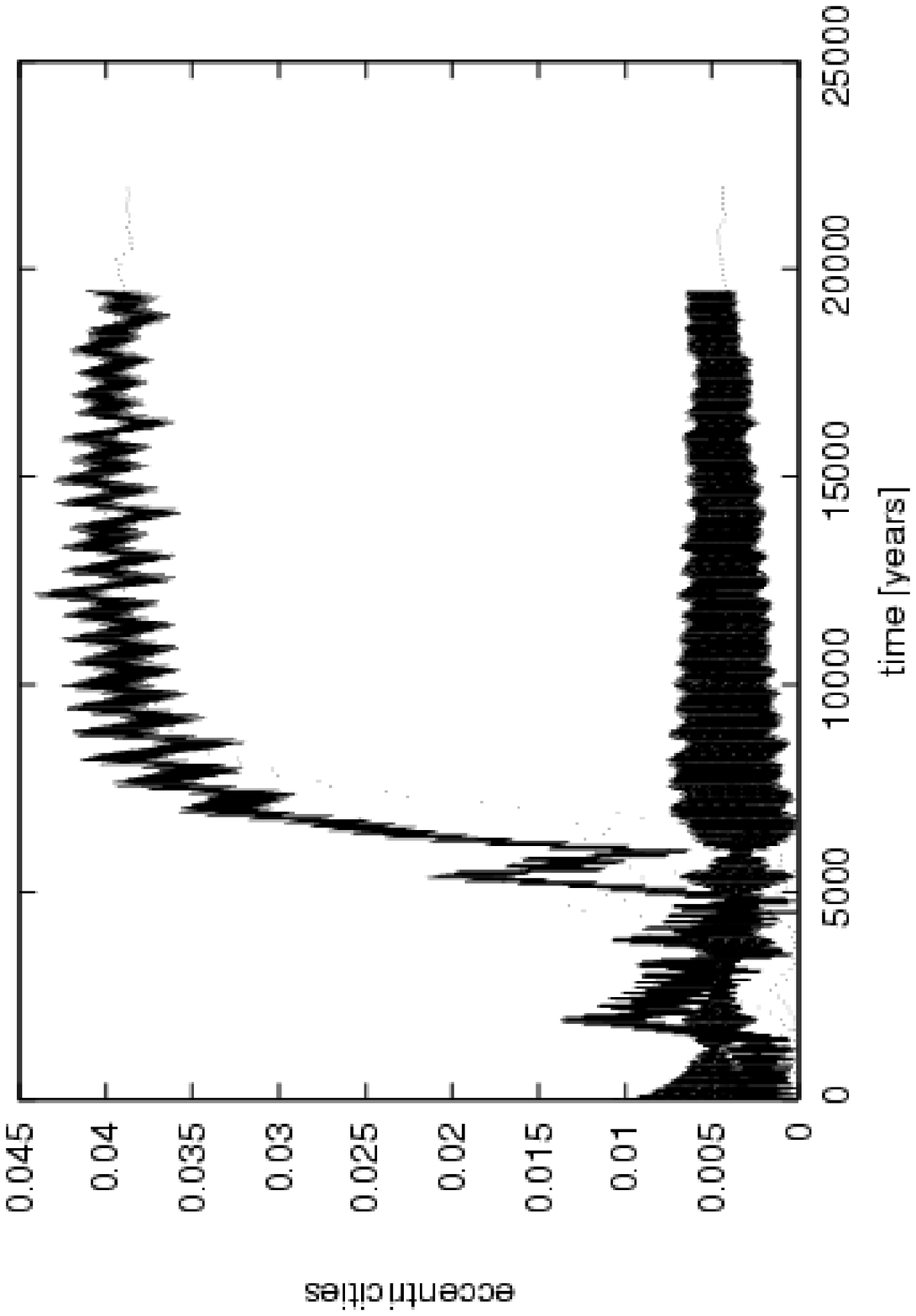}
\includegraphics{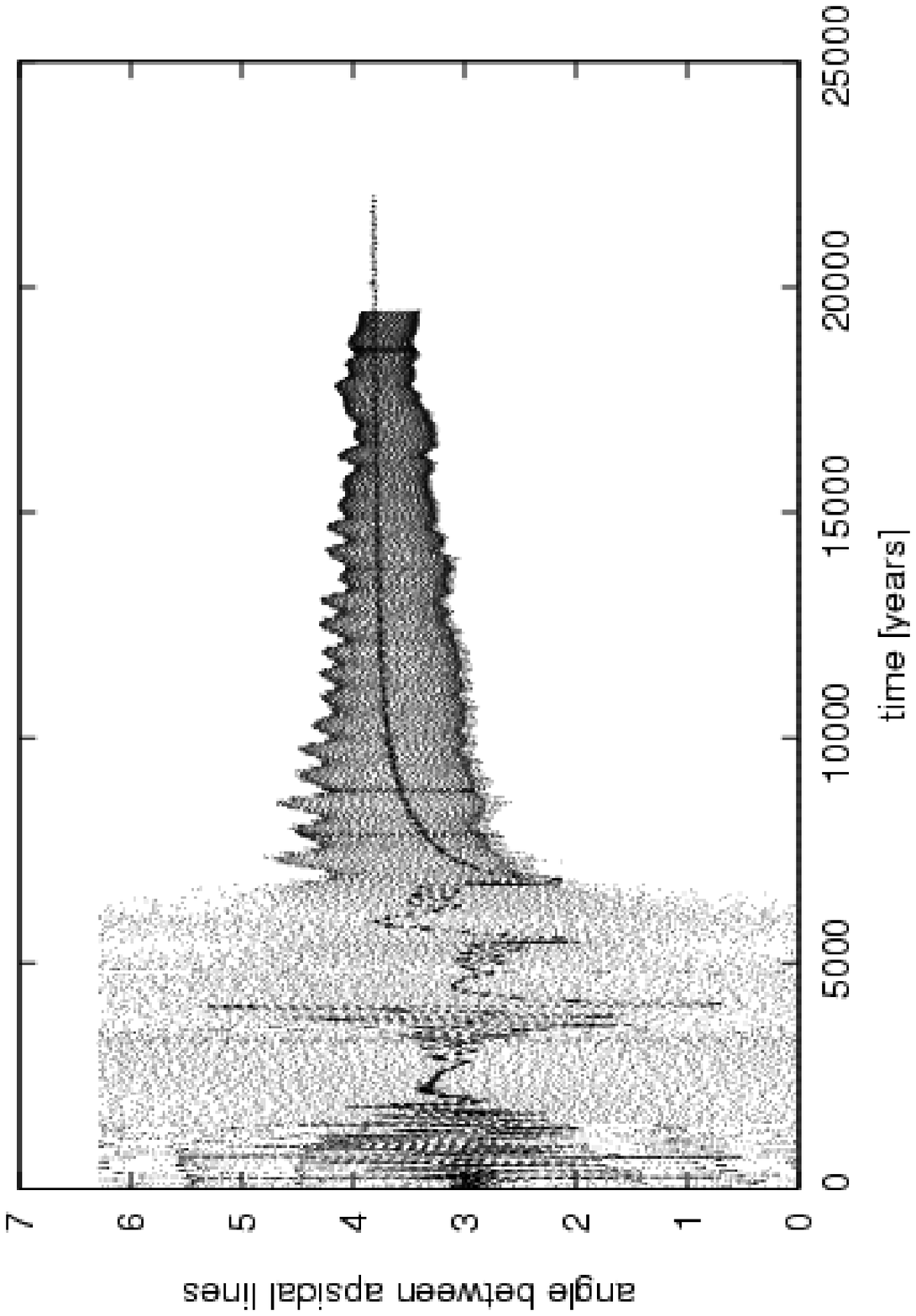}
\includegraphics{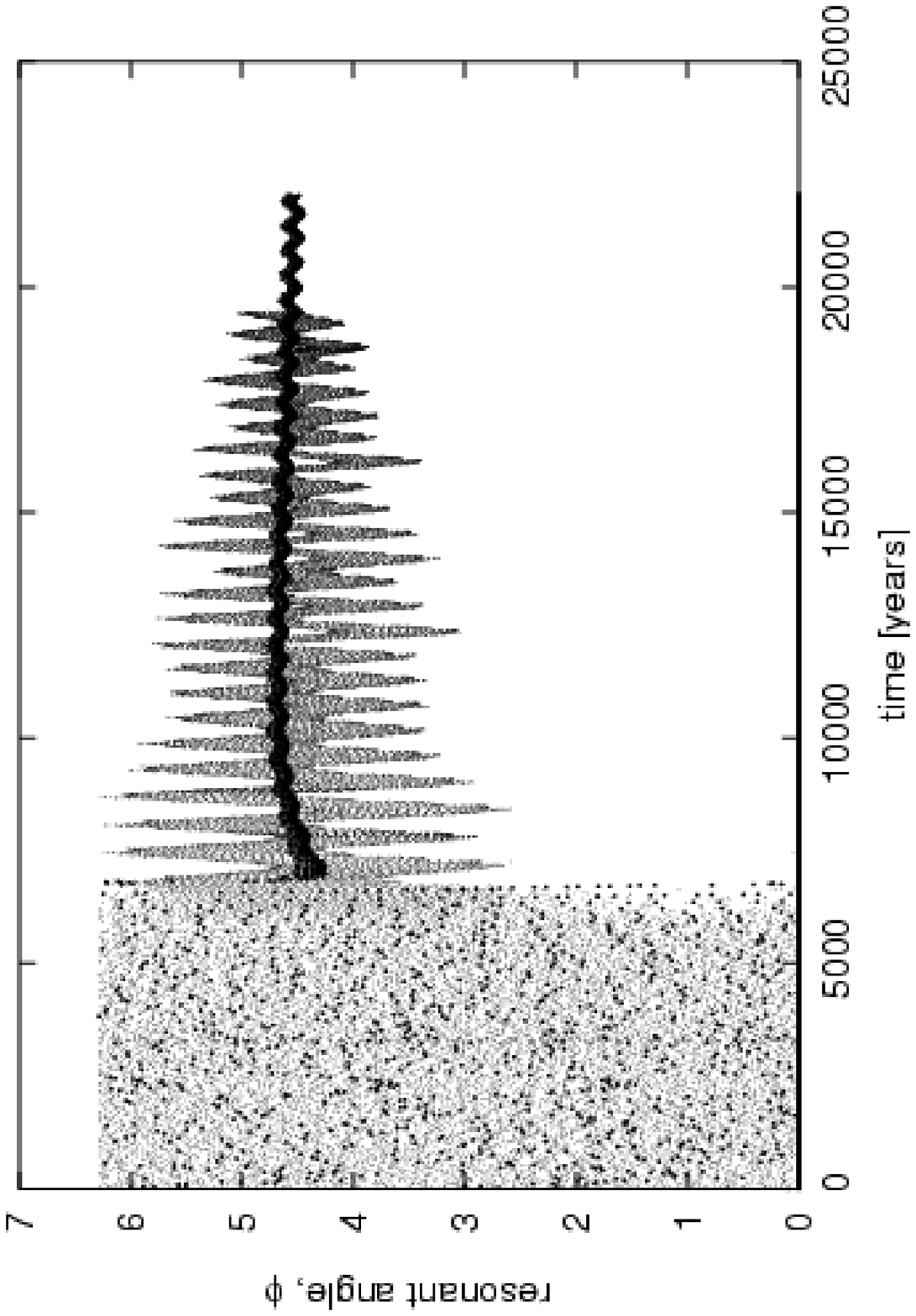}
\caption{\label{fig18}{
The evolution of semi-major axes, eccentricities, 
angle between apsidal lines and resonant angle
for two planets with masses, $m_{1} = 4 M_{\oplus}$ and
$m_{2} = 1 M_{\oplus}$ migrating towards
a central star embedded in the disc with $\Sigma =\Sigma _4$
obtained by   N-body simulation (dotted lines in two upper panels
and dark lines in two lower panels) superimposed onto  
the hydrodynamic simulation shown already in Figure~\ref{fig3}.
}}
\end{minipage}
\end{figure*}

\begin{figure*}
\begin{minipage}{175mm}
\vspace{140mm}
\includegraphics{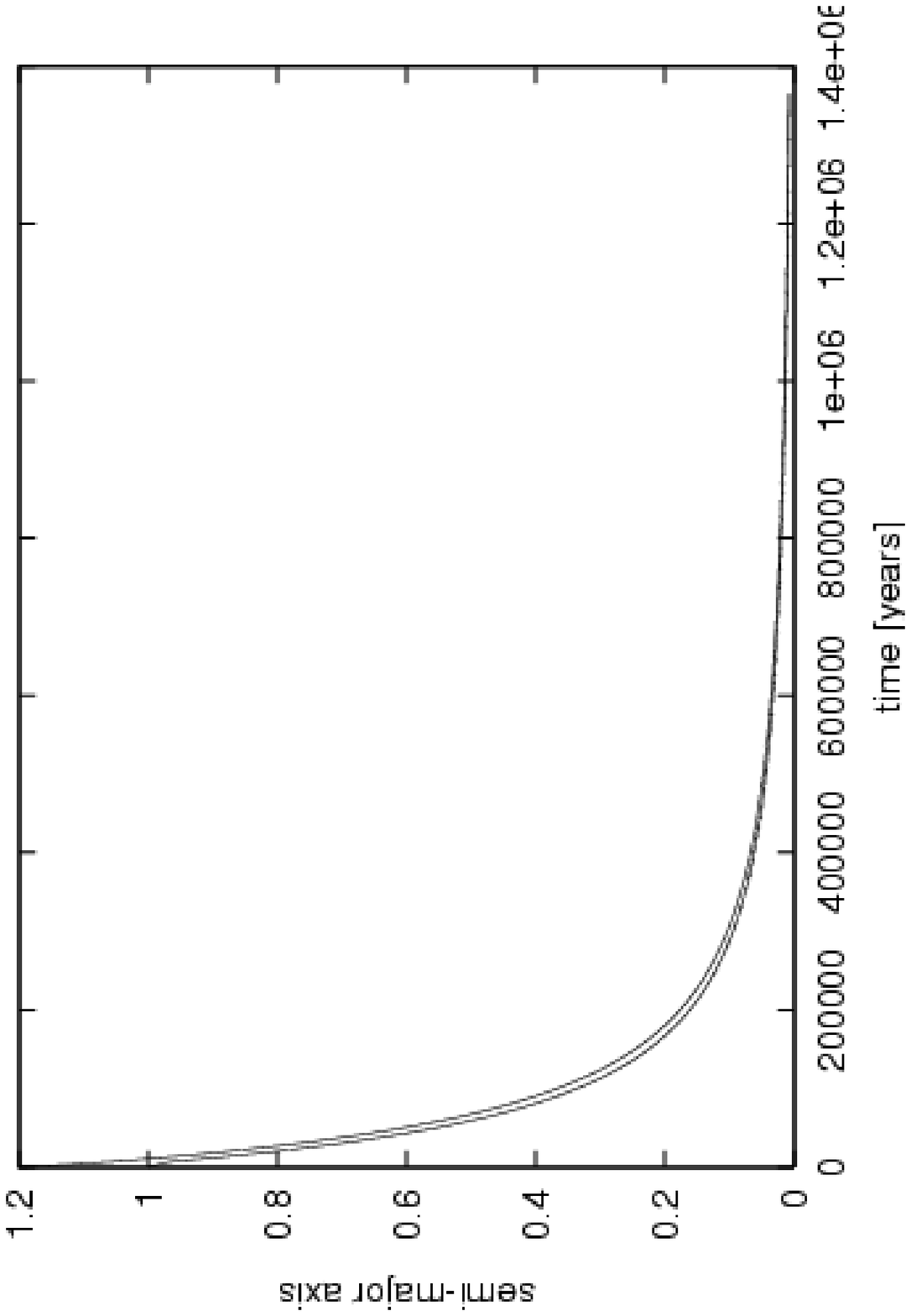}
\includegraphics{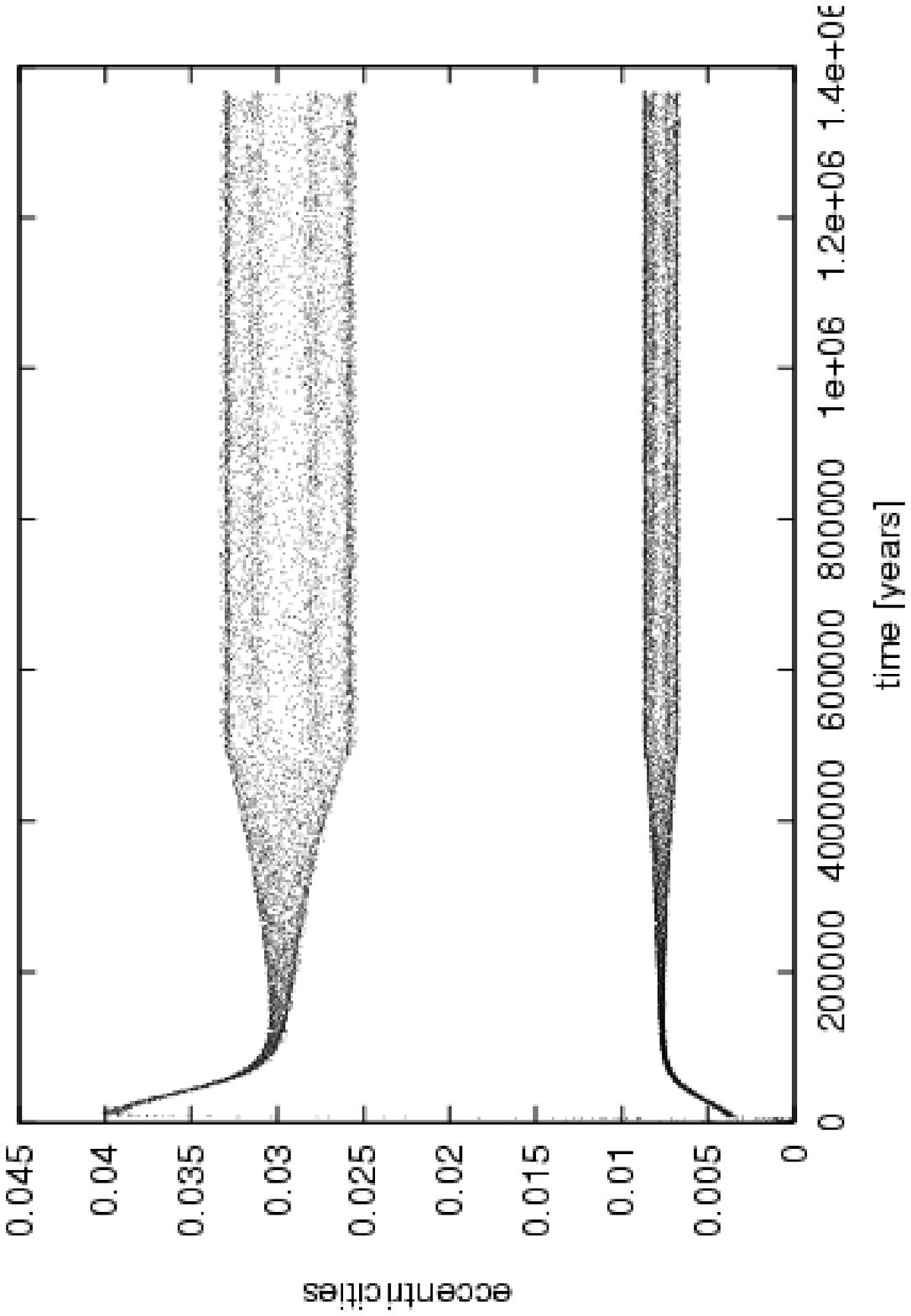}
\includegraphics{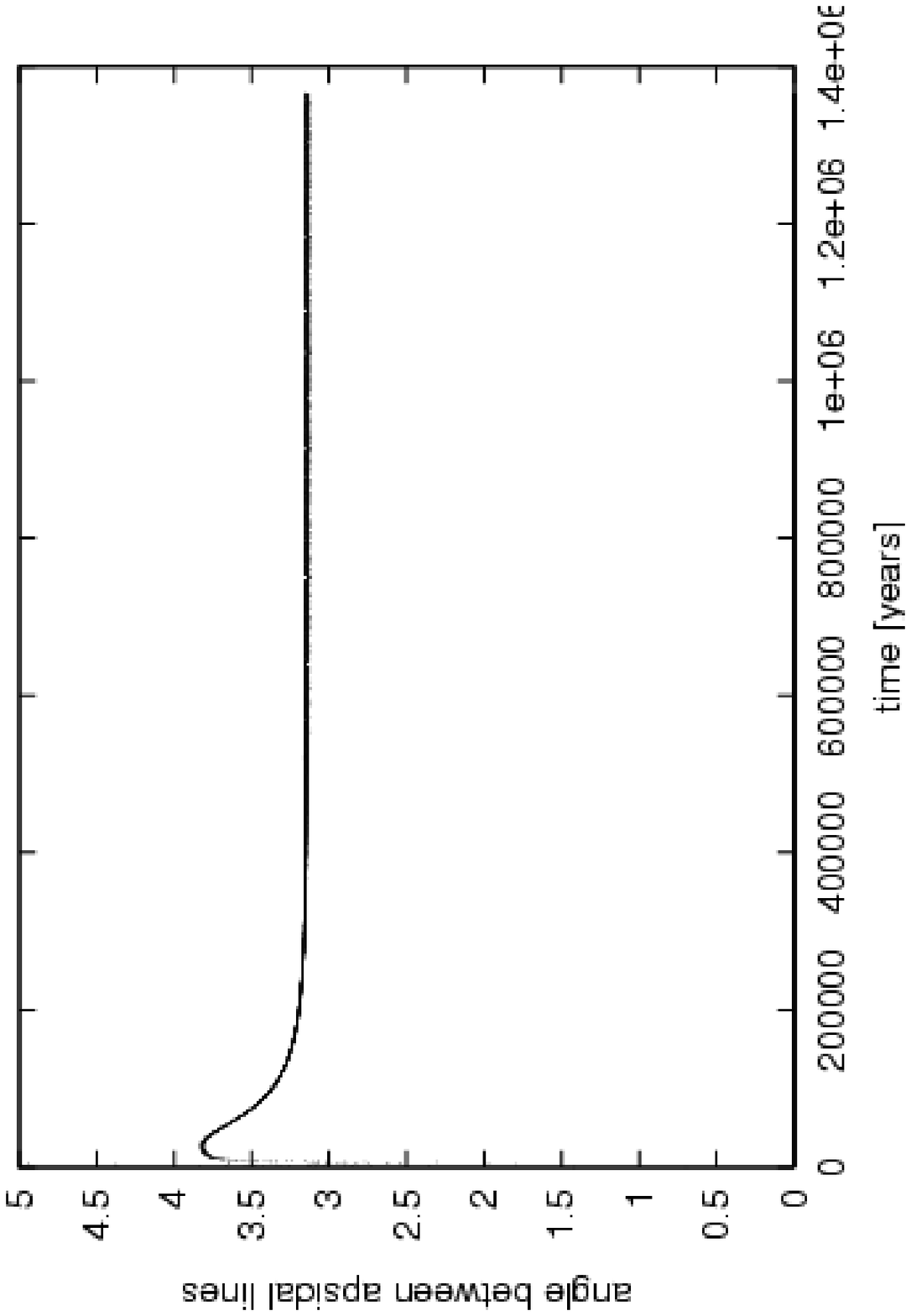}
\includegraphics{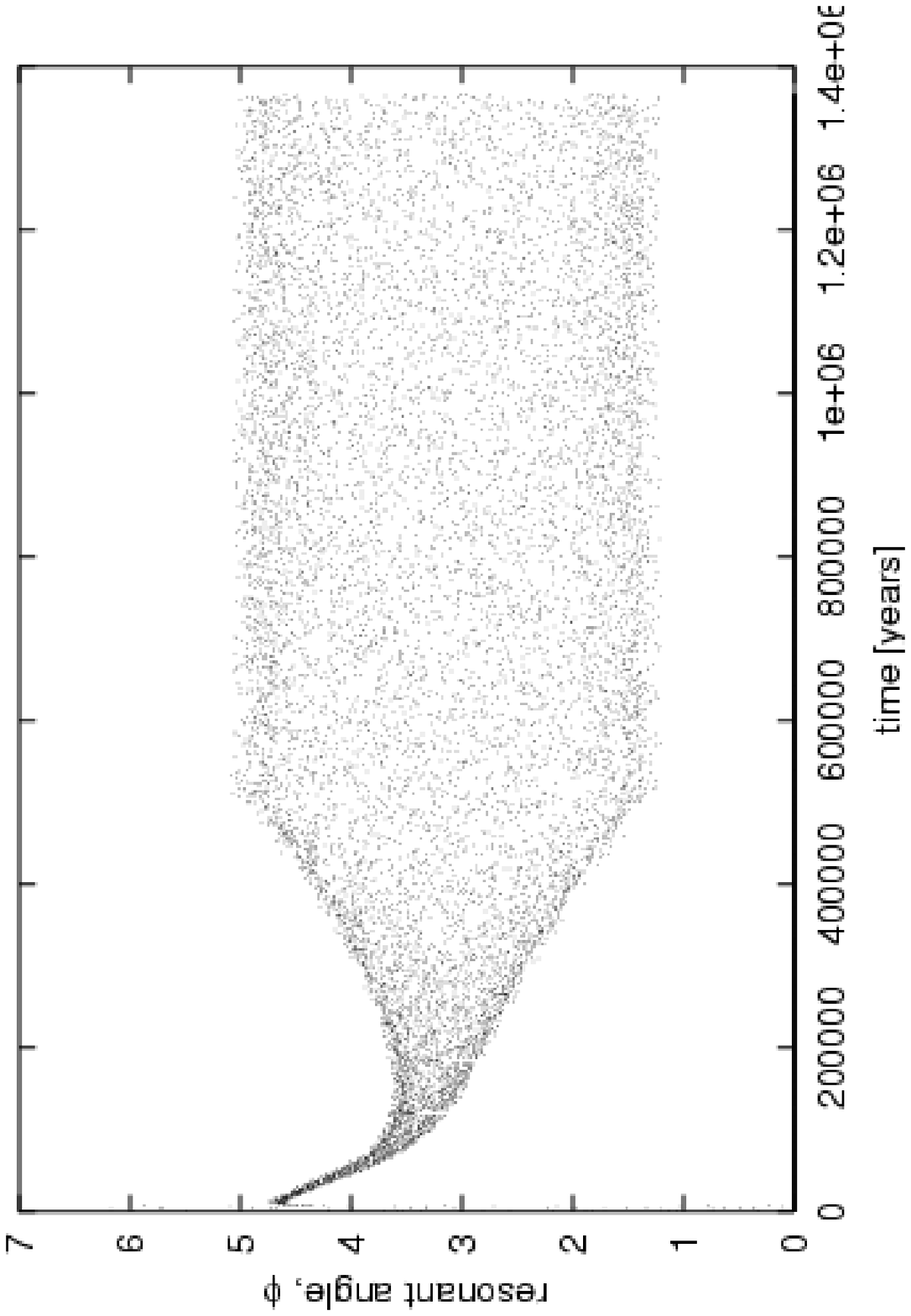}
\caption{\label{fig19}{
The evolution of the  semi-major axes, eccentricities, 
angle between apsidal lines and resonant angle
for two planets with masses, $m_{1} = 4 M_{\oplus}$ and
$m_{2} = 1 M_{\oplus}$ migrating towards
a central star embedded in the disc with $\Sigma =\Sigma _4$ 
obtained with N-body simulations.
}}
\end{minipage}
\end{figure*}

\noindent 
As an example in Figure  \ref{fig18} we show the results of 
the comparison for the case of two planets  with masses  
1 $M_{\oplus}$ and 4 $M_{\oplus}$ respectively embedded 
in a  disc with initial  surface density scaling parameter 
$\Sigma_0 = \Sigma_4$.
The numerical coefficients adopted were   $W_m = 0.3647$  
and $W_c = 0.225.$
These results show good agreement with the hydrodynamic 
results (see Figure \ref{fig3}) and display the same 8:7 commensurability.
The fitted coefficients were also  reasonably close to those expected 
from the analytic disc planet interaction theory, confirming 
both the migration and eccentricity damping times. 
Though the latter were typically $\sim 40 \%$ longer than expected
from the analytic theory.
The evolution was followed using the N-body approach for 
an additional 1.2 $\times 10^{6}$ years.
The two planets remain in the 8:7 resonance during this 
time and there is no indication of any significant changes 
in the  monitored quantities for the last 9 $\times 10^{5}$ 
years. The long term evolution of this
system is shown in Figure  \ref{fig19}. It is interesting 
to note that the equilibrium value of eccentricity of the 
inner planet is around $e_2=0.03$, in accordance with the 
value predicted by the simple analytic
model discussed in Section 2 and derived in the Appendix. 
The angle between apsidal lines is close to 180$^{\circ}$  
the other resonant angles occupy
a wide  band between 80$^{\circ}$ and 287$^{\circ}$.
A similar procedure has been applied to other cases presented 
in Section 3. The values of the numerical coefficients 
$(W_m , W_c)$  used in equations (\ref{MIG}-\ref{CIRC}) obtained 
by comparing  and matching the planet evolution calculated by 
hydrodynamic and N-body codes are given in Table \ref{table2}.

\begin{table*}
 \centering
 \begin{minipage}{140mm}
  \caption{\label{table2} The fitting parameters obtained by 
comparing hydrodynamic and N-body calculations }

\begin{tabular}{|r|r|r|c|c|}
\hline
           &           &            &     &   \\
$r_{1}/r_{2}$      &$m_{1}/m_{2}$ &
$\Sigma_{0}$  &
$W_m$ & $W_c$
 \\
            &           &            &     &   \\
\hline
1.2       &4&$\Sigma_{0.5}$ &0.3134     &0.140 \\
1.2       &4&$\Sigma_{1}$  &0.3419    &0.175 \\
1.2       &4&$\Sigma_{2}$  &0.3611     &0.175 \\
1.2       &4&$\Sigma_{4}$  &0.3647     &0.225 \\
\hline
1.2       &1&$\Sigma_{0.5}$&0.3277     &0.080 \\
1.2       &1&$\Sigma_{4}$  &0.3706    &0.080 \\
\hline
\hline
1.32       &4&$\Sigma_{0.5}$&0.3134     &0.190 \\
1.32       &4&$\Sigma_{4}$  &0.3604    &0.210 \\
\hline
1.32       &1&$\Sigma_{1}$  &0.3619      &0.200 \\
\hline
\hline
1.23       &4&$\Sigma_{0.5}$ &0.3277     &0.170 \\
1.23       &4&$\Sigma_{1}$  &0.3131    &0.190 \\
\hline
1.23       &1&$\Sigma_{0.5}$&0.3419     &0.200 \\
1.23       &1&$\Sigma_{4}$  &0.4274     &0.200 \\
\hline

\end{tabular}
\end{minipage}
\end{table*}

In the case of two planets with masses, $m_{1} = 4 M_{\oplus}$ 
and $m_{2} = 1 M_{\oplus}$ migrating towards
a central star embedded in the disc with $\Sigma_0 =\Sigma _2$ 
the situation is less stable. The 7:6
resonance found in the hydrodynamic simulations is maintained 
for around $2\times 10^5$ years, later there is a shift into 
a 10:9 commensurability. Finally at around $6\times 10^5$ years 
scattering occurred. Also the 5:4 commensurability in the lower 
surface density case ($\Sigma_0 =\Sigma _1$) was maintained 
only until $1.2\times 10^5$ years, then there is a transition 
to 6:5 which lasted till  $2.1\times 10^5$ years. Next, the  
planets evolved till $6\times 10^5$ years locked in 7:6 resonance 
and after that continued in 8:7. That was the status of the 
evolution at the end of our calculations at $1.3\times 10^6$ 
years. The same pair of planets embedded in the disc with even 
lower surface density ($\Sigma_0 =\Sigma _{0.5}$) migrated together 
in 5:4 resonance till $1.15\times 10^6$ years, then they moved into 
a 6:5 commensurability. These outcomes are suggestive of the long 
term instability of the high $p$ commensurabilities 
\citep{klg93}.

We have noticed a strong sensitivity of outcomes to the fitting
parameters (especially in the numerical coefficient $W_m.$
This  is symptomatic of chaotic motion. A very small change
in the values of the numerical coefficients given in 
Table \ref{table2} can produced
large variations  in outcomes.
The value  $W_c$ is difficult to fit
for pairs of two equal mass planets because of smaller and 
less regular changes in eccentricities obtained in the 
hydrodynamic simulations of such systems (see for example 
Figure   \ref{fig13}).

\subsection{N-body resonance survey}

We applied the much faster N-body calculations to extend our 
hydrodynamic simulations to a wider range of planet masses and  
disc surface densities.
The results are summarized in Table \ref{table1}.  We allowed 
the   numerical coefficients $W_m$ and $W_c$ in
equations (\ref{MIG}-\ref{CIRC}), when used in the N body 
evolution, to be free parameters that we could choose
to provide the best match between the hydrodynamic and N body 
methods. In this way we could test the applicability of the 
results of \citet{ttw02} and \citet{tw04}  to this problem.

\noindent 
When we chose $W_m = 0.3704$   in equation (\ref{MIG}) 
and $W_c = 0.289$ in equation (\ref{CIRC}) for the case when
the planets have equal masses, ($m_1/m_2)=1$, their initial 
separation in circular orbits in dimensionless units is such 
that $r_1/r_2=1.4$ and they are embedded in a disc  which 
extends from $r= 0.3$ to $r = 1.4,$ 
than the most likely outcome of their orbital
evolution caused by the disc-planet interaction,
in agreement with the hydrodynamic simulations,
is attainment of a 3:2 commensurability.
This is the case  for a wide range of disc surface density 
scalings. For all values of $\Sigma_0$ considered here,
this commensurability was attained. A transition from the 3:2 
commensurability  to the 4:3 commensurability for a given  
mass ratio (bigger than 1) is expected to occur when either 
the surface density scaling  and/or the mass ratio is large 
enough.  The value of $p$ for the resonance which might be 
established accordingly increases.

\noindent 
When $m_1/m_2 >6$ scattering is likely to occur rather than 
stable resonance trapping, especially for higher surface 
densities. This is clearly indicated in Table \ref{table1}.

\section{Conclusions}

We have investigated migration induced resonant capture in a system
of two planets for the most part in the mass range $1 - 4M_{\oplus}$
embedded in a gaseous disc. We also considered a system with equal 
masses of $30M_{\oplus}.$
As the disc was laminar, apart from in the $30M_{\oplus}$ case,
the planets underwent type I migration which 
occurs through the linear response of the disc to perturbation by the planets.
We considered 
disc surface densities in the range $0.5 - 4$ times that expected 
for the minimum mass
solar nebula at $5.2$~AU. Thus migration rates should thus be 
typical of those expected
for standard protoplanetary discs. 

\noindent Our conclusions are based on the results of two surveys 
in which we  studied the 
 evolution of a pair of planets with a range of planetary
masses (1 - 30$M_{\oplus}$), disc surface densities and initial planet 
separations. 
The smaller survey  used  
hydrodynamic simulations and because of computational requirements
was limited to run lengths of $\sim 2\times 10^4$ inner planet orbits.
Similarly initial conditions which started with the planets
too far apart were precluded. A more extensive survey  used a
simplified N-body approach where the influence of the disc 
was modelled through the addition of external torques and  eccentricity 
damping terms 
(see equations (\ref{MIG}) and (\ref{CIRC})).

\noindent The numerical calculations are supplemented with an  analytic model
describing two planets migrating in resonance with arbitrary 
eccentricities. We  found a very good agreement between numerical
and analytical results for an established commensurability.

\noindent Both our hydrodynamic and N-body approaches give clear evidence of
an interesting relation between the mass ratio in a planetary pair
and the type of the resonance which is expected to be established
as a consequence of tidally induced orbital migration.
When both planets have similar masses, provided the disc surface density
at their locations is very similar, the rate of convergent or relative 
migration is small.
This favours attainment of low $p$ commensurabilities. An example of 
this occurred
in the equal $4M_{\oplus}$ case we considered.
There,  the planets became locked in the nearest first order commensurability
lying between them. Thus if they
were assembled with   $ 1.55 > r_1/r_2 > 1.31$
 the migration  led
to  trapping in 3:2 resonance.

\noindent Here, we comment that the very well known example of such
a system, namely the two largest  mass planets orbiting 
a millisecond radio pulsar PSR B1257+12,
are close to a 3:2 commensurability.
The most recent determination of the  masses of these planets 
 gives them as  $4.3\pm 0.2 M_{\oplus}$ and $3.9\pm 0.2 M_{\oplus}$ 
respectively,
\citep{kw03}. The mass ratio is thus  very close to unity and 
according to our 
findings if  migration in a standard quiescent protoplanetary 
gaseous disc is responsible 
for the evolution then a possible
outcome could be attainment of a 3:2 commensurability.
The orbital periods for
both planets are 66.5419 days and 98.2114 days \citep{kw03} which 
places them near 3:2
resonance. Such a resonance could be formed and maintained
through the processes discussed in this paper, with the planets
later moving slightly out of resonance.
More specific detailed modeling  will be presented elsewhere. 

\noindent For more disparate mass ratios, our simulations 
indicate the attainment of
first order commensurabilities with higher $p$ (for example 
4:3, 5:4, 7:6 and 8:7).  The dynamics develops a stochastic character,
and these may be maintained for a different periods of
time before becoming unstable.   Then often a passage from one resonance 
to another takes place.
The short duration of the hydrodynamic simulations precluded extensive
exploration of stochastic behaviour or the long term stability
of resonances.  However, we could investigate these issues employing 
the simplified N-body approach.
 
\noindent In this context we comment on a result we have 
obtained for an 8:7 commensurability, being
the stable end state of the evolution of two planets 
in circular orbit with the initial
separation ratio $r_1/r_2 = 1.2$, and mass ratio  $m_1/m_2 = 4$ 
and the highest
 disc surface density scaling considered here with $\Sigma_0 = \Sigma_4$. 
 The stable
configuration was  maintained
until $1.4 \times 10^6$ orbits at which point our calculations were stopped.
Of course this result should be considered in the context 
of extreme sensitivity
to initial conditions and other parameters. Nonetheless,
the final value of the inner planet eccentricity is in good agreement
with the prediction of our analytic model for established commensurabilities.

\noindent Thus while stochastic behaviour tends to disrupt higher $p$
commensurabilities, the calculations presented here do not enable us
to rule them out entirely.

\noindent Simple N-body calculations allowed us to extend our 
investigations not only
to consider  longer term behaviour but also  a wider range 
of initial conditions.
These calculations strengthen the conclusions derived from the hydrodynamic
simulations.

\noindent Finally, we  comment further on the onset of stochastic behaviour
 which we encountered
both in our hydrodynamic simulations and N-body calculations. The reason
for this phenomenon can be related to resonance overlap. According
to our estimates we might expect isolated resonances in which a  system of
planets in the mass range considered can be locked and migrate together to be such that the integer $p$ in the
expression $p+1 : p$ for a first order commensurability
 is $\ls 8$. For larger $p$, as we can see from Table
\ref{table1} a system is likely to undergo a scattering and exchange
of  orbits of the two planets.  However, stochastic behaviour 
may extend to  values as small as 4.
  Stable commensurabilities with
sufficiently  small $p$ correspond to physically possible planetary 
configurations and
they should be  the centre of further investigation.

\noindent Among all planned future space missions COROT will be the first 
with the capability to observe planets
in the  mass range  relevant to our investigation. Studies of 
commensurabilities have already been applied successfully to analyse 
the motion of the pulsar 
planets, and they proved to be a powerful tool in that context.
They have the  potential to be similarly useful in our search for Earth-like
planets in other systems. Detection of such resonances can also yield useful
 information about orbital migration as a process operating during 
planet formation.

\section*{Acknowledgments}
E.S. would like to express  her gratitude for support
through its PPARC funded visitors grant
and for the provision of
computer facilities                     
to the  Astronomy
Unit,
Queen Mary, University of London.
E.S. acknowledges the hospitality of the Kavli Institute for Theoretical
Physics, University of California, Santa Barbara.
This research was supported in part by the National Science Foundation under
Grant No. PHY99-0794.

\section*{Appendix}

\subsubsection*{A simple model}

The equations of motion for a system consisting of 
$2$ planets and  primary star, modeled as point masses
moving under their mutual gravitational attraction,
are conveniently expressed in Hamiltonian form using 
Jacobi coordinates (see \citet{si75}).
Then the  radius vector ${\bf r}_2,$ of the inner planet 
of reduced  mass $m_2$ is measured from the  primary star 
of mass $M_*$ and that of the  outer planet, ${\bf r }_1,$
of reduced  mass $m_1$ is referred to the  centre of mass 
of the   primary star and inner planet.
The Hamiltonian can be written  correct to second order 
in the planetary masses as
\begin{eqnarray} 
H & = &  {1\over 2} ( m_1 | \dot {\bf r}_1|^2 +m_2| \dot {\bf r}_2|^2)
- {GM_{*1}m_1\over  | {\bf r}_1|} - {GM_{*2}m_2\over  | {\bf r}_2|} 
\nonumber \\
& - &{Gm_{1}m_2\over  | {\bf r}_{12}|}
 +  {Gm_{1}m_2 {\bf r}_1\cdot {\bf r}_2
\over  | {\bf r}_{1}|^3}.
\end{eqnarray}
Here $M_{*1}=M_*+m_1, M_{*2}= M_* + m_2 $ and
$ {\bf r}_{12}= {\bf r}_{2}- {\bf r}_{1}.$
The Hamiltonian can be expressed in terms of the  osculating  
semi-major axes, eccentricities and longitudes of periastron 
$a_i,e_i,\varpi_i, i=1,2$ respectively   as well as the
longitudes $\lambda_i,$  and the time
$t.$ We recall that $\lambda_i = n_i(t-t_{0i}) + \varpi_i,$ 
with $n_i = \sqrt{GM_{*i}/a_i^3}$ being the mean
motion and $t_{0i}$ giving the time of periastron passage.
The energy is given by  $E_i = -Gm_iM_{*i}/(2a_i),$
and the angular momentum $h_i = m_i\sqrt{GM_{*i}a_i(1-e_i^2)}$  
which may  be used  to describe the motion  instead of $a_i$ 
and $e_i.$

\noindent 
The Hamiltonian may quite generally  be expanded in 
a Fourier series involving linear combinations of 
the three angular differences
$\lambda_i - \varpi_i, i=1,2$ and $\varpi_1 -\varpi_2$
(eg. \citet{bc61}).

Near a first order  $p+1 : p $ resonance, we expect that both
$\phi = (p+1)\lambda_1-p\lambda_2-\varpi_1, $ and
$\psi = (p+1)\lambda_1-p\lambda_2-\varpi_2,$ 
will be slowly varying. When  resonances  are non overlapping
so that  the  $p+1 : p $ resonance may  be isolated, in its 
neighborhood, terms in the Fourier expansion involving only 
linear combinations of $\phi$ and $\psi$  as argument are 
expected to produce the largest perturbations, while the remainder,
which will have  rapidly varying argument,
may be averaged out.

\noindent 
The resulting  Hamiltonian $(\propto m_1 m_2)$  may then 
be written in the general form
\be 
H _{12}= -{Gm_1m_2\over a_1}\sum C_{k,l}( a_1/a_2, e_1,e_2
) \cos (k\phi +l\psi), 
\label{Hamil} 
\ee
where in the above and similar summations below, the sum   
ranges over all positive and negative integers $(k,l)$  
and  the dimensionless  coefficients  $C_{k,l}$
depend on $e_1,e_2$ and  the ratio $a_1/a_2$ only.
Here we shall not need to specify these further.
We also  make the inconsequential simplification
of replacing $M_{*i}$ by $M_*.$

\subsubsection*{Basic Equations}

We take the Hamiltonian system derived from (\ref{Hamil}) 
and modify it by incorporating torques and rates of change 
of  energy that act on each of the protoplanets and which 
are presumed to be derived from interaction with the disc.
  
The equations of motion are very similar to those
given in \citet{pa03}. They differ only in that the additional 
forces derived from disc interaction are allowed to act on both 
protoplanets rather than only the outer one in this case.
  \\
${dE_i}/{dt} =-n_i\partial H_{12} / \partial \lambda_i
 -(n_i T_i /\sqrt{1-e_i^2} +D_i),
$ \\
${dh_i}/{dt}=
-{\partial H_{12} / \partial \lambda_i} -
{\partial H_{12} / \partial \varpi_i} -T_i,$ \\
$ {d\lambda_i/dt} = n_i +n_i {\partial  H_{12}/ \partial E_i} + 
{\partial  H_{12} / \partial h_i}, $ \\
${d\varpi_i /dt} ={\partial  H_{12} / \partial h_i}, $

The  external disc torque acting on  the  planet $m_i$ is 
$-T_i .$  Associated with this, we   remove  orbital energy 
at a rate $n_iT_i/\sqrt{1-e_i^2}$ from $m_i$ which would
correspond to the action of a disc density perturbation
rotating with a pattern speed $n_i /\sqrt{1-e_i^2}$ 
(see also \citet{spn01,np02}).
We include an additional energy loss rate $D_i$ for $m_i$  
which  leads to an orbital circularization time for $m_i$ 
given by 
\be 
t_{ci}  ={ n_im_ie_i^2\sqrt{GM_* a_i} \over D_i(1-e_i^2)} .
\ee
This circularization time is such that if the the other 
planet was absent such that $m_i$  was affected only by 
the central star and disc, $e_i$ would decay according to
\be 
{d e_i \over dt} = {-e_i \over t_{ci}}. 
\ee
We presume that $T_i,$ and $D_i$ are given as functions
of the orbital parameters $a_i, e_i $ associated with $m_i.$
In addition we suppose that they lead to orbital evolution 
on a timescale sufficiently long compared to other effects 
that they may always be considered to  act as small 
perturbations to the equations of motion.   However,  
we do not assume an expansion in terms of small eccentricities.

We thus obtain to  lowest order in the  perturbing masses.
\begin{eqnarray} 
{dn_1\over dt} & = 
& {3(p+1)n_1^2 m_2 \over M_*  } \sum C_{k,l}(k+l)\sin (k\phi +l\psi)
\nonumber \\
& + &{3 n_1 a_1 \over G M_*m_1 }
\left[{n_1 T_1\over\sqrt{1-e_1^2}} +D_1 \right]
\label{first}
\end{eqnarray}

\begin{eqnarray} 
{dn_2\over dt} &  = 
& -{3pn_2^2 m_1 a_2\over M_* a_1 }\sum C_{k,l}(k+l)\sin (k\phi +l\psi)
\nonumber \\
& + &{3 n_2 a_2 \over G M_*m_2 }
\left[{n_2 T_2\over\sqrt{1-e_2^2}} +D_2 \right]
\label{first1}
\end{eqnarray}

\begin{eqnarray}
{de_1\over dt} = -{e_1\over t_{c1}} -
{m_2 n_1 \sqrt{1-e_1^2} \over  e_1 M_* } 
\times  \ \ \ \ \ \ \ \ \ \ \ \ \ \ \ \ \ \ \ \ \ \ \ \ \ \ \ \ \ \ \ \ \  
\nonumber \\
\sum C_{k,l}\sin (k\phi +l\psi) 
\left(k-(p+1)(k+l)(1-\sqrt{1-e_1^2})\right)
\label{first2} 
\end{eqnarray}

\begin{eqnarray} {de_2\over dt}  =  -{e_2\over t_{c2}} -
{m_1 a_2  n_2 \sqrt{1-e_2^2} \over a_1 e_2 M_* } 
\times  \ \ \ \ \ \ \ \ \ \ \ \ \ \ \ \ \ \ \ \
\nonumber \\
\sum C_{k,l}\sin (k\phi +l\psi)
\left(l+p(k+l)(1-\sqrt{1-e_2^2})\right)   
\label{first3}
\end{eqnarray}

\be 
{d\phi \over dt} = (p+1)n_1- pn_2 -
\sum  (D_{k,l} +E_{k,l})\cos(k\phi +l\psi).
\label{last2}
\ee

\be 
{d\psi \over dt} =  (p+1)n_1- pn_2 -
\sum ( D_{k,l} + F_{k,l} )\cos(k\phi +l\psi).
\label{last}
\ee
Here
\begin{eqnarray} 
D_{k,l} &=& {2(p+1)n_1a_1^2m_2\over M_*}{\partial \over \partial a_1}
\left(  C_{k,l}/a_1 \right)
\nonumber \\
&-& {2pn_2a_2^2m_1 \over M_*}{\partial \over \partial a_2}
\left(  C_{k,l}/a_1 \right),
\end{eqnarray}

\begin{eqnarray} 
E_{k,l} & = & {n_1m_2 ((p+1)(1 -e_1^2)-p\sqrt{1-e_1^2}) \over e_1 M_*}
{\partial  C_{k,l}\over \partial e_1}
\nonumber \\
&+& {pn_2a_2m_1(\sqrt{1-e_2^2}-1+e_2^2) \over a_1 e_2 M_*}
{\partial  C_{k,l} \over \partial e_2} 
\end{eqnarray}
and
\begin{eqnarray} 
F_{k,l} & = & {(p+1)n_1m_2 (1 -e_1^2-\sqrt{1-e_1^2}) \over e_1 M_*}
{\partial  C_{k,l}\over \partial e_1} \nonumber \\
& +&  {n_2a_2m_1((p+1)\sqrt{1-e_2^2}-p(1-e_2^2)) \over a_1 e_2 M_*}
{\partial  C_{k,l} \over \partial e_2}.
\end{eqnarray}
We further note that $\phi - \psi = \varpi_2 - \varpi_1 $
is the  angle between the two  apsidal lines of the two 
planetary orbits.

\subsubsection*{Time Independent Solutions}
When no  external disc torques or dissipation act ($T_i = D_i =0$)
time independent or stationary  solutions may  occur for which 
$\psi,$ $\phi$ and  each of $n_1, n_2, e_1, e_2$ are constant.
In principle any stationary  values of $\psi = \psi_0$ and
$\phi = \phi_0 $ might be possible.  Obvious possibilities,  
as also indicated in our numerical
simulations, are values of  $\phi_0$ and $\psi_0$
that are either $0,$ or $\pi.$
In that case it follows directly from equations 
(\ref{first}-\ref{first3}) with $T_i = D_i = 0$
that there are stationary solutions with  $n_1,n_2,e_1,e_2$   
being constant.
In other cases, consideration of equations (\ref{first} - \ref{last})
indicates that we should regard
$\phi_0$ and $\psi_0$
as functions of  $e_1,e_2$ and the ratio $a_2/a_1$.

\noindent A relation between  $e_1,e_2$ and the ratio $a_2/a_1$ for 
steady solutions  follows  by
subtracting equations (\ref{last2}) and (\ref{last})
in the form
\be
\sum (E_{k,l}\cos(k\phi_0 +l\psi_0)
= \sum  F_{k,l}\cos(k\phi_0 +l\psi_0) =  S_p.
\label{lst}
\ee
This condition in fact  matches the precession rates of the 
orbits of the two planets such that they maintain a fixed 
orientation. It then additionally follows from equations 
(\ref{last2}) and (\ref{last}) that
\be 
(p+1)n_1- p n_2 =
\sum  D_{k,l}\cos(k\phi_0 +l\psi_0) + S_p.
\label{lst1}
\ee
This gives a further relation between $e_1,e_2$ and the 
ratio $a_2/a_1.$ We could for example specify $e_1$ and then 
both $e_2$ and $a_2/a_1$ would be specified but in any case 
the  quantity $((p+1)n_1)/(p n_2) - 1,$ supposing $m_1$ and 
$m_2$ to be comparable, is at least  of order  of smallness 
of the mass ratio $m_1/M_* \ll 1.$ Thus in a perturbation  
quantity that is already first order in the perturbing
masses we may set $a_2/a_1 = ((p+1)/p)^{-2/3}$ which gives 
the condition for a  $p+1:p$ commensurability.

\subsubsection*{Time Dependent Solutions with Migration 
and Circularization}

We now generalize the above solution  to incorporate 
the effects of small non zero disc torques $T_i$ and 
dissipation $D_i.$ In this case the two planets
migrate inwards locked in resonance with $n_1/n_2$ 
maintained nearly equal to $p/(p+1).$
In the absence of any tidal effects which could act to
circularize the orbits the eccentricities are found to 
increase monotonically with time.
We note that equations (\ref{first}-\ref{first1}) define  inward  
migration timescales or $e$ folding rates for $n_i$ as
$t_{migi}= GM_* m_i \sqrt{1-e_i^2}/(3 T_i a_in_i).$
We  suppose  that in order to accommodate
small values of $T_i$ and $D_i,$  the angles  
$\psi_1 = \psi - \psi_0$ and  $\phi_1 = \phi - \phi_0,$ 
which measure the departures from the  values appropriate 
to the steady state solution, take on non zero  values 
of small magnitude which can also change
very slowly with time. These  magnitudes are presumed
sufficiently small that we may  employ a first order Taylor 
expansion so that the perturbation to $\sin (k\phi +l\psi)$ 
is $ (k\phi_1 +l\psi_1)\cos (k\phi_0 +l\psi_0)$
and similarly the perturbation to  $\cos (k\phi +l\psi)$ 
is $-(k\phi_1 +l\psi_1)\sin(k\phi_0 +l\psi_0).$

\noindent When the migration time is very long compared to
any other time scale in the problem, 
  equations (\ref{last2}) and (\ref{last}) are of the 
same form as in the time independent case, apart from small corrections
proportional to the migration rate which we neglect.
These then lead to the same conclusions as in the time 
independent case. Accordingly we conclude that there are 
relationships between $e_1,e_2$ and $a_2/a_1$ which specify 
the latter two once $e_1$ is specified but that always 
$a_2/a_1$ satisfies the condition for $(p+1):p$ resonance 
with a correction of order of the perturbing masses.

Equations (\ref{first}-\ref{first3}) then become
\begin{eqnarray} 
{dn_1\over dt} & = & {3(p+1)n_1^2 m_2 \over M_* } 
\sum {\cal C}_{k,l}(k+l)(k\phi_1 +l\psi_1)
\nonumber \\
& + &{3 n_1 a_1 \over G M_*m_1}
\left[{n_1 T_1\over \sqrt{1-e_1^2}} + D_1\right]
\label{fist}
\end{eqnarray}

\begin{eqnarray} 
{dn_2\over dt} & = & -{3pn_2^2 m_1 a_2\over M_* a_1 }
\sum {\cal C} _{k,l}(k+l)(k\phi_1 +l\psi_1)
\nonumber \\
& + &{3 n_2 a_2 \over G M_*m_2}\left[{n_2 T_2\over \sqrt{1-e_2^2}} + D_2\right]
\label{fist1}
\end{eqnarray}

\begin{eqnarray} 
{de_1\over dt} =  -{e_1\over t_{c1}} -{m_2 n_1 \sqrt{1-e_1^2} \over  e_1 M_* }
 \ \ \ \ \ \ \ \ \ \ \ \ \ \ \ \ \ \ \ \ \ \ \ \ \ \ \ \ \ \ \ \ \ \ \ \  
\nonumber \\ 
\times \sum {\cal C}_{k,l}(k\phi_1 +l\psi_1)
\left(k-(p+1)(k+l)(1-\sqrt{1-e_1^2})\right) 
\label{fist2}
\end{eqnarray}

\begin{eqnarray} 
{de_2\over dt}= -{e_2\over t_{c2}} -
{m_1 a_2  n_2 \sqrt{1-e_2^2} \over a_1 e_2 M_* } 
\ \ \ \ \ \ \ \ \ \ \ \ \ \ \ \ \ \ \ \ \ \ \ \ \ \ \ \ \ \  
\nonumber \\ 
\times\sum {\cal C}_{k,l} (k\phi_1 +l\psi_1)
\left(l+p(k+l)(1-\sqrt{1-e_2^2})\right).
\label{fist3}
\end{eqnarray}
Here ${\cal C}_{k,l} = C_{k,l} \cos(k\phi_0+l\psi_0).$

\noindent 
Taking the ratio of (\ref{fist}) and (\ref{fist1}) gives
\begin{eqnarray} 
{dn_1\over dn_2}  = -
\ \ \ \ \ \ \ \ \ \ \ \ \ \ \ \ \ \ \ \ \ \ \ \ \ \ \ \ \ \ \ 
\ \ \ \ \ \ \ \ \ \ \ \ \ \ \ \ \ \ \ \ \ \ \ \ \ \ \ \  
\nonumber \\ 
{(p+1)n_1^2 m_2  a_1 \sum {\cal H}_{k,l}(k+l)
+{n_1 a_1^2 \over G m_1}
\left({n_1 T_1\over \sqrt{1-e_1^2}} + D_1\right)\over
pn_2^2 m_1 a_2 \sum {\cal H}_{k,l}(k+l)-{n_2 a_1 a_2 \over  G m_2}
\left({n_2 T_2\over \sqrt{1-e_2^2}} + D_2\right) }, 
\label{rat}
\end{eqnarray}
with ${\cal{H}}_{k,l} = {\cal C}_{k,l}(k\phi_1 +l\psi_1).$

\noindent 
The ratio of (\ref{fist2}) and (\ref{fist3}) 
gives another  relation of the form 
\begin{eqnarray} 
{m_2  e_2 a_1 n_1  \sqrt{1-e_1^2} \over  
m_1 e_1  a_2  n_2 \sqrt{1-e_2^2}}{de_2\over de_1} =    
\ \ \ \ \ \ \ \  \ \ \ \ \ \ \ \ \ \ \ \ \ \ \ 
\ \ \ \ \ \ \ \ \ \ \ \ \ \ \ \ \  \ \ \ 
\nonumber \\
{ \sum {\cal{H}}_{k,l}
(l+p(k+l)(1-\sqrt{1-e_2^2} )) + {e_2^2  M_*a_1 
\over \sqrt{1-e_2^2} m_1 a_2  n_2 t_{c2}} \over
\sum  {\cal{H}}_{k,l} (k-(p+1)(k+l)(1-\sqrt{1-e_1^2})) 
+ {e_1^2  M_* \over  \sqrt{1-e_1^2} m_2  n_1 t_{c1}}} 
\label{rat1},
\end{eqnarray}

\noindent 
Using $pn_2 =(p+1)n_1,$ equation (\ref{rat}) gives
\begin{eqnarray}  
\phi_1 \sum k{\cal C}_{k,l}(k+l) +\psi_1 \sum l{\cal C} _{k,l}(k+l)
=  \ \ \ \ \ \ \ \ \ \ \ \ \ 
\nonumber \\
{-(p+1)n_1 a_1^2 m_2 
\left({n_1 T_1 \over \sqrt{1-e_1^2}} + D_1\right)+ 
p m_1n_2a_1 a_2 \left({n_2 T_2\over  \sqrt{1-e_2^2}}  + D_2\right)
\over G m_1 m_2 \left((p+1)^2n_1^2 m_2  a_1 + p^2n_2^2 m_1 a_2\right)} 
\label{rat2}
\end{eqnarray}

\noindent 
Given that we may regard $e_2$ and $a_2/a_1$ as functions 
of $e_1,$ we may also  regard  equations (\ref{rat}) and 
(\ref{rat1}) as determining $\phi $ and $\psi $ as functions 
of $ a_1,$ and $e_1.$
Using this information, equations (\ref{fist}) and (\ref{fist2}) 
may be used to determine $n_1$ and $e_1$ as functions of time 
so completing the solution.

\noindent One readily finds that these satisfy
\begin{eqnarray} 
{1\over n_1}{dn_1\over dt} = 3  
\ \ \ \ \ \ \ \ \ \ \ \ \ \ \ \ \ \ \  \ \ \ \ \ 
\ \ \ \ \ \ \ \ \ \ \ \ \ \ \ \ \ \ \ \ \ \ \ \ \  
\nonumber \\
\times{   m_1a_2\left({1\over 3 t_{mig1}}+
 {e_1^2\over (1-e_1^2)t_{c1}}\right)
+a_1m_2\left({1\over 3t_{mig2}}+ {e_2^2\over (1-e_2^2)t_{c2}}\right)
\over \left[m_2a_1 +  m_1 a_2 \right]}
\label{fe} 
\end{eqnarray}

\begin{eqnarray}  
e_1{de_1\over dt} = - {e_1^2\over \Lambda t_{c1}}
- {e_1e_2(\Lambda -1)\over \Lambda t_{c2}}{de_1\over de_2} + 
\ \ \ \ \ \ \ \ \  \ \ \ \ \ \ \ \
\nonumber \\
{ \left(\left({1\over 3 t_{mig1}}+ {e_1^2\over (1-e_1^2)t_{c1}}\right)
-\left({1\over 3t_{mig2}}+ {e_2^2\over (1-e_2^2)t_{c2}}\right)\right)
{\cal T}
\over \Lambda \left[m_2a_1 +  m_1 a_2 \right]}
\label{fe1} 
\end{eqnarray}
where
\be 
\Lambda = 1 + {e_2\over e_1} {de_2\over de_1}
\left[{\sqrt{1-e_1^2} m_2n_1a_1 \over \sqrt{1-e_2^2} m_1n_2a_2}\right ]
\label{fex}
\ee
and
\begin{eqnarray} 
{\cal T} & = &   
 {m_2a_1\sqrt{1-e_1^2}\over p+1}  \nonumber \\
 & & \times \left(1- (p+1)(1 - \sqrt{1-e_1^2}) -
 p\sqrt{1-e_2^2}+p\right).  
\end{eqnarray}
Equation (\ref{fe}) implies that $n_1$ always
increases with time corresponding to inward migration.
Notably this equation does not depend on the order 
of the resonance.

\noindent 
Equation (\ref{fe1}) governs the evolution of the 
eccentricities. It  indicates that  when $e_1 = e_2 = 0,$
given that the derivative $de_2/de_1$ remains finite, 
the eccentricities will increase with time 
provided that $1/ t_{mig1}  > 1/ t_{mig2}.$
This is just the condition for the migration of the uncoupled 
two planet system  
to be convergent, in the sense that the relative separation should
decrease. When the circularization times 
are finite, there is the possibility of  steady state 
eccentricities found by equating the right hand side 
of (\ref{fe1}) to zero.
In the limit that they are small, the equilibrium eccentricities  
must thus satisfy:
\begin{eqnarray}
 {e_1^2\over t_{c1}} + {e_2^2\over t_{c2}}{m_2n_1a_1
\over m_1n_2a_2} 
 -\left({e_1^2\over t_{c1}}-{e_2^2\over t_{c2}}\right)f
=& & \nonumber \\
\left({1\over t_{mig1}}-{1\over t_{mig2}}\right){f\over 3}, & & 
\label{ejcons}
\end{eqnarray}
where  $ f =m_2a_1/((p+1)(m_2a_1+m_1a_2)).$
Note  that when the migration and circularization
times  $t_{mig}$ and $t_c$ scale in the same way 
as the mean motions at an orbital location, the 
equilibrium eccentricities are  
independent of orbital location.

\bsp

\label{lastpage}
\end{document}